\documentclass[prd,12pt,nofootinbib,preprint]{revtex4}
\usepackage[T1]{fontenc}
\usepackage[utf8x]{inputenc}
\usepackage{amsmath,amssymb}
\usepackage{epsfig}
\usepackage{url}
\usepackage{subfigure}
\usepackage{graphicx}
\usepackage{grffile}
\usepackage[usenames,dvipsnames]{color}
\usepackage{slashed}
\usepackage{array}
\usepackage{multirow}
\usepackage{xcolor}
\usepackage[colorlinks,citecolor=blue]{hyperref}
\usepackage{color}
\usepackage{cancel}
\usepackage{gensymb}
\usepackage{soul}

\def \met{\not \! E_T }

\def\a {\alpha}
\def\b {\beta}

\def\l {\lambda}

\def\bar {\overline}

\def\be {\begin{equation}}
\def\ee {\end{equation}}
\def\beq {\begin{equation}}
\def\eeq {\end{equation}}
\def\bea {\begin{eqnarray}}
\def\eea {\end{eqnarray}}

\newcommand{\besub}{\begin{subequations}}
\newcommand{\eesub}{\end{subequations}}

\def\beq{\begin{equation}}
\def\eeq{\end{equation}}
\def\barr{\begin{array}}
\def\earr{\end{array}}

\def\a{\alpha}

\def\b{\beta}

\def\l{\lambda}

\def\s{\sigma}

\def\q2 {q^2}

\def\bt{\begin{table}}
\def\et{\end{table}}

\def\mET{E_T \hspace{-1.0em}/\;\:}

\begin{document}

\title{Muon $g - 2$ in a 2HDM assisted by inert scalars: probing
at the ILC}

\author{Nabarun Chakrabarty}
\email{nabarunc@iitk.ac.in}
\affiliation{Department of Physics, Indian Institute of Technology Kanpur, Kanpur, Uttar Pradesh-208016, India} 

\author{Indrani Chakraborty}
\email{indranic@iitk.ac.in, indrani300888@gmail.com}
\affiliation{Department of Physics, Indian Institute of Technology Kanpur, Kanpur, Uttar Pradesh-208016, India}

\begin{abstract}

A Two-Higgs doublet model (2HDM) can predict the observed muon $g-2$ for an appropriately light pseudoscalar that now faces tight constraints. However, It was shown in past that augmenting the 2HDM by an additional inert doublet can 
lead to an explanation to the muon $g-2$ anomaly for a much heavier pseudoscalar. 
In this study, we probe such a framework at the proposed International Linear Collider (ILC) using beam polarization for $\sqrt{s}$ = 1 TeV. Using multivariate techniques, we analyse the signals $e^+ e^- \rightarrow \tau^+ \tau^- +$ missing transverse energy  and $e^+ e^- \rightarrow \mu^+ \mu^- + $ missing transverse energy in the lepton- and muon-specific versions of the framework respectively.
Our analysis reveals that the $e^+ e^-$ machine operating at a 3000 fb$^{-1}$ luminosity predicts a 5$\sigma$ discovery of a pseudoscalar as heavy as 400 GeV. Comparing with the previous study, it is concluded that the ILC is a much more potent machine than the LHC in this regard.

\end{abstract} 
\maketitle
\section{Introduction}
\label{intro}

A point of contention within the Standard Model (SM) is its failure to account for the observed value of the muon anomalous magnetic moment \cite{Peskin:1995ev,Blum:2013xva,RBC:2018dos,Keshavarzi:2018mgv,Davier:2019can,Aoyama:2020ynm,Colangelo:2018mtw,Hoferichter:2019mqg,Melnikov:2003xd,Hoferichter:2018kwz,Blum:2019ugy,ParticleDataGroup:2020ssz}. A combination of the results reported by BNL \cite{Muong-2:2006rrc} and FNAL \cite{Muong-2:2021ojo} shows that the discrepancy is
\bea
\Delta a_\mu \equiv a^{\text{exp}}_\mu - a^{\text{SM}}_\mu = 
251(59) \times 10^{-11}.
\eea  
A Two-Higgs doublet model (2HDM) featuring the flavour conserving Yukawa interactions of the Type-X texture has long been known to address this anomaly~\cite{Branco:2011iw,Broggio:2014mna,Cao:2009as,Wang:2014sda,Han:2015yys,Ilisie:2015tra,Abe:2015oca,Crivellin:2015hha,Chun:2016hzs,Cherchiglia:2016eui,Han:2018znu,Chun:2019oix,Dey:2021pyn,Chowdhury:2017aav,Wang:2018hnw,Chun:2017yob,Chun:2018vsn}. And this happens for a high tan$\beta$ ($\gtrsim$ 20) and a low pseudoscalar mass ($\lesssim$ 70 GeV). 
However, lepton universality constraints tend to rule out tan$\beta \gtrsim 50$~\cite{Chun:2016hzs}. Also, the Large Hadron Collider (LHC) search $h_{125} \to A A \to 4\tau,2\tau2\mu$~\cite{CMS:2018qvj} stringently constrains the Type-X 2HDM parameter region for $M_A < M_h/2$ = 62.5 GeV. In all, the model parameter space favouring the observed muon $g-2$ is driven to a corner with such constraints. 

Reference \cite{Chakrabarty:2021ztf} proposed augmenting the Type-X 2HDM with another \emph{inert} scalar doublet with the aim to enlarge the parameter region compatible with muon $g-2$. The resulting framework was dubbed as the (2+1)HDM. The scalars emerging from the inert multiplet were shown to induce sizeable contributions to muon $g-2$ through 2-loop Barr-Zee (BZ) amplitudes. And large amplitudes were reported to be consistent with various constraints from theory and experiments such as perturbative unitarity, Higgs signal strengths and dark matter direct detection. It was established that the region in the $M_A-\tan\beta$ plane corroborating the observed muon $g-2$ and other constraints enlarges significantly upon the introduction of the inert scalar doublet. In this study, we perform a similar exercise for the \emph{muon-specific} variant \cite{Johansen:2015nxa, Kajiyama:2013sza, Abe:2017jqo} of the (2+1)HDM, i.e, where the muons have enhanced Yukawa couplings while the taus have suppressed ones. We dub the (2+1)HDM with the Type-X texture as introduced in \cite{Chakrabarty:2021ztf} as a \emph{lepton-specific} (2+1)HDM for clarity.

What phenomenologically sets apart the (2+1)HDM from the Type-X 2HDM from the perspective of muon $g-2$ is the possibility of having a heavy pseudoscalar. And this can lead to interesting collider signatures through the  $A \to \tau^+ \tau^-$ decay.
A heavier $A$ would accordingly lead to more boosted $\tau^+\tau^-$ pair. Also, a final state involving the DM candidate $\eta_R$ can have a very different spectrum of missing transverse energy ($\mET$) compared the SM or even the Type-X 2HDM. With such considerations, \cite{Chakrabarty:2021ztf} probed the signal
$p p \to \eta_R \eta_I \to \eta_R \eta_R A \to \tau^+ \tau^- + \mET$ at the 14 TeV LHC.
Fully hadronic decays of the $\tau^+\tau^-$ pair were looked at. Encouraged by the ensuing results, in this work, we take up to probe the $e^+ e^- \to \eta_R \eta_I \to \eta_R \eta_R A \to \tau^+ \tau^- + \mET$ signal at the proposed International Linear Collider (ILC) operating at $\sqrt{s}$ = 1 TeV. We aim to explore all three possibilities: (i) both $\tau$ decay leptonically, (ii) one $\tau$ decays leptonically and the other hadronically, (iii) both $\tau$ decay hadronically. An $e^+ e^-$ collider is expected to offer a much higher sensitivity in probing a hadronic final state than what does the LHC given the tiny hadronic background in the former compared to in the latter. As for the muon-specific (2+1)HDM, the $A \to \mu^+ \mu^-$ decay mode can have a sizeable branching ratio. Therefore, the channel we choose to investigate for this case is $e^+ e^- \to \eta_R \eta_I \to \eta_R \eta_R A \to \mu^+ \mu^- + \mET$. We plan to analyse the signals and the backgrounds using sophisticated multivariate techniques.

The study is structured as follows. We describe the details of the framework in section \ref{model}. The relevant theoretical and experimental constraints are discussed briefly in  section \ref{muong-2}. The same section also outlines explanation of the muon $g-2$ anomaly in the present setup. In section \ref{collider}, we present exhaustive analyses of the aforementioned signals using multivariate techniques. Finally, we summarize and conclude in section \ref{conclusions}.

\section{Theoretical framework: The (2+1)HDM}
\label{model}

The (2+1)HDM \cite{Chakrabarty:2021ztf} is an extension of the 2HDM, comprising the scalar doublets $\phi_1$ and $\phi_2$, by an additional scalar doublet $\eta$. A $\mathbb{Z}_2$ symmetry is imposed under which $(\phi_1, \phi_2) \to (\phi_1, \phi_2)$, while $\eta \to -\eta$. We quote below the most general scalar potential compatible with the gauge and discrete symmetries,
\bea
V (\phi_1,\phi_2,\eta) &=& V_2^{\{\phi_1,\phi_2,\eta\}} + V_4^{\{\phi_1,\phi_2\}} + V_4^{\{\phi_1,\phi_2,\eta\}}, 
\label{eqn:pot}
\eea
with 
\bea
V_2^{\{\phi_1,\phi_2,\eta\}} &=& - m_{11}^2 |\phi_1|^2 - m_{22}^2 |\phi_2|^2
+ m_{12}^2 (\phi_1^\dagger \phi_2 + \text{h.c.})
+ \mu^2 |\eta|^2, \nonumber
\eea
\bea
V_4^{\{\phi_1,\phi_2\}} &=& \frac{\l_1}{2}|\phi_1|^4 + \frac{\l_2}{2}|\phi_2|^4
 + \l_3 |\phi_1|^2 |\phi_2|^2 
+ \l_4 |\phi_1^\dagger \phi_2|^2 + \frac{\l_5}{2} [(\phi_1^\dagger \phi_2)^2 + h.c.] \nonumber \\
&&
+ \l_6 [(\phi_1^\dagger \phi_1)(\phi_1^\dagger \phi_2) + h.c.]
+ \l_7 [(\phi_2^\dagger \phi_2)(\phi_1^\dagger \phi_2) + h.c.], \nonumber
\eea
\bea
V_4^{\{\phi_1,\phi_2,\eta\}} &=& \frac{\l^\prime}{2}|\eta|^4
+ \sum_{i=1,2} \bigg\{  \nu_i |\phi_i|^2 |\eta|^2
+ \omega_i |\phi_1^\dagger \eta|^2
+ \Big[ \frac{\kappa_i}{2} (\phi_i^\dagger \eta)^2 + h.c. \Big] \bigg\} 
\nonumber \\
&&
+ \Big[ \sigma_1 |\eta|^2 \phi_1^\dagger \phi_2 + \sigma_2 \phi_1^\dagger \eta \eta^\dagger \phi_2
+ \sigma_3 \phi_1^\dagger \eta
\phi_2^\dagger \eta + h.c. \Big]. 
\label{eqn:pot1}
\eea

Here the subscripts in Eq.(\ref{eqn:pot}) denote the dimensions of the respective terms while the superscripts denote the scalar doublets involved. All parameters in Eq.(\ref{eqn:pot}) are taken to be real to avoid CP-violation. The particle content of the scalar doublets after electroweak symmetry breaking (EWSB) can be expressed as
\bea
\phi_i = \begin{pmatrix}
\phi_i^+ \\
\frac{1}{\sqrt{2}} (v_i + h_i + i z_i)
\end{pmatrix} , (i = 1,2) ,~~~ 
\eta = \begin{pmatrix}
\eta^+ \\
\frac{1}{\sqrt{2}} (\eta_R + i \eta_I)
\end{pmatrix}.
\eea
Here $v_i$ denotes the vacuum expectation value (VEV) of doublet $\phi_i$ with $i=1,2$ and one defines tan$\beta = \frac{v_2}{v_1}$. 
The scalar doublet $\eta$ is therefore \emph{inert}, its component scalars do not mix with those coming from $\phi_1$ and $\phi_2$ on account of the $\mathbb{Z}_2$ symmetry. It then follows that the physical scalar spectrum from these two doublets is identical to the pure 2HDM. However, we mention for completeness that such a spectrum comprises the CP-even $h,H$, the CP-odd $A$ and one charged Higgs $H^+$. Of these, $h$ is identified with the discovered Higgs having mass 125 GeV. We refer to \cite{Branco:2011iw} for details. On the other hand, the inert sector is composed of three scalars 
$\eta_R, \eta_I$ and $\eta^+$. Their masses in terms of quartic couplings and mixing angles can be found in \cite{Chakrabarty:2021ztf}.

For the Yukawa interactions,  we take the two following cases, i.e., (i) \emph{lepton-specific}~\cite{Branco:2011iw,Broggio:2014mna,Cao:2009as,Wang:2014sda,Han:2015yys,Ilisie:2015tra,Abe:2015oca,Chun:2016hzs,Cherchiglia:2016eui,Chun:2019oix,Dey:2021pyn,Chowdhury:2017aav}: the quarks get their masses from $\phi_2$ while the all the leptons do from $\phi_1$, and, 
(ii) \emph{muon-specific}~\cite{Kajiyama:2013sza,Abe:2017jqo,Johansen:2015nxa}: the quarks and the $e$-,$\tau$-leptons get their masses from $\phi_2$  while the $\mu$-lepton does from $\phi_1$. The lepton-specific case is canonically known as the Type-X 2HDM. The Yukawa Lagrangian in either case can be expressed as
\bea
-\mathcal{L}_Y &=& y_u \bar{Q_L} \tilde{\phi}_2 u_R + y_d \bar{Q_L} \phi_2 d_R + \sum_{\ell=e,\mu,\tau} \Big[ n^1_\ell y_\ell \bar{Q_L} \phi_1 \ell_R + n^2_\ell y_\ell \bar{Q_L} \phi_2 \ell_R \Big] + \text{h.c.} .
\label{Lagrangian-flav}
\eea
Here $y_u, y_d, y_\ell$ are the Yukawa coupling matrices for the up-type quarks, down-type quarks and charged leptons respectively. We have taken the entries of these Yukawa coupling matrices to be real to avoid CP-violation. The integers $n^1_\ell$ and $n^2_\ell$ are tabulated in Table \ref{tab:xi} for the lepton- and muon-specific cases.
We can rewrite the Lagrangian for the leptonic part in Eq.(\ref{Lagrangian-flav}) in terms of the physical scalars as:
\bea
\mathcal{L}^\text{lepton}_Y &=& \sum_{\ell=e,\mu,\tau} \frac{m_\ell}{v} \bigg(\xi_\ell^h h \bar{\ell} \ell + \xi_\ell^H H \bar{\ell} \ell - i \xi_\ell^A A \bar{\ell} \gamma_5 \ell + \Big[ \sqrt{2} \xi^A_\ell H^+ \bar{\nu_\ell} P_R \ell + \text{h.c.} \Big] \bigg).
\label{Lagrangian-mass}
\eea
In the above equation, $m_\ell$ is the mass of the lepton $\ell$, $P_R$ is the projection operator, {\em i.e.} $P_R = \frac{( 1 + \gamma_5)}{2}$. The various $\xi_\ell$ factors are also quoted in Table \ref{tab:xi} for the lepton-specific and muon-specific cases.

\begin{table}
\centering
\begin{tabular}{ | c | c | c | c | c | c | c | c | c | c | c | c | c | c| } 
\hline
& $n^{1}_{e,\tau}$ & $n^{2}_{e,\tau}$ & $n^{1}_{\mu}$ & $n^{2}_{\mu}$ & $\xi^h_e$ & $\xi^h_\mu$ & $\xi^h_\tau$ & $\xi^H_e$ & $\xi^H_\mu$ & $\xi^H_\tau$ & $\xi^A_e$ & $\xi^A_\mu$ & $\xi^A_\tau$ \\ \hline
Lepton-specific & 1 & 0 & 1 & 0 & $-\frac{\text{sin}\a}{\text{cos}\b}$ & 
 $-\frac{\text{sin}\a}{\text{cos}\b}$ &
$-\frac{\text{sin}\a}{\text{cos}\b}$ &
 $\frac{\text{cos}\a}{\text{cos}\b}$ &
 $\frac{\text{cos}\a}{\text{cos}\b}$ &
 $\frac{\text{cos}\a}{\text{cos}\b}$ &
 tan$\beta$ & tan$\beta$ & tan$\beta$ \\ \hline
Muon-specific & 0 & 1 & 0 & 1 & $\frac{\text{cos}\a}{\text{sin}\b}$ &
 $-\frac{\text{sin}\a}{\text{cos}\b}$ &
 $\frac{\text{cos}\a}{\text{sin}\b}$ &
$\frac{\text{sin}\a}{\text{sin}\b}$ &
$\frac{\text{cos}\a}{\text{cos}\b}$ &
 $\frac{\text{sin}\a}{\text{sin}\b}$ &
 $-\cot\beta$ & tan$\beta$ & $-\cot\beta$ \\ \hline
\end{tabular}
\caption{Leptonic scale factors for the lepton- and muon-specific cases.}
\label{tab:xi}
\end{table}


\section{Constraints and the muon $g-2$ anomaly}
\label{muong-2}

We first describe in a nutshell the constraints applicable on this framework. The scalar quartic couplings are subject to the theoretical requirements of perturbativity, unitarity and a bounded-from-below scalar potential. Several crucial restrictions follow from experiments. First, the electroweak oblique parameters $S,T,U$ must lie within their stipulated limits \cite{10.1093/ptep/ptaa104}. Secondly, the framework must pass the Higgs signal strength constraints for various channels. In this study, we adhere to the 2HDM \emph{alignment limit} in which tree-level couplings of $h$ to fermions and gauge bosons become identical to the corresponding SM values. And the only non-trivial signal strength constraint in this limit comes from the $h \to \gamma \gamma$ channel. The oblique parameter and $h \to \gamma \gamma$ signal strength constraints are imposed at 2$\sigma$ in this analysis.

The $\mathbb{Z}_2$ symmetry used in this framework renders the lighter of $\eta_R$ or $\eta_I$ as a DM candidate. We take $\eta_R$ to be the one in this analysis. However, instead of demanding that $\eta_R$ entirely accounts for the observed DM relic density, we allow for DM under-abundance in this scenario. That is, we demand the predicted  relic density of $\eta_R$ should not exceed the latest Planck data at the
2$\sigma$ level that reads $\Omega_{\text{Planck}} h^2 = 0.120 \pm 0.001$ \cite{Aghanim:2018eyx}. The DM relic density is computed in this study by sequentially using the publicly available tools \texttt{LanHEP}~\cite{Semenov:2008jy} and \texttt{micrOMEGAs}~\cite{Belanger:2018ccd}.
In addition, upper limits are put on DM-nucleon scattering rates by direct detection experiments with the most stringent bound for sub-TeV DM comes from XENON-1T \cite{Aprile:2018dbl}.

We shall not discuss detailed calculation of $\Delta a_\mu$ here. Elaborate computation and description regarding this can be found in \cite{Chakrabarty:2021ztf}. For convenience, we have provided the mathematical expressions and corresponding Feynman diagrams of one-loop and two-loop Barr-Zee contributions to $\Delta a_\mu$ coming from BSM scalars occurring in the loop in Appendix \ref{App:A}. The BSM contributions to $\Delta a_\mu$ can be divided into two parts : (i) contribution coming from pure 2HDM, (ii) contribution coming from the inert sector.

\begin{figure}
\centering
\subfigure[]{
\includegraphics[scale=0.48]{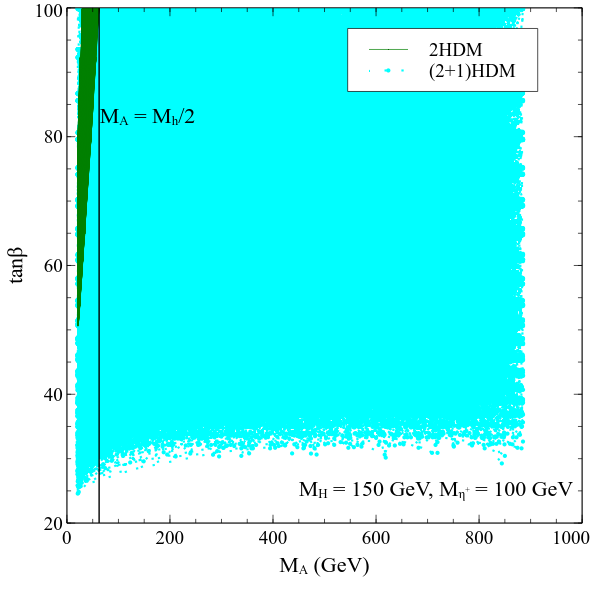}}
\subfigure[]{
\includegraphics[scale=0.48]{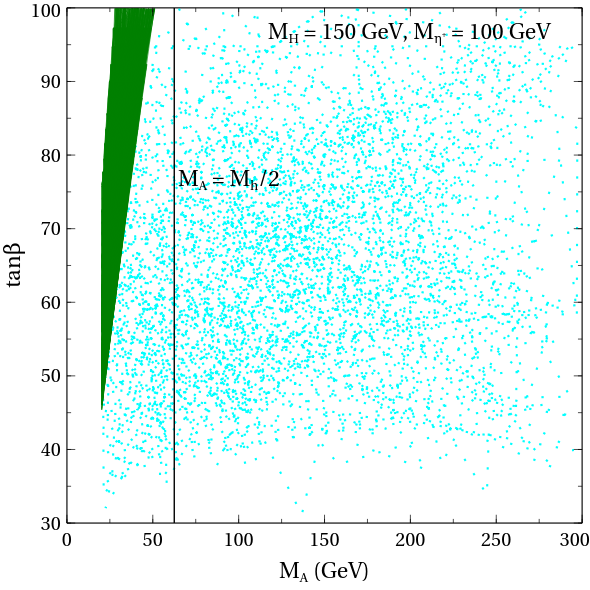}}
\caption{The parameter space compatible with 
the observed $\Delta a_\mu$ in the $M_A-$tan$\beta$ plane for $M_{\eta^+}$ = 100 GeV 
in case of (a) lepton-specific 2HDM and (b) muon-specific 2HDM. The color-coding is explained in the legends. The region to the left of the vertical line is tightly constrained by BR($h \to A A$) measurements. The plot in the left panel is taken from \cite{Chakrabarty:2021ztf}.}
\label{f:MA-tb_ls}
\end{figure}

We scan over the model parameters and filter out particular parameter points which are compatible with the theoretical and experimental constraints and also obey the observed muon $g-2$ anomaly. At the alignment limit, we consider the following parameters as independent in the (2+1)HDM framework : $\{m_{12}, M_H, M_A, M_{H^+}, M_{\eta_R}, M_{\eta_I}, M_{\eta^+}, 
\tan \beta, \alpha, \lambda_6, \lambda_7, \omega_1, \kappa_1, \sigma_1, \sigma_2, \sigma_3, \lambda_{L_1}, \lambda_{L_2}\}$, with $\lambda_{L_{1(2)}} = \nu_{1(2)} + \omega_{1(2)} + \kappa_{1(2)}$. To minimize the number of input parameters, we fix $M_H = M_{H^+} = 150$ GeV, $M_{\eta^+} = M_{\eta_R} + 1$ GeV = 100 GeV \footnote{1 GeV mass gap between $M_{\eta^+}$ and $M_{\eta_R}$ prohibits $W$-mediated direct detection inelastic scattering. } and $\lambda_6 = \lambda_7 = \lambda_{L_{1(2)}} =  0.01$. Low mass splittings between the neutral and charged scalars are consistent with the $T$-parameter constraint. Other independent input parameters are varied between the following window :
\bea
&& 0< m_{12} < 1 ~ {\rm TeV}, ~~20 ~ {\rm GeV} ~ M_A < 1 ~ {\rm TeV}, ~~ 10 < \tan \beta < 100, \nonumber \\
&& |\omega_1|,~|\kappa_1| < 4\pi,  ~~|\sigma_1|,~|\sigma_2|,~|\sigma_3| < 2\pi, ~~M_{\eta_R} < M_{\eta_I} < 500~\text{GeV}
\eea
After validating all parameter points successfully with all the constraints, we plot $\tan \beta$ against $M_A$ for $M_H = 150$ GeV and $M_{\eta^+} = 100$ GeV for purely 2HDM (Type-X and muon specific) and (2+1)2HDM in Fig.\ref{f:MA-tb_ls}(a) and Fig.\ref{f:MA-tb_ls}(b). One can conclude that for both variants of 2HDM, the parameter space consistent with the observed muon anomaly in the $\tan \beta$ vs. $M_A$ plane is enlarged in presence of the inert sector (cyan colored region) with respect to a pure 2HDM (green colored region).

\section{Collider analysis}
\label{collider}

In this section, we present exhaustive
probes in context of a 1 TeV ILC of a signal topology arising in the (2+1)HDM.
Before going to the details of the analysis, we reiterate that compared to the pure Type-X 2HDM, the (2+1)HDM allows for much a heavier $A$ that is consistent with the observed $\Delta a_\mu$. Therefore, this finding motivates to probe these heavier pseudoscalars through their decays to $\tau^+\tau^-$ or $\mu^+\mu^-$.

In one of the previous studies \cite{Chakrabarty:2021ztf}, a signature involving the pair production of $\eta_R, \eta_I$, followed by their subsequent decay into two $\tau$-hadrons ($\tau_h$) along with missing transverse energy ($\mET$) was explored at the high-luminosity 14 TeV Large Hadron Collider (HL-LHC). We reckon that the same final state could turn out to be more promising at the ILC owing to the much hadronically cleaner environment. We also plan to include the leptonic and semi-leptonic decay modes of $\tau $ to draw comparisons. Thus in this paper, we shall study the following channel for the lepton-specific (2+1)HDM.
\bea
e^+ e^- \to \eta_R \eta_I \to \eta_R \eta_R A \to \tau^+ \tau^- + \mET 
\label{channel-type-X}
\eea 
The following are the possibilities vis-a-vis $\tau$-decays:

\begin{itemize}
\item Both $\tau$s decay leptonically leading to the final state $2 \tau_\ell + \mET$ with $\tau_\ell= \tau_e, \tau_\mu$.

\item A semi-leptonic decay (one of the $\tau$ decays leptonically, another hadronically) leading to the final state $1 \tau_\ell + 1 \tau_h+ \mET$.

\item Both $\tau$s decaying hadronically leading to the final state $2 \tau_h+ \mET$. Here, $\tau_h$ denotes the visible hadronic decay product of the $\tau$, or, a $\tau$-jet.
\end{itemize}

The branching ratio BR$(A \rightarrow \mu^+ \mu^-)$ can be the dominant one for the muon-specific (2+1)HDM. The signal we take up for this case is:
\bea
e^+ e^- \to \eta_R \eta_I \to \eta_R \eta_R A \to \mu^+ \mu^- + \mET 
\label{channel-muonsp}
\eea 

\begin{figure}[htpb!]{\centering
\subfigure[]{
\includegraphics[height = 4.5 cm, width = 8.0 cm]{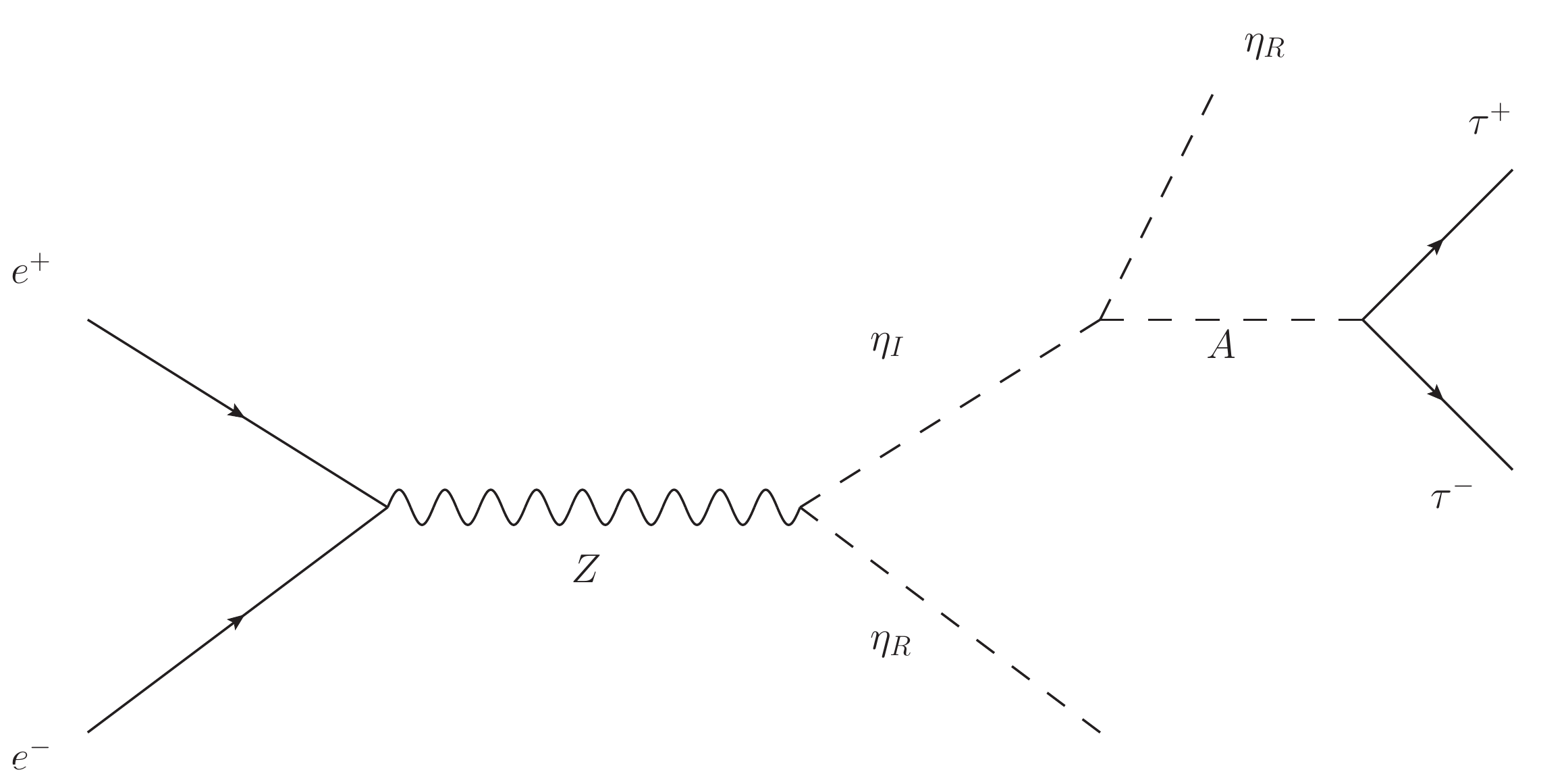}}
\subfigure[]{
\includegraphics[height = 4.5 cm, width = 8.0 cm]{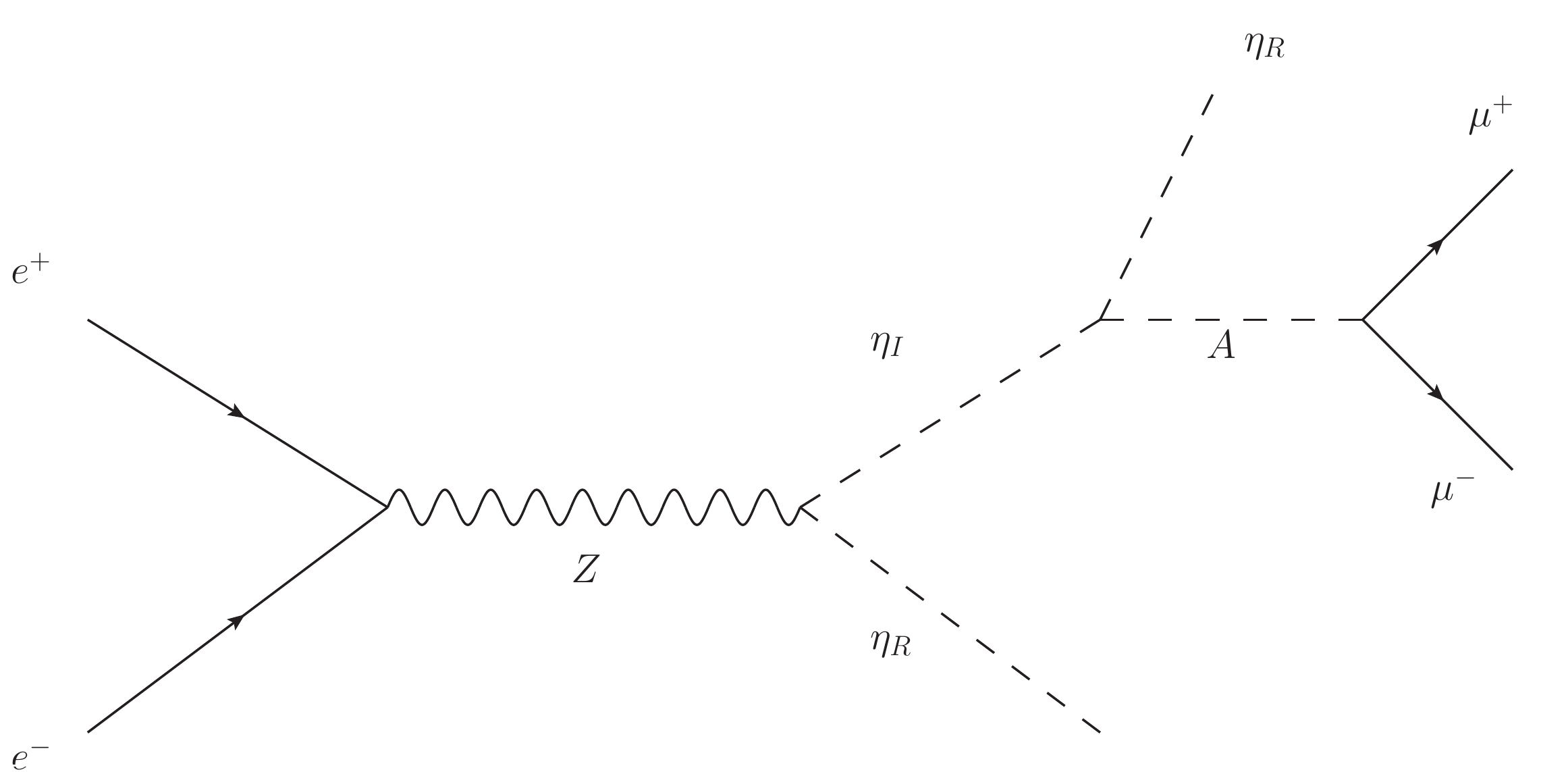}} \\
\subfigure[]{
\includegraphics[height = 4.5 cm, width = 8.0 cm]{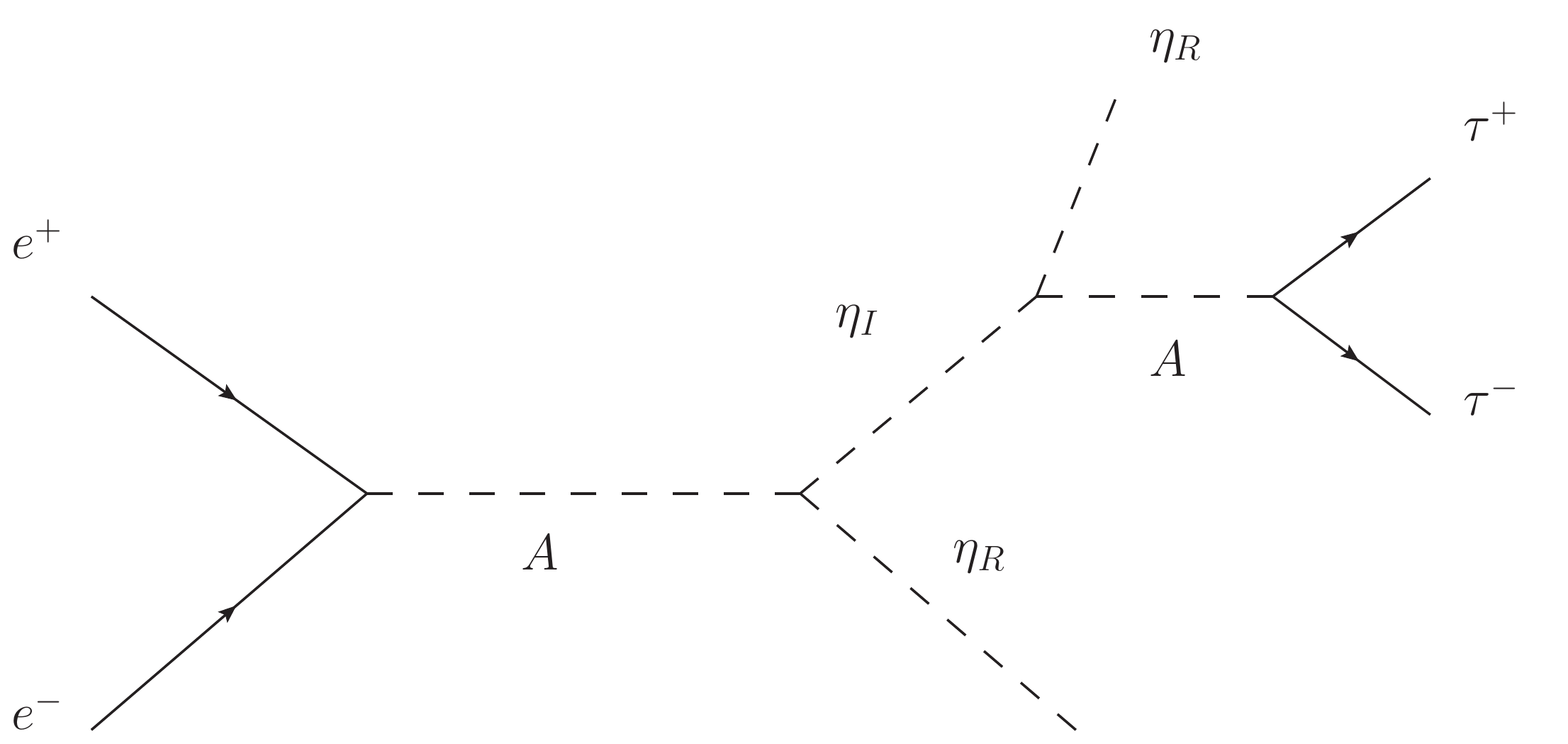}}
\subfigure[]{
\includegraphics[height = 4.5 cm, width = 8.0 cm]{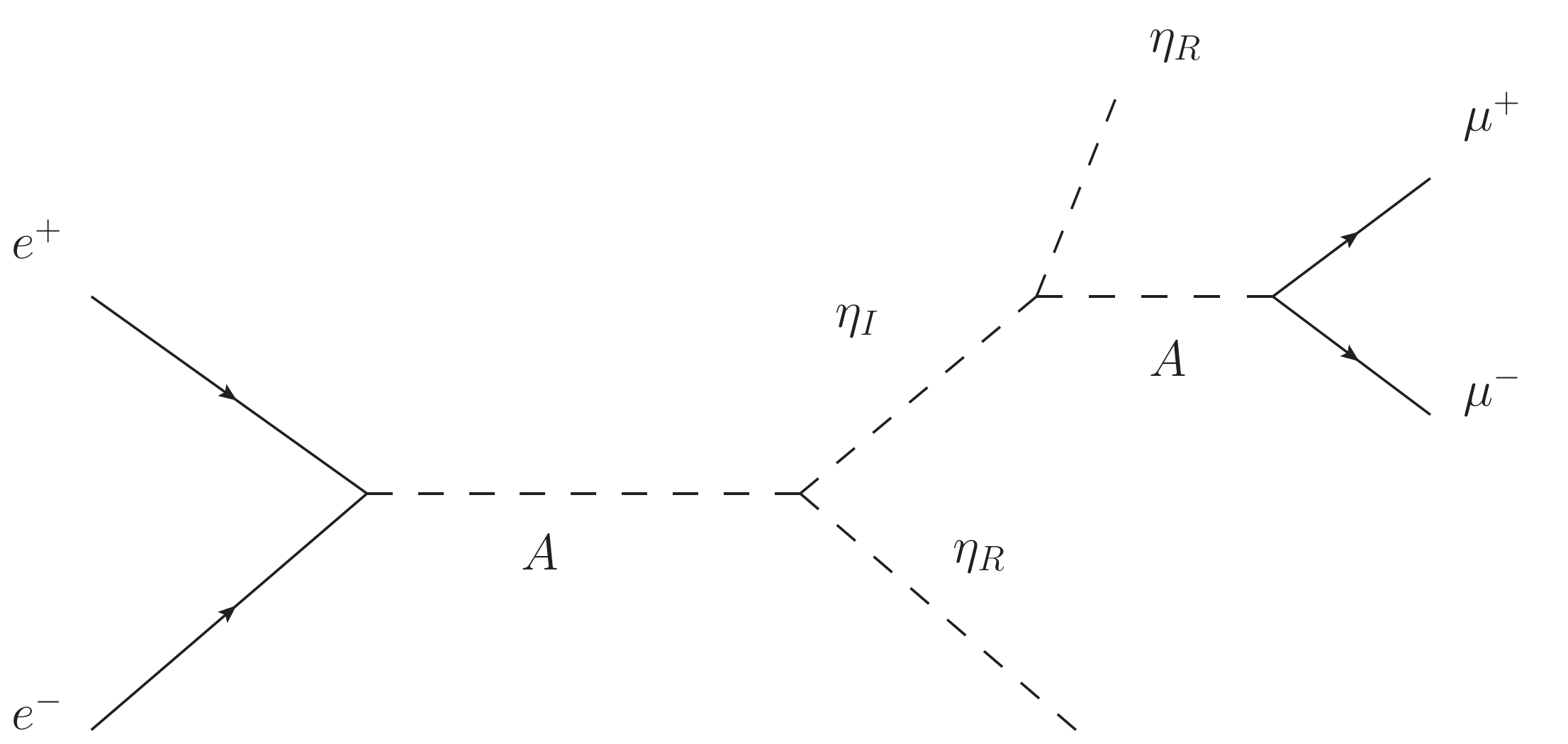}}
} \\
\caption{ Feynman diagrams for the signal channels : (a) for lepton specific 2HDM mediated by $Z$, (b) for muon specific 2HDM mediated by $Z$, (c) for lepton specific 2HDM mediated by $A$, (d) for muon specific 2HDM mediated by $A$.}
\label{diagrams}
\end{figure}

Here we would like to mention that the dominant mode of $\eta_R, \eta_I$ pair production is an $s$-channel $Z$-mediated process. There can be similar diagrams mediated by A instead of $Z$. The Feynman diagrams corresponding to two signals in lepton- and muon-specific 2HDMs are depicted in Fig.\ref{diagrams}. 

In hindsight, we would like to make an overall comment on the choice of the signal topology. First, the $A \to \tau^+ \tau^-,\mu^+ \mu^-$ channels become the natural choices to look for $A$ given the healthy branching ratios and also the possibility to reconstruct the pseudoscalar mass. Secondly, involving the
inert scalars in the signals should ultimately lead to a modified $\met$
which in turn could be a discerning kinematical feature. 

We state the common parameter choices made in the lepton- and muon-specific cases. We take $M_H = M_{H^+}$ = 150 GeV and $M_{\eta^+} = M_{\eta_I}$ + 1 GeV = 100 GeV throughout the analyses. This mass splitting of 1 GeV between the two inert scalars is large enough to avert inelastic DM-nucleon scattering that would otherwise immediately get ruled out by the direct detection data. As for the subsequent collider analyses, we divide it into the following two subsections.

\subsection{Signatures in the lepton-specific case}

We propose a few benchmark points (BP1-4 in the increasing order of $M_A$) in Table~\ref{tab:BP} for the lepton-specific case. All four BPs are consistent with the constraints imposed and predict $\Delta a_\mu$ in the 2$\sigma$ band. One further notes that $M_{\eta_I} > M_{\eta_R} + M_A$ holds for all the BPs so that the $\eta_I \to \eta_R A$ mode is kinematically open.
$M_{\eta_I}$ increases in going from BP1 to BP4 and branching fractions for the $\eta_I \to \eta_R Z, \eta^\pm W^\mp$ modes accordingly increase thereby explaining the 
the observed drop in the $\eta_I \to \eta_R A$ branching ratio. We discuss the decays of $A$ next. While $A \to \tau^+ \tau^-$ is the dominant mode for BP1, the larger values of $M_A$ taken for BP2-4 imply that $A \to Z H, W^\pm H^\mp$ also open up. And BR($A \to \tau^+ \tau^-$) accordingly diminishes.

We now discuss the possible backgrounds for hadronic, semi-leptonic and leptonic decays of the $\tau$ pair. First, $e^+ e^- \to \tau^+ \tau^- + \mET$ 
can be a common source of background in all the three cases. The dominant contributors to a $\tau^+ \tau^- + \mET$ process are: $W^+W^- ~(W^+ \to \tau^+ \nu_\tau, W^- \to \tau^- \bar{\nu}_\tau), ZZ~( Z \to \tau^+ \tau^-, Z \to \nu_\tau \bar{\nu}_\tau), W^+ W^- Z~(W^+ \to \tau^+ \nu_\tau, W^- \to \tau^- \bar{\nu}_\tau, Z \to \nu_\tau \bar{\nu}_\tau), ZZZ ~( Z \to \tau^+ \tau^-, Z \to \nu_\tau \bar{\nu}_\tau,Z \to \nu_\tau \bar{\nu}_\tau ), Zh ~ ( h \to \tau^+ \tau^-, Z \to \nu_\tau \bar{\nu}_\tau)$. In addition, one also needs to consider  $e^+ e^- \to 2j + \mET$ and $e^+ e^- \to \ell^+ \ell^- + \mET$ where $j$ denotes a light jet and $\ell = e,\mu$. The former is important while analysing $2\tau_h + \mET$ since light jets can get mis-tagged as $\tau$-hadrons. On the other hand, $\ell^+ \ell^- + \mET$ becomes the principal background for $2\tau_\ell + \mET$. The dominant contributors to this are: $W^+W^- ~(W^+ \to l^+ \nu_l, W^- \to l^- \bar{\nu}_l), ZZ~( Z \to l^+ l^-, Z \to \nu_l \bar{\nu}_l), W^+ W^- Z~(W^+ \to l^+ \nu_l, W^- \to l^- \bar{\nu}_l, Z \to \nu_l \bar{\nu}_l), ZZZ ~( Z \to l^+ l^-, Z \to \nu_l \bar{\nu}_l,Z \to \nu_l \bar{\nu}_l ), Zh ~ ( h \to l^+ l^-, Z \to \nu_l \bar{\nu}_l)$ etc. Finally, possible backgrounds to $1\tau_h + 1\tau_\ell + \mET$ can come only through mis-tagging in an $e^+ e^-$ environment. This might include a $\tau_h$ from $2\tau_h + \mET$ faking as an 
$\ell$ or an $\ell$ from $2\tau_\ell + \mET$ getting mis-identified as a $\tau_h$.

 \begin{table}
\centering
\begin{tabular}{ |c|c|c|c|c| } 
\hline
& BP1 & BP2 & BP3 & BP4 \\ \hline
$m_{12}$ & 21.60 GeV & 20.4 GeV & 21.0 GeV & 22.2 GeV  \\
tan$\b$ & 47.8 & 53.83 & 50.77 & 45.46  \\
$M_A$ & 152.3 GeV & 253.24 GeV & 353.20 GeV & 404.16 GeV \\
$M_{\eta_I}$ & 270.5 GeV & 397.00 GeV & 492.0 GeV  & 547.0 GeV  \\
$k_1$ & -4.12177 & -2.07345 & -1.4954 & -0.615752  \\
$\omega_1$ & -5.50407 & -0.125664 & -5.93133 &  -2.01062  \\
$\s_1$ & -4.24743 & -5.70513 & -5.31557 & -5.17734  \\
$\s_2$ & 4.1469 & -0.263894 & 5.81823 & 5.98159  \\
$\s_3$ & 6.05699 & 5.44124 & 6.19522 & 6.06956  \\
$\Delta a_\mu \times 10^9$ & 1.57538 & 1.51138 & 2.05397 & 1.85177  \\
$\sigma^{eff}_{SI}$ & $2.73 \times 10^{-48}$ cm$^2$ &  $3.81 \times 10^{-50}$ cm$^2$  & $2.03 \times 10^{-48}$ cm$^2$ & $2.96 \times 10^{-48}$ cm$^2$  \\ \hline
BR$(\eta_I \to \eta_R A)$ & 0.97475 & 0.822958 & 0.642342 &  0.513814  \\ 
BR$(A \to \tau^+ \tau^-)$ & 0.99 & 0.7983 &  0.199417 &  0.104883  \\ \hline
\end{tabular}
\caption{Benchmark points used for studying the discovery prospects of an $A$ in the lepton-specific (2+1)HDM. The values for the rest of the masses are 
$M_H = M_{H^+} = 150$ GeV, $M_{\eta^+} = M_{\eta_R} + 1$ GeV = 100 GeV.}
\label{tab:BP}
\end{table}

We display the signal and background cross sections at the leading order (LO) for unpolarized (P0) and polarized (P1,P2,P3) $e^+$ and $e^-$ beams in Table \ref{tab:cs-BP}. The polarizations P1,P2,P3 are defined as follows \cite{Behnke:2013lya}:

P1 $\equiv$ 80$\%$ left-handed $e^-$ and 30$\%$ right-handed $e^+$ beam  ($P_{e^-},P_{e^+} = 80\%L,30\%R)$

P2 $\equiv$ 80$\%$ right-handed $e^-$ and unpolarized $e^+$ beam ($P_{e^-},P_{e^+} = 80\%R, 0)$

P3 $\equiv$ 80$\%$ right-handed $e^-$ and 30$\%$ left-handed $e^+$ beam ($P_{e^-},P_{e^+} = 80\%R, 30\%L)$

To incorporate the interaction vertices of the model, the Lagrangian is first implemented in {\texttt{FeynRules}} \cite{Alloul:2013bka}. LO signal and background cross sections with unpolarized and polarized $e^+-e^-$ beams are computed through {\texttt{MG5aMC@NLO}} \cite{Alwall:2014hca,Frederix:2018nkq} using the {\em Universal Feynrules Output (UFO)}  file which is generated as the output from {\texttt{Feynrules}}. Parton showering and hadronization are performed through {\texttt{Pythia8}} \cite{Sjostrand:2014zea}. The detector effects are included in the analysis via passing the signal and backgrounds through {\texttt{Delphes-3.4.1}} \cite{deFavereau:2013fsa}. For this purpose, we use the default ILD detector simulation card present in Delphes-3.4.1.
To obtain the best possible results we refrain from doing a traditional cut-based analysis but rather perform the more sophisticated multivariate analysis using Decorrelated Boosted Decision
Tree (BDTD) algorithm embedded in TMVA (Toolkit for Multivariate Data Analysis) \cite{Hocker:2007ht}
platform. We refer to \cite{Hocker:2007ht} for the detailed description of this algorithm. The signal significance is computed using $\mathcal{S} = \sqrt{2 \left[(S+B) {\rm log} \left(\frac{S+B}{B}\right)-S\right]}$, where $S$ and $B$ are number of signal and background events left after imposing cuts on pertinent kinematic variables \cite{Adhikary:2020cli}.

While generating signal and background cross sections in {\texttt{MG5aMC@NLO}}, we first impose the following preliminary cuts:
\bea
&& p^j_T > 20~ {\rm GeV}, ~~ |\eta_j| < 5.0, \nonumber \\
&& p^\ell_T > 10~ {\rm GeV}, ~~ |\eta_\ell| <2.5, ~~ \Delta R_{mn} > 0.4, {\rm where} ~ m,n = \ell, {\rm jets}
\label{prelim-cut}
\eea

Here $p^{j(\ell)}_T$ and $|\eta_{j(\ell)}|$ denote the transverse momentum and pseudorapidity of the final state jets (leptons)  respectively. One defines $\Delta R_{mn} = \sqrt{\Delta \eta^2_{mn} + \Delta \phi^2_{mn}}$, $\Delta \eta_{mn}$ and $\Delta \phi_{mn}$ being the difference between pseudorapidity and azimuthal angles of $m$th and $n$th particles respectively.

\begin{table}[h]
\centering
\begin{tabular}{ | c | c | c | c | c | c | } 
\hline
Signal/Backgrounds & Process & P0  & P1 & P2 & P3 \\ 
 &   & & &  & \\ 
& & (fb) & (fb) & (fb)   & (fb) \\ \hline
Signal & & & & &  \\
BP1 & & 8.687  & 12.6  & 8.23 & 10.045\\
BP2 & $e^+ e^- \to \eta_R \eta_I \to \eta_R \eta_R A \to \tau^+\tau^- + \met$
 & 4.992  & 6.42  & 4.14 & 5.125\\ 
 BP3 &  & 0.645  & 0.9   & 0.57 & 0.707\\
BP4 & & 0.2118  & 0.29 & 0.189 & 0.235 \\ \hline
Background & $e^+ e^- \to \tau^+ \tau^- + \met$ & 55.76  & 127.9  & 12.26 & 9.405\\ 
 & $e^+ e^- \to 2j + \met$ & 414.6  & 949.8  & 96.78 & 76.94 \\ 
 & $e^+ e^- \to 2\ell + \met$ & 419.0  & 915.2  & 115.8 & 89.34\\ \hline
\end{tabular}
\caption{Signal and background cross sections for lepton specific (2+1)HDM at the 1 TeV ILC.}
\label{tab:cs-BP}
\end{table}

To avoid repetition, we shall only tabulate some relevant parameters used in the BDTD algorithm for all channels. A naive estimation of the signal-to-background ratio at this level (from Table \ref{tab:cs-BP}) for all the polarizations shows P3 to be the most prospective in this regard. Therefore, we pick up P3 for the multivariate analyses in case of all the final states. The analyses for the three cases is divided into the three following subsections for clarity.

\subsubsection{$2 \tau_\ell+ \mET$ final state}
\label{2lmet}

\begin{figure}[htpb!]{\centering
\subfigure[]{
\includegraphics[height = 6 cm, width = 8.0 cm]{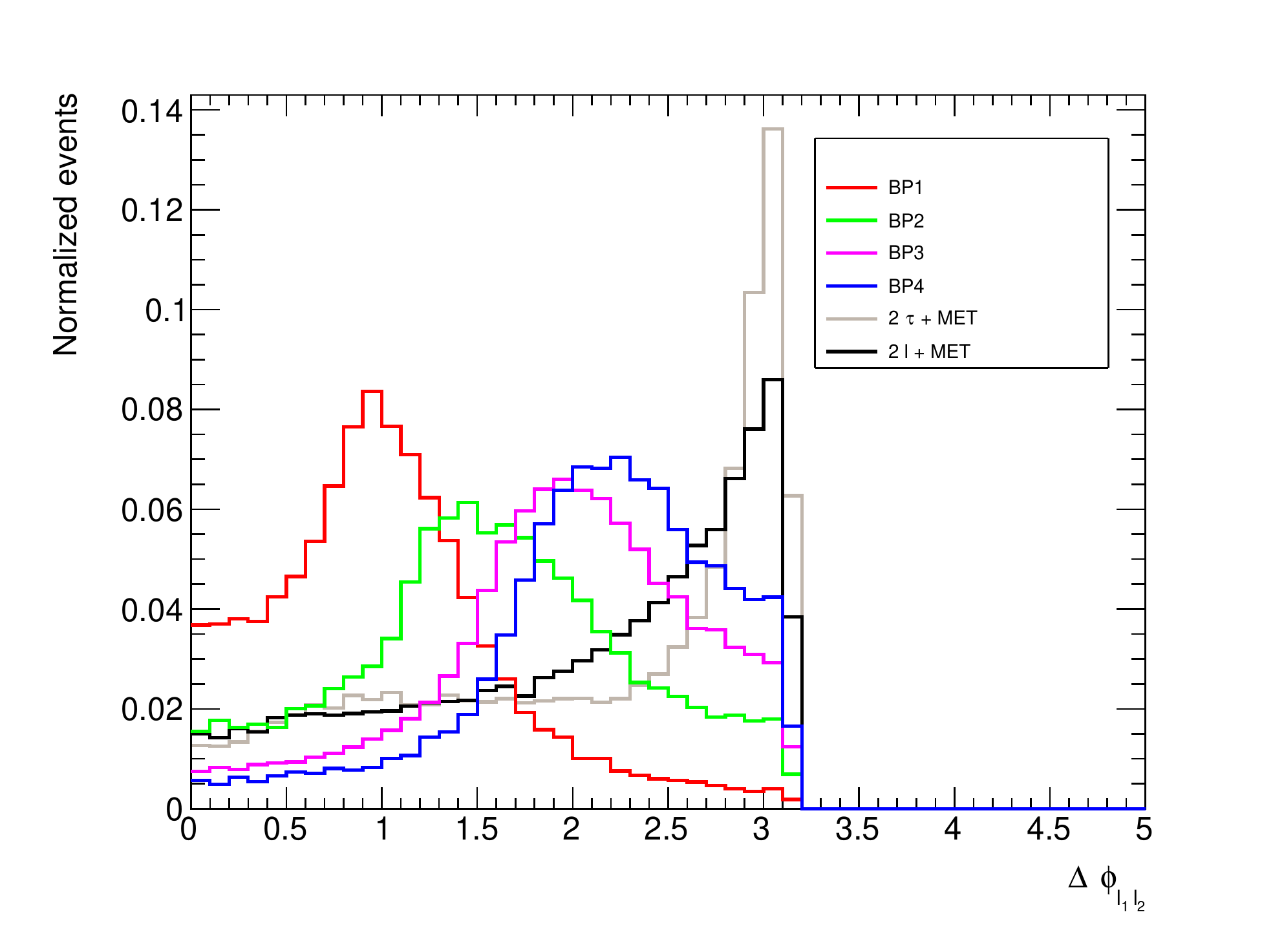}} 
\subfigure[]{
\includegraphics[height = 6 cm, width = 8 cm]{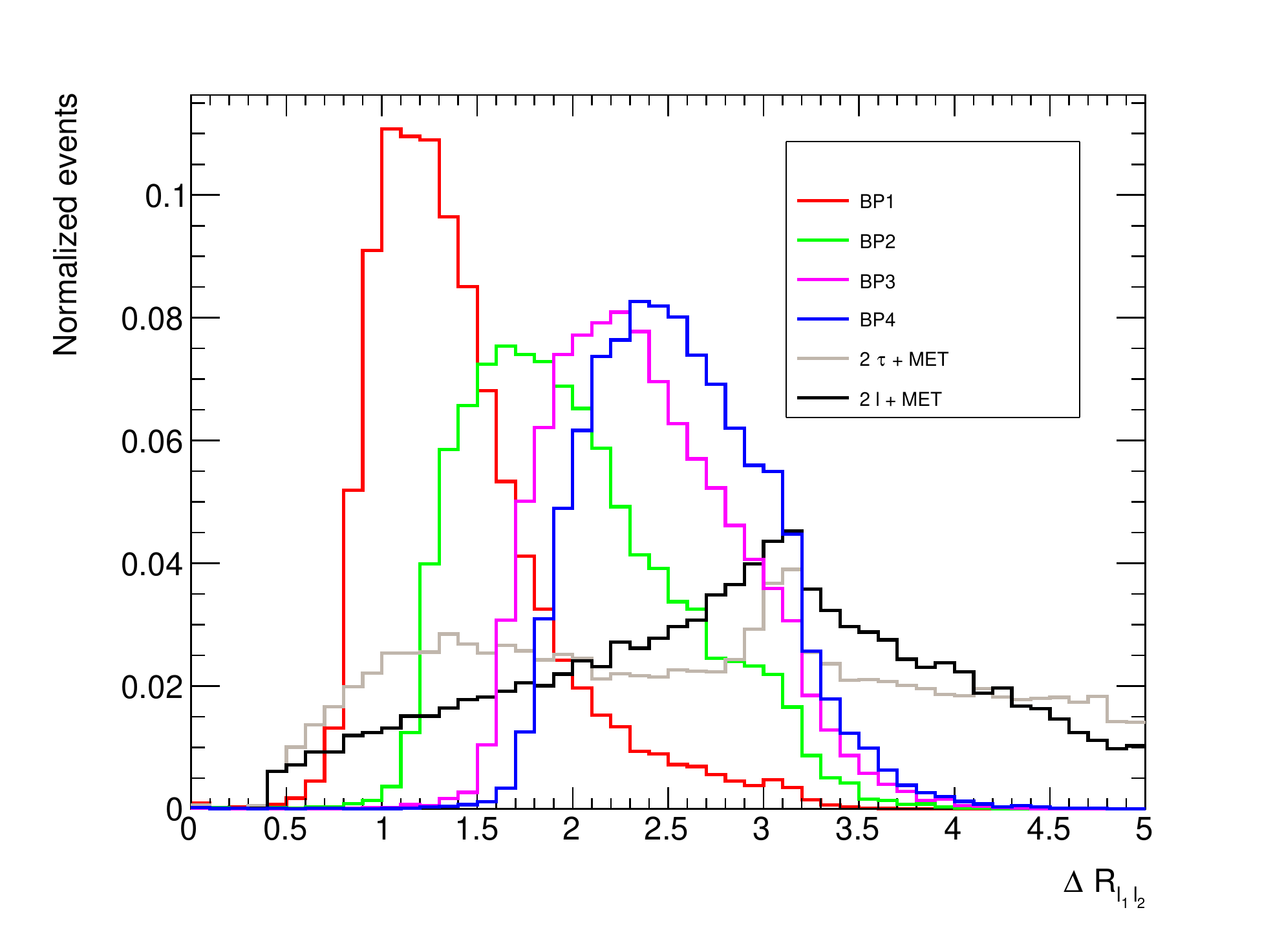}} \\
\subfigure[]{
\includegraphics[height = 5.5 cm, width = 8 cm]{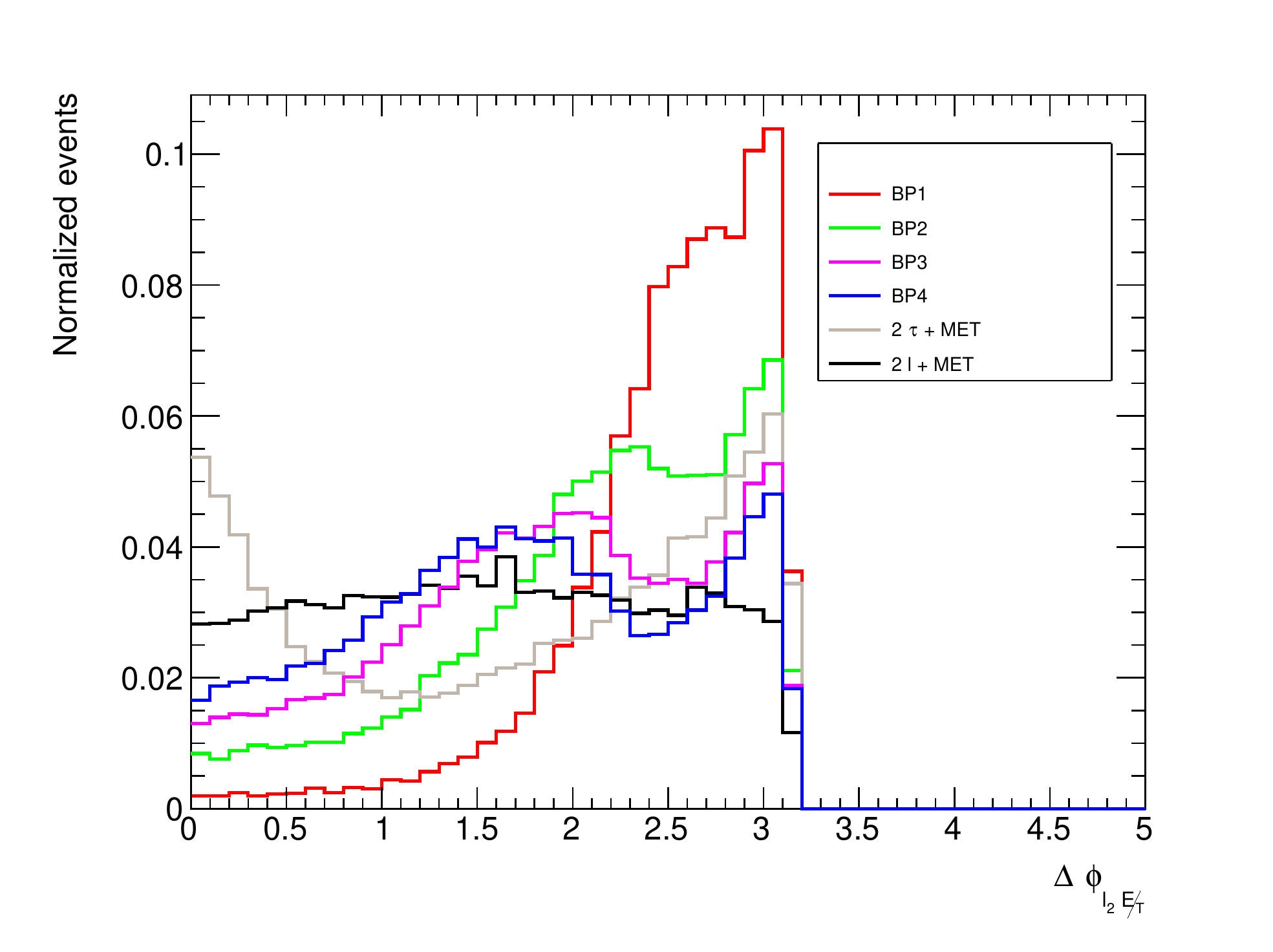}}
\subfigure[]{
\includegraphics[height = 5.5 cm, width = 8 cm]{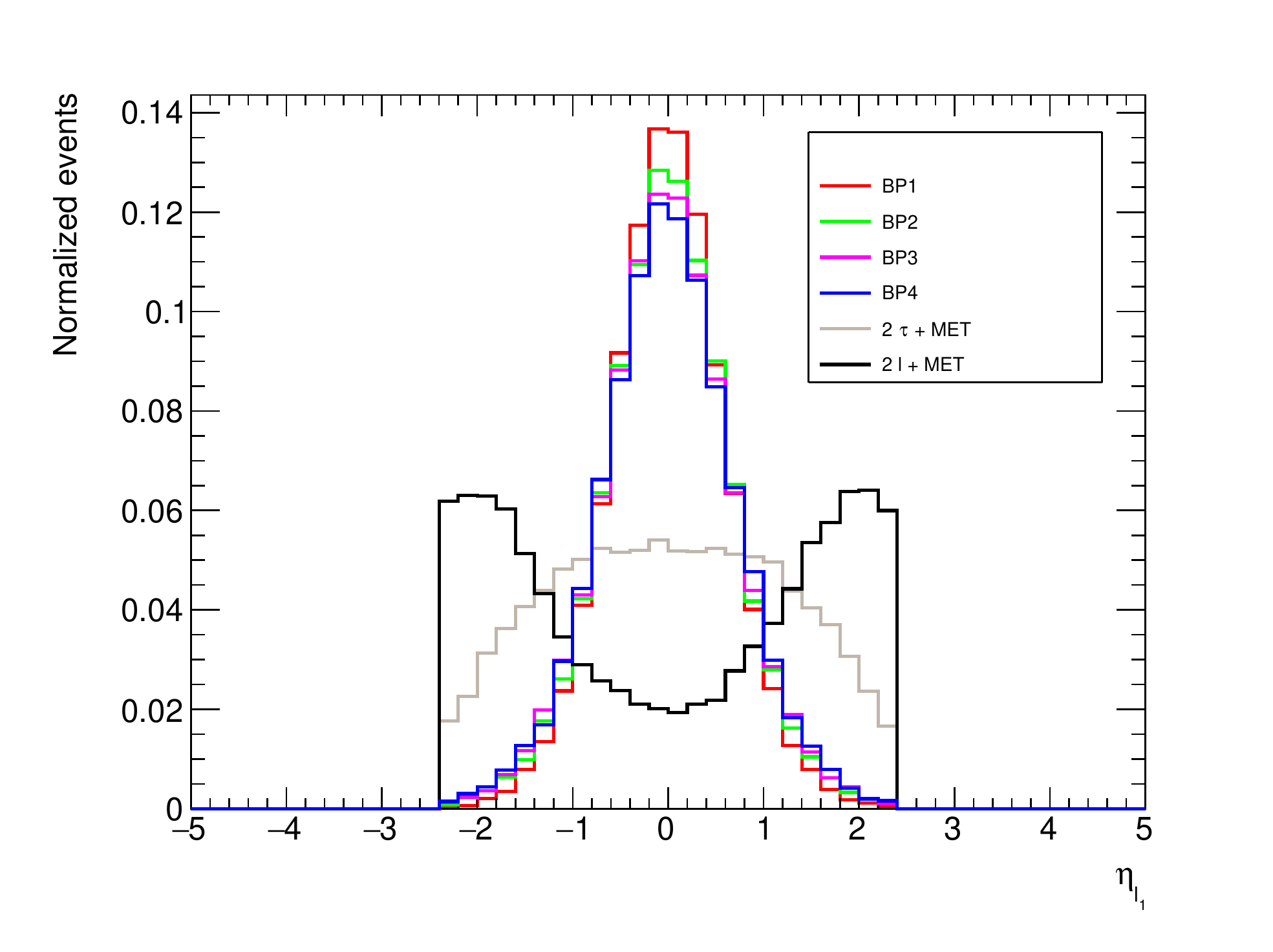}}
\subfigure[]{
\includegraphics[height = 5.5 cm, width = 8 cm]{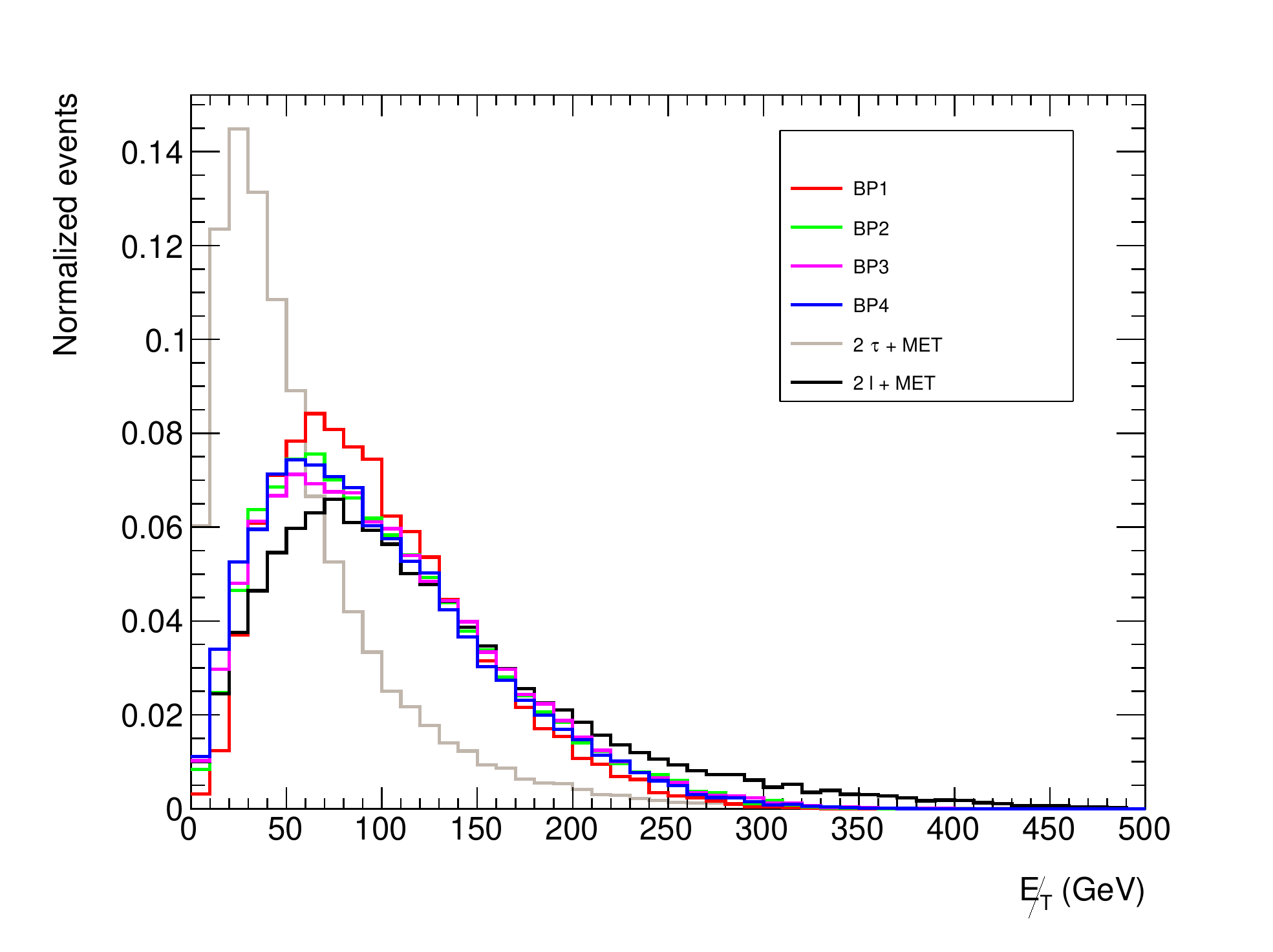}} 
\subfigure[]{
\includegraphics[height = 5.5 cm, width = 8 cm]{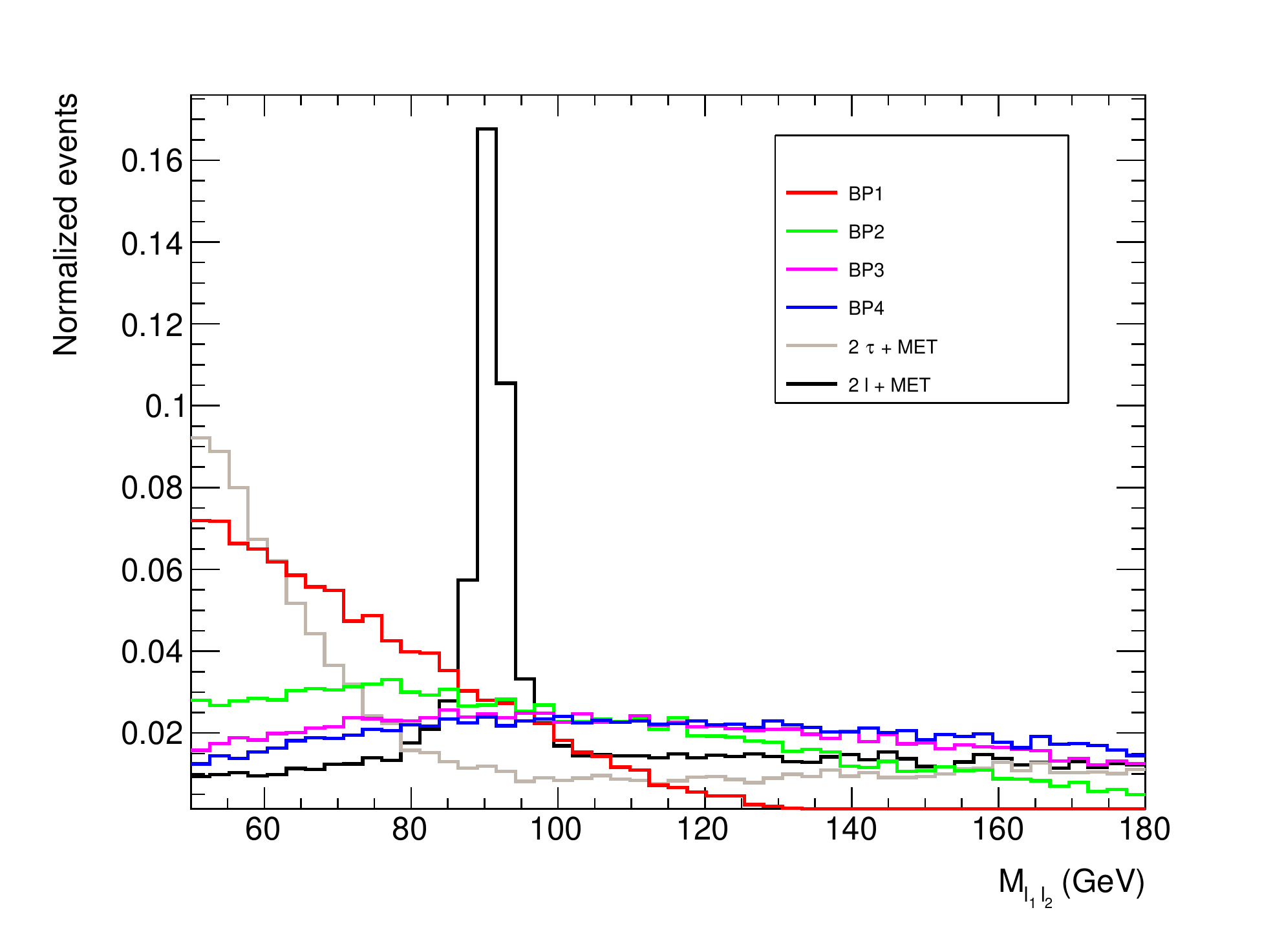}} 
} \\
\caption{ Normalized distributions of $\Delta \phi_{\ell_1 \ell_2},~ \Delta R_{\ell_1 \ell_2},~\Delta \phi_{\ell_2 \mET}~,~ \eta_{\ell_1},~ \mET,~ M_{\ell_1 \ell_2} $ for $2 \tau_\ell + \mET$ channel at 1 TeV ILC. }
\label{distribution-2lmet-1}
\end{figure}

\begin{figure}[htpb!]{\centering
\subfigure[]{
\includegraphics[height = 5.5 cm, width = 8 cm]{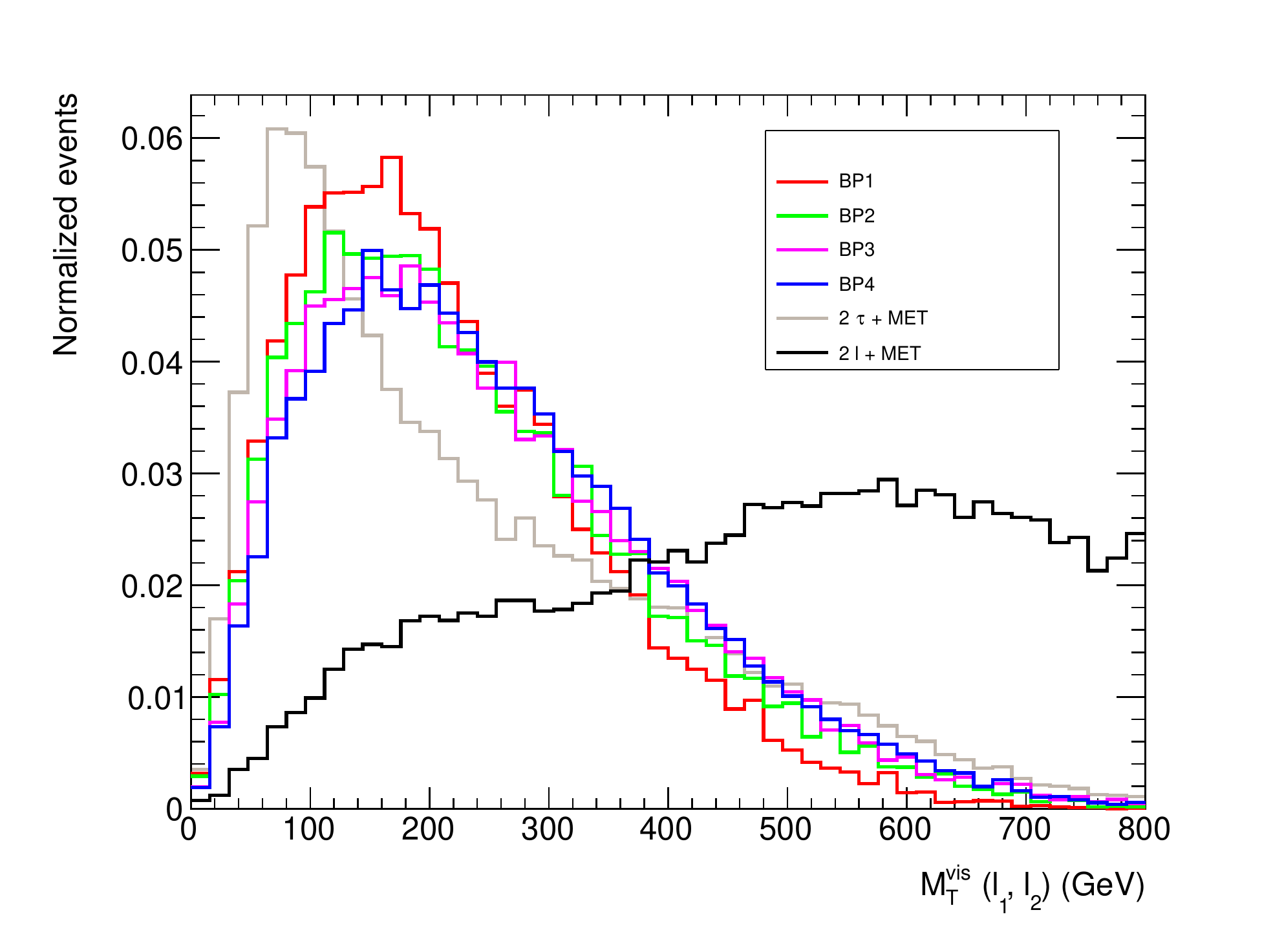}}
\subfigure[]{
\includegraphics[height = 5.5 cm, width = 8 cm]{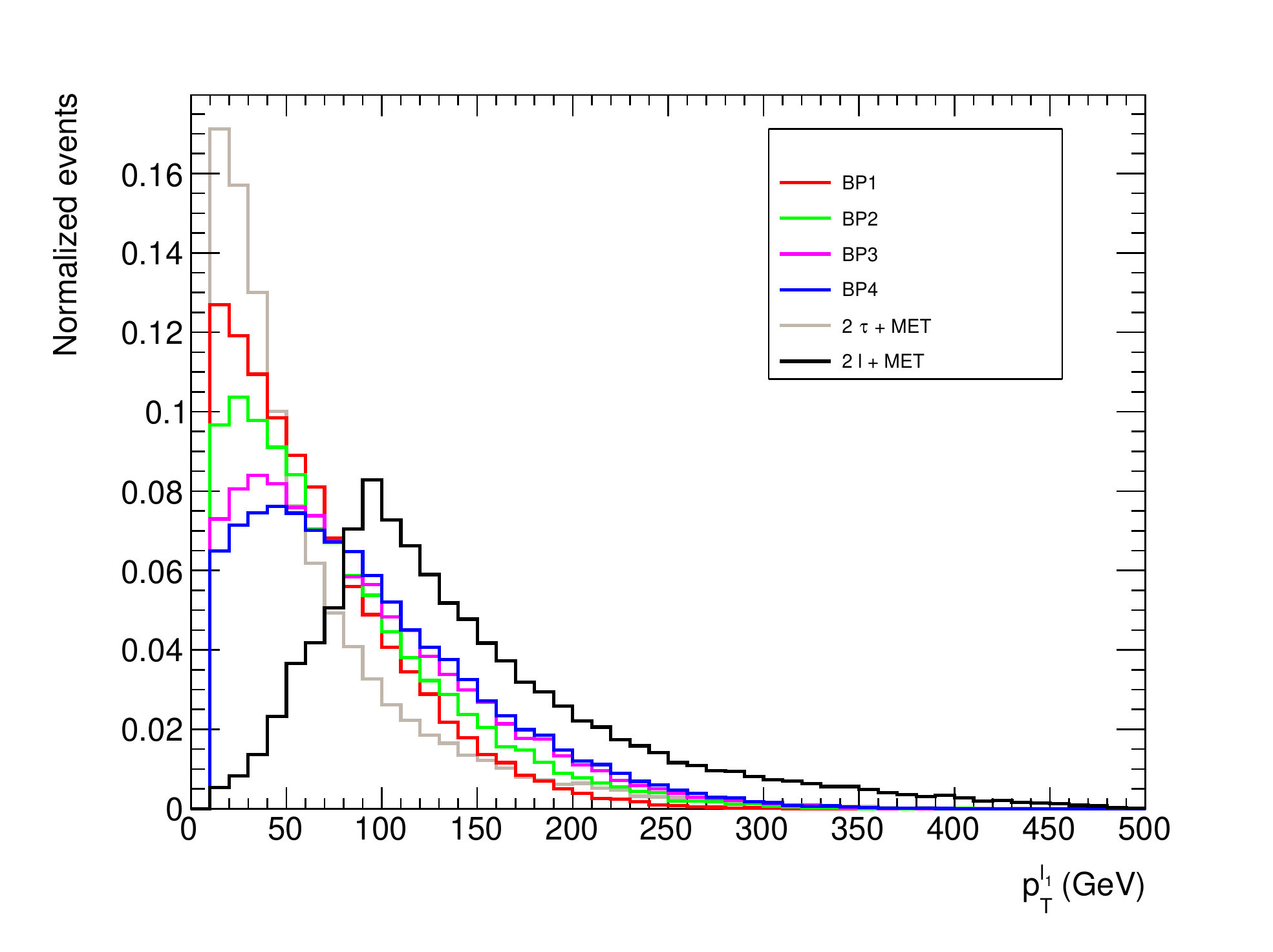}}
} \\
\caption{ Normalized distributions of $M_T^{\rm vis} (\ell_1, \ell_2),~ p_T^{\ell_1} $ for $2 \tau_\ell + \mET$ channel at 1 TeV ILC. }
\label{distribution-2lmet-2}
\end{figure}

The two $\tau$ decay leptonically leading to $\tau_{e^+} \tau_{e^-},\tau_{e^\pm} \tau_{\mu^\mp}, \tau_{\mu^+} \tau_{\mu^-}$ along with $\mET$. We denote the two daughter leptons to be $\ell_1$ and $\ell_2$ in the decreasing order of their $p_T$. We use the following kinematic variables while training the signal and background samples using the BDTD algorithm :
\bea
\Delta {\phi}_{{\ell_1 \ell_2}},~~\Delta {\phi}_{{\ell_2 \met}},~~ \Delta {R}_{{\ell_1 \ell_2}}, ~~ \eta_{\ell_1}, ~~\mET, ~~M_{\ell_1 \ell_2}, ~~M_T^{\rm vis}(\ell_1, \ell_2),~~ p_T^{\ell_1} \nonumber
\eea
Let us define the aforementioned kinematic variables for completeness. Here, $\Delta {\phi}_{{\ell_1 \ell_2}} ~(\Delta {\phi}_{{\ell_2 \met}})$ is the difference in azimuthal angles between $\ell_1, \ell_2$ ($\ell_2$ and the missing transverse energy vector $\mET$) in the final state. Also, $\Delta R_{\ell_1 \ell_2} $ \footnote{Here $\Delta R_{ij}  = \sqrt{(\eta_{i}-\eta_{j})^2 + (\phi_{i}-\phi_{j})^2}$, where $\eta_i$ and $\phi_i$ are the pseudo-rapidity and azimuthal angle of the $i$-th particle.} is the separation between the two leptons. While $M_{\ell_1 \ell_2}$ is the invariant mass of the two daughter leptons, $M_T^{\rm vis} (\tau_{\rm vis1}, \tau_{\rm vis2})$ is the cluster transverse mass \cite{Hou:2022nyh} which can be constructed out of the visible decay products of $\tau$s $(\tau_{\rm vis1}, \tau_{\rm vis2})$ in the final state and $\mET$ as follows:
\bea
[M_T^{\rm vis} (\tau_{\rm vis1}, \tau_{\rm vis2})]^2 &=& (\sqrt{|\vec{p_T}(\tau_{\rm vis1}, \tau_{\rm vis2})|^2 + M(\tau_{\rm vis1}, \tau_{\rm vis2})^2} + \mET )^2 \nonumber \\
&& -(\vec{p_T}(\tau_{\rm vis1}, \tau_{\rm vis2}) + \vec{\mET} )^2
\label{MTvis}
\eea 
Here $M(\tau_{\rm vis1}, \tau_{\rm vis2})$ and $\vec{p_T}(\tau_{\rm vis1}, \tau_{\rm vis2})$ are the invariant mass and the vector transverse momentum of the two visible $\tau$ decays respectively. We note that $M_T^{\rm vis} (\tau_{\rm vis1}, \tau_{\rm vis2})$ becomes relevant when there are more than one sources of missing transverse energy and the collinear approximation \cite{Ellis:1987xu} is no longer valid. As discussed earlier, the main background in this case comes from 
$e^+ e^- \to \ell^+ \ell^- + \mET$ with a sub-leading component contributed by $e^+ e^- \to \tau^+ \tau^- + \mET$ when both $\tau$ decay leptonically. Fig.\ref{distribution-2lmet-1} and Fig.\ref{distribution-2lmet-2} shows normalized distributions of the most important variables for the signal BPs and the backgrounds.

Fig.\ref{distribution-2lmet-1}(a),(b) display the normalized distributions of $\Delta \phi_{\ell_1 \ell_2}$ and $\Delta R_{\ell_1 \ell_2}$ respectively. Since for the signal, two $\tau$ originate from a single mother particle $A$, the daughter leptons $\ell_1,\ell_2$ are not as widely separated as in case of the backgrounds.
Thus the distributions of $\Delta \phi_{\ell_1 \ell_2}$ and $\Delta R_{\ell_1 \ell_2}$ for the four signal benchmark points peak at lower values compared to the backgrounds. However, the larger the $M_A$, the higher is the $\Delta R_{\ell_1 \ell_2}$ value where the signal distribution peaks. This can be understood from the fact that a lighter $A$ would be somewhat more boosted than a heavier $A$. And the decay products of a more boosted object would be more collimated accordingly. Hence the observed pattern in $\Delta \phi_{\ell_1 \ell_2}$ and $\Delta R_{\ell_1 \ell_2}$ for BP1 to BP4.

In Fig.\ref{distribution-2lmet-1}(d), we have drawn the pseudo-rapidity distribution of the leading lepton $\ell_1$. It is seen that this distribution peaks at zero for both signal and the $2\tau + \mET$ background. Such a semblance is expected given $\ell_1$ comes from a $\tau$ in all such cases. On the other hand, the $2\ell+\mET$ background is generated via inclusive mode through $s$-channel exchanges of $\gamma,Z$ and a $t$-channel exchange of $\nu$'s. Thus, owing to the different kinematics of this background, the 
$\eta_{\ell_1}$ distribution accordingly is distinct with the peaks placed symmetrically about zero. We also point out that the invariant mass ($M_{\ell_1 \ell_2}$) distribution for $2\ell+\mET$ background in Fig.\ref{distribution-2lmet-1}(f) has a sharp peak around Z-boson mass $M_Z$. This is due to the fact that an contributor to the $2\ell+\mET$ background is $ZZ$ with one $Z$ decaying to two opposite sign same flavour leptons and other decaying into neutrinos.

Normalized distributions of $p_T^{\ell_1}$ and $\mET$ are shown in Figs.\ref{distribution-2lmet-2}(b) and \ref{distribution-2lmet-1}(e) and we first discuss the the distribution of $p_T^{\ell_1}$. The most boosted leptons are seen in case of the $2\ell + \mET$ background. This can be attributed to the fact that $\ell_1,\ell_2$ in this case are produced directly  
through $s$- and $t$-channel scatterings.
The next \emph{hardest} $p_T^{\ell_1}$-spectrum is that of the signal BP4. It is expected that the heavier the decaying pseudoscalar, the more boosted are the daughter $\tau$-leptons. This is why the peak of $p_T^{\ell_1}$-distribution progressively shifts towards lower values from BP4 to BP1. The \emph{softest} distribution of all is that of the $2\tau_\ell + \mET$ background where the leptons come from $\tau$-decay only.

Neutrinos are the only source of missing transverse energy for the backgrounds. The neutrinos are the most boosted for the $2\ell + \mET$ background since these come from $W$- and $Z$-decays. Hence, the hardest $\mET$-spectrum of all is seen in case of the $2\ell + \mET$ background. However, the very low $p_T$ neutrinos coming from the $\tau$-decays contaminate the sample in case of $2\tau_\ell + \mET$ background and hence the soft distribution observed. The $\mET$-distribution for the signals is a measure of how boosted is the undetected $\eta_R$ and the events peak around $p^{\eta_R}_T \sim \frac{(M^2_{\eta_I} - M^2_A)}{2 M_{\eta_R}}$. It is inferred for the $M_{\eta_I}$ and $M_A$ values for BP1-4 that the peaks of the $\mET$-distribution for these benchmarks are not far apart from one another. Overall, the $\mET$-distribution in Fig.\ref{distribution-2lmet-1}(e) for BP1-4 are harder than that of $2\tau_\ell + \mET$ to an extent. In Fig.\ref{distribution-2lmet-1}(c) and Fig.\ref{distribution-2lmet-2}(b), normalized distributions of $\Delta \phi_{\ell_2 \mET}$ and $M_T^{\rm vis}(\ell_1, \ell_2)$ are drawn.

Having described the primary features of the kinematic distributions, we now proceed to perform the BDTD analysis. We refer to \cite{Hocker:2007ht} for details of the BDTD methodology. Different BDTD parameters like \texttt{NTrees, MinNodesize, MaxDepth, nCuts} and \texttt{KS-scores } (for signal and backgrounds) for each benchmark are presented in Table \ref{BDT-param-2lmet}. One can stabilize the \texttt{KS-scores} simultaneously for signals and backgrounds upon tuning these parameters.

\begin{table}[htpb!]
\begin{center}
\begin{tabular}{|c|c|c|c|c|c|}
\hline
 &  \hspace{5mm} {\texttt{NTrees}} \hspace{5mm} & \hspace{5mm} {\texttt{MinNodeSize}} \hspace{5mm} & \hspace{5mm} {\texttt{MaxDepth}}~~ \hspace{5mm} & \hspace{5mm} {\texttt{nCuts}} ~~\hspace{5mm} & \hspace{5mm} {\texttt{KS-score for}}~~\hspace{5mm}\\
 & & & & & {\texttt{Signal(Background)}} \\
\hline
\hline
\hspace{5mm} BP1 \hspace{5mm} & 120 & 3 \% & 2.0 & 55 & 0.661~(0.119) \\ \hline
\hspace{5mm} BP2 \hspace{5mm} & 110 & 3 \% & 2.0 & 50 & 0.579~(0.023) \\ \hline
\hspace{5mm} BP3 \hspace{5mm} & 120 & 4 \% & 2.0 & 55 & 0.134~(0.908) \\ \hline
\hspace{5mm} BP4 \hspace{5mm} & 120 & 4 \% & 2.0 & 55 & 0.104~(0.315) \\ \hline
\end{tabular}
\end{center}
\caption{Tuned BDT parameters for BP1, BP2, BP3, BP4 for the $2 \tau_\ell + \mET$ channel.}
\label{BDT-param-2lmet}
\end{table}

\begin{figure}[htpb!]{\centering
\subfigure[]{
\includegraphics[width=3in,height=2.45in]{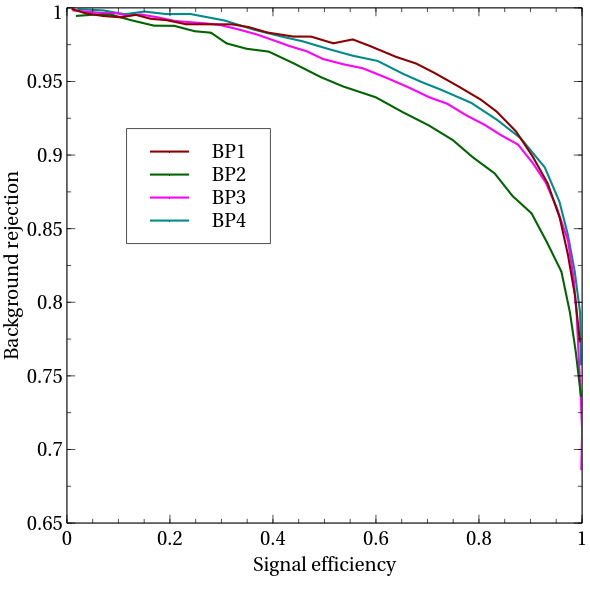}}
\subfigure[]{
\includegraphics[width=3.1in,height=2.46in]{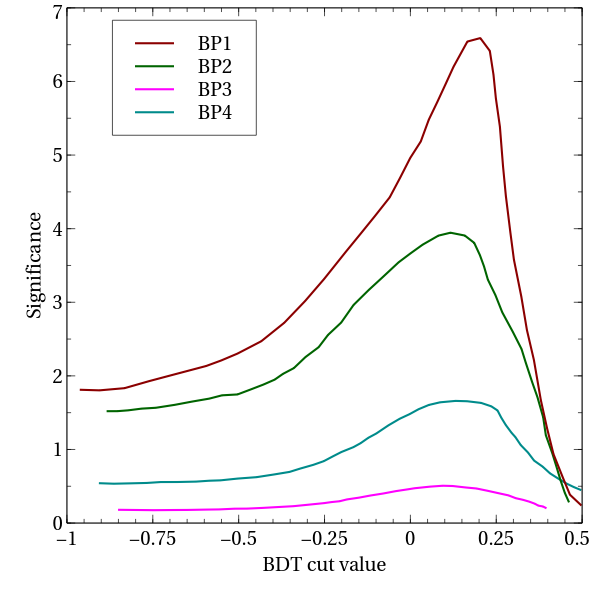}}}
\caption{ (a) ROC curves for chosen benchmark points for $2 \tau_\ell + \mET$ channel. (b) BDT-scores corresponding 
to BP1, BP2, BP3 for $2 \tau_\ell + \mET$ channel.}
\label{ROC-BDTScore-2lmet}
\end{figure}

To get an idea on the efficiency of distinguishing signals from backgrounds, one can plot background rejection against signal efficiency for the BPs in what is called the {\em Receiver’s Operative Characteristic} (ROC) curve. From Fig.\ref{ROC-BDTScore-2lmet}(a), one can see that the background rejection is maximum for BP1 and minimum for BP4. This pattern will be reflected while computing the signal significance, as we shall see shortly. Next to achieve maximum possible significance, we have to regulate the {\em BDT cut value} or the {\em BDT score}. Fig.\ref{ROC-BDTScore-2lmet}(b) depicts the variation of significance with BDT cut value. For different signal benchmarks, the significance attains the maximum value for a particular BDT score. 

In Table \ref{BDTD-2lmet}, we have tabulated the signal and the background yields at a reference integrated luminosity 4000 fb$^{-1}$. In the same table we quote the luminosity required to achieve  a 5$\sigma$ significance for each BP. The maximum observability is obtained for BP1. In fact, an $M_A \simeq 250$ GeV in BP2 is also found within the reach of the proposed 4000 fb$^{-1}$ integrated luminosity. Higher $M_A$ values however remain beyond  this reach.
\begin{center}
\begin{table}[htb!]
\begin{tabular}{|c|c|c|c|}\hline
 Benchmark Point & Signal Yield & Background Yield & $\mathcal{L}_{5\sigma}$ (fb$^{-1}$) \\ 
 & at 4000 fb$^{-1}$  &at 4000 fb$^{-1}$ &  \\ \hline 
BP1 & 1396 & 2412 & 146 \\ \hline
BP2 & 1066  & 15161 & 1366\\ \hline
BP3 & 126 &  8598 & $\sim 5.4 \times 10^{4}$ \\ \hline
BP4 & 53 &  10709 & $\sim 3.8 \times 10^{5}$ \\ \hline
 \end{tabular}
\caption{The signal and background yields at 1 TeV ILC with 4000 fb$^{-1}$ integrated luminosity for BP1,BP2, 
BP3,BP4 for the $ e^+ e^- \rightarrow 2 \tau_\ell +\mET$ channel after performing 
the BDTD analysis.}
\label{BDTD-2lmet}
\end{table}
\end{center}

\subsubsection{$1 \tau_\ell+ 1 \tau_h + \mET$ final state}
\label{1l1taumet}

The main background in this case is the common process $e^+ e^- \to \tau^+ \tau^{-} + \mET \to 1 \tau_\ell + 1 \tau_h + \mET$. Given the miniscule $j \to \tau_h$ and $j \to \ell$  misidentification rates in an $e^+ e^-$ environment, we find that the $e^+ e^- \to l^+ l^{-} + \mET$ and $e^+ e^- \to j j + \mET$ backgrounds become negligible in this case. We denote the lepton coming from the decay of one $\tau$-lepton as $\ell_1$ and the $\tau$-hadron as $\tau_{h_1}$. The following kinetic variables for the multivariate analysis
\bea
&& {M}_{{\ell_1 {\tau}_{{h_1}}}},~ \mET, ~ {p_T}_{{\ell_1 {\tau}_{{h_1}}}}^{\rm vect},~ M_T^{\rm vis}(\ell_1,{\tau}_{{h_1}}),~ {\eta}_{{\ell_1}}, ~ \Delta {\phi}_{{\ell_1 {\tau}_{{h_1}}}},~ p_T^{\ell_1 {\tau}_{{h_1}}}, \nonumber \\
&& \Delta {R}_{{\ell_1 {\tau}_{{h_1}}}},~ p_T^{{\tau}_{{h_1}}},~  \Delta {\eta}_{{\ell_1 {\tau}_{{h_1}}}},~ \Delta {\phi}_{{\ell_1 \mET}},~\Delta {\phi}_{{{\tau}_{{h_1}} \mET}}~,~ {\eta}_{{\tau}_{{h_1}}}
\eea

We define some of the variables not defined earlier. The invariant mass of the $\ell_1$ and ${\tau}_{{h_1}}$ in the final state is ${M}_{{\ell_1 {\tau}_{{h_1}}}}$. Next, ${p_T}_{{\ell_1 {\tau}_{{h_1}}}}^{\rm vect},~p_T^{\ell_1 {\tau}_{{h_1}}}$ and $p_T^{{\tau}_{{h_1}}}$ are the vector sum of the transverse momenta of $\ell_1$ and ${\tau}_{{h_1}}$, scalar sum of the transverse momenta of $\ell_1$ and ${\tau}_{{h_1}}$ and transverse momentum of the ${\tau}_{{h_1}}$ respectively. As mentioned in subsection \ref{2lmet}, $M_T^{\rm vis}(l_1,{\tau}_{{h_1}})$ can be defined following Eq.(\ref{MTvis}), where two visible decay products of the two $\tau$-leptons are $\ell_1$ and  ${\tau}_{{h_1}}$ for the present channel. The definition of the other variables can be understood from their notation and the previous subsection can also be referred to for clarification.

\begin{figure}[htpb!]{\centering
\subfigure[]{
\includegraphics[height = 6 cm, width = 8 cm]{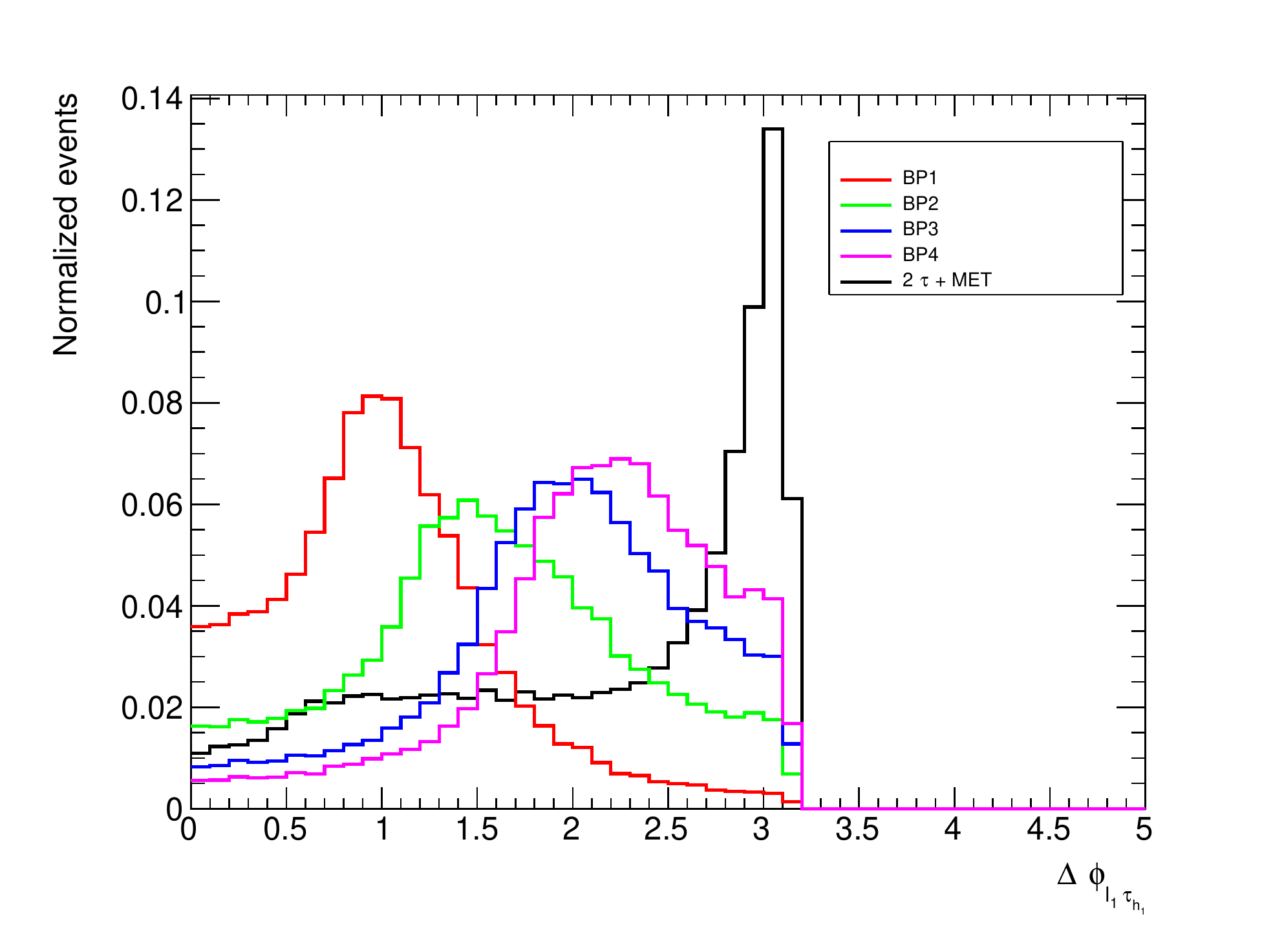}} 
\subfigure[]{
\includegraphics[height = 6 cm, width = 8 cm]{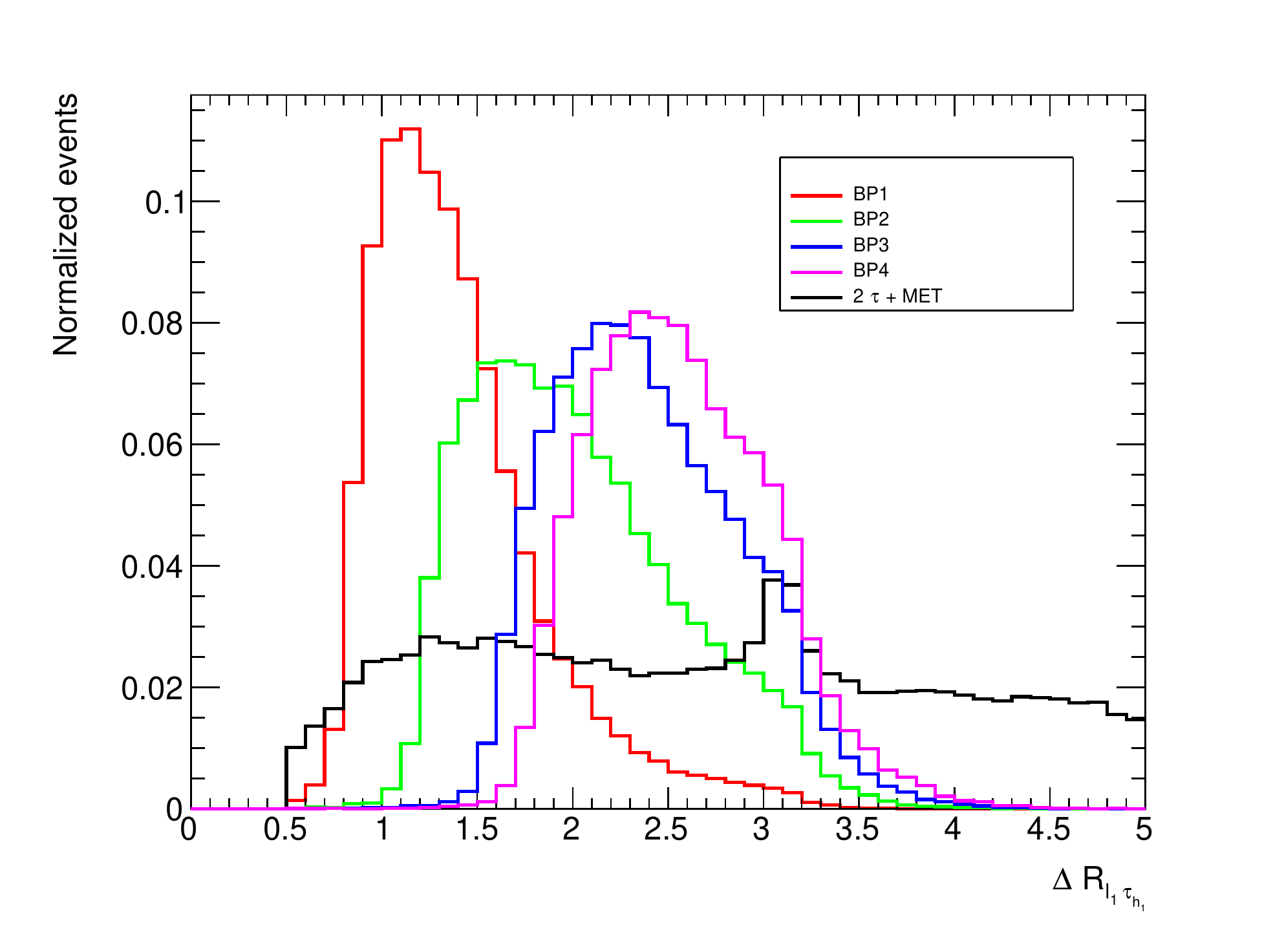}} \\
\subfigure[]{
\includegraphics[height = 6 cm, width = 8 cm]{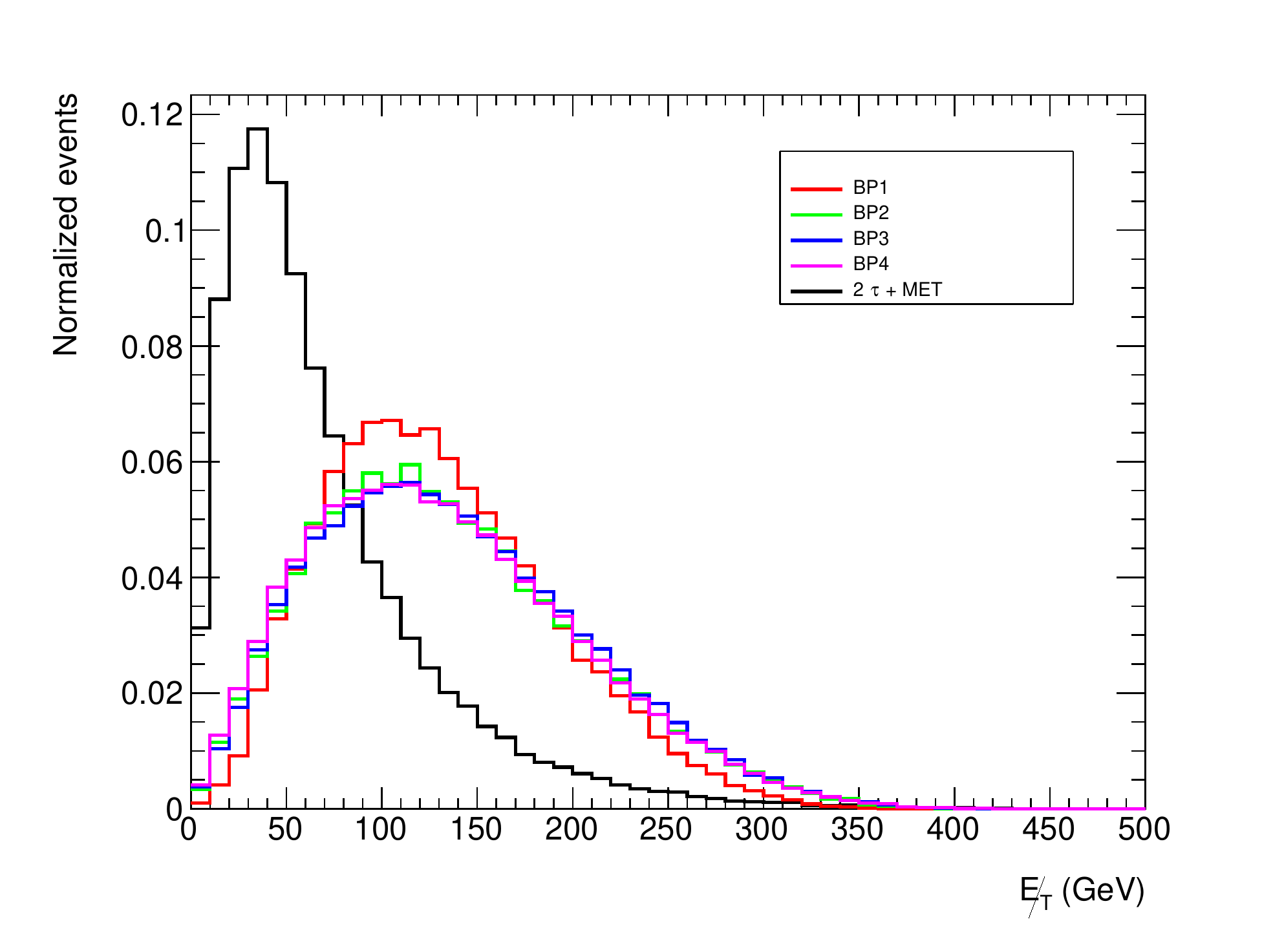}}
\subfigure[]{
\includegraphics[height = 6 cm, width = 8 cm]{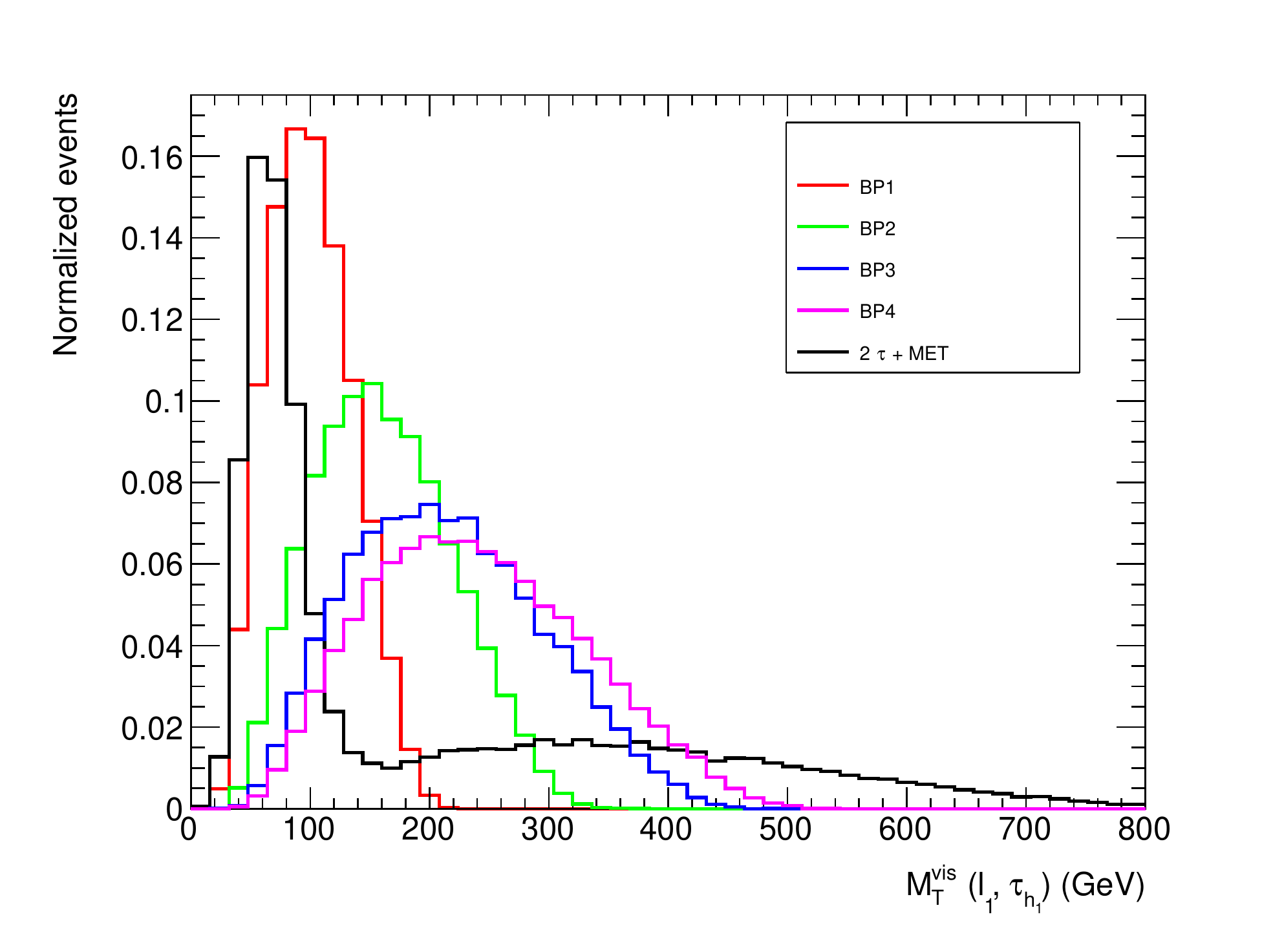}}
} \\
\caption{Normalized distributions of $\Delta {\phi}_{{\ell_1 {\tau}_{{h_1}}}},~ \Delta {R}_{{\ell_1 {\tau}_{{h_1}}}},~\mET,~ M_T^{\rm vis}(\ell_1,{\tau}_{{h_1}})$ for signal and backgrounds for $1 \tau_\ell+ 1 \tau_h + \mET$ final state.}
\label{distribution-1l1taumet}
\end{figure}

 Among the aforementioned important variables used in BDTD analysis, we present the normalized distributions of $\Delta {\phi}_{{\ell_1 {\tau}_{{h_1}}}},~ \Delta {R}_{{\ell_1 {\tau}_{{h_1}}}},~\mET,~ M_T^{\rm vis}(\ell_1,{\tau}_{{h_1}})$ in Fig.\ref{distribution-1l1taumet}(a),\ref{distribution-1l1taumet}(b),\ref{distribution-1l1taumet}(c) and \ref{distribution-1l1taumet}(d), where the distributions of the signal and backgrounds are maximally separated. The distributions of $\Delta {\phi}_{{\ell_1 {\tau}_{{h_1}}}},~ \Delta {R}_{{\ell_1 {\tau}_{{h_1}}}}$ for signal benchmarks peak at lower values, whereas the background distribution has a peak at higher values in Fig.\ref{distribution-1l1taumet}(a),\ref{distribution-1l1taumet}(b). The reason being the generation of two final state particles ($\ell_1$ and ${\tau}_{{h_1}}$) from a single mother particle $A$ for signal. For background the two $\tau$-leptons not necessarily originate from a single source, providing a large separation. The nature of the $\mET$ distribution has already been explained in subsection \ref{2lmet}. From Fig.\ref{distribution-1l1taumet}(d), one can see that $M_T^{\rm vis}(\ell_1,{\tau}_{{h_1}})$ has the ability to differentiate the signal and background efficiently.

\begin{table}[htpb!]
\begin{center}
\begin{tabular}{|c|c|c|c|c|c|}
\hline
 &  \hspace{5mm} {\texttt{NTrees}} \hspace{5mm} & \hspace{5mm} {\texttt{MinNodeSize}} \hspace{5mm} & \hspace{5mm} {\texttt{MaxDepth}}~~ \hspace{5mm} & \hspace{5mm} {\texttt{nCuts}} ~~\hspace{5mm} & \hspace{5mm} {\texttt{KS-score for}}~~\hspace{5mm}\\
 & & & & & {\texttt{Signal(Background)}} \\
\hline
\hline
\hspace{5mm} BP1 \hspace{5mm} & 110 & 4 \% & 2.0 & 55 & 0.235~(0.378) \\ \hline
\hspace{5mm} BP2 \hspace{5mm} & 110 & 4 \% & 2.0 & 55 & 0.074~(0.363) \\ \hline
\hspace{5mm} BP3 \hspace{5mm} & 110 & 4 \% & 2.0 & 50 & 0.018~(0.304) \\ \hline
\hspace{5mm} BP4 \hspace{5mm} & 110 & 4 \% & 2.0 & 50 & 0.908~(0.131) \\ \hline
\end{tabular}
\end{center}
\caption{Tuned BDT parameters for BP1, BP2, BP3, BP4 for the $1 \tau_\ell + 1 \tau_h + \mET$ channel.}
\label{BDT-param-1l1taumet}
\end{table}

\begin{figure}[htpb!]{\centering
\subfigure[]{
\includegraphics[width=3in,height=2.45in]{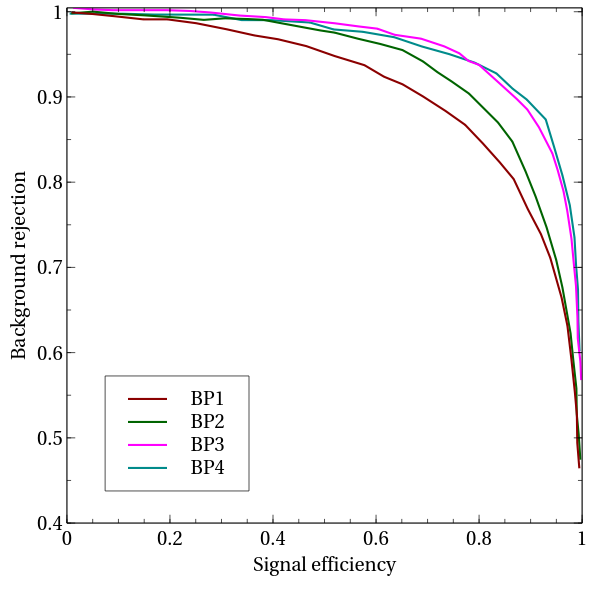}}
\subfigure[]{
\includegraphics[width=3.1in,height=2.46in]{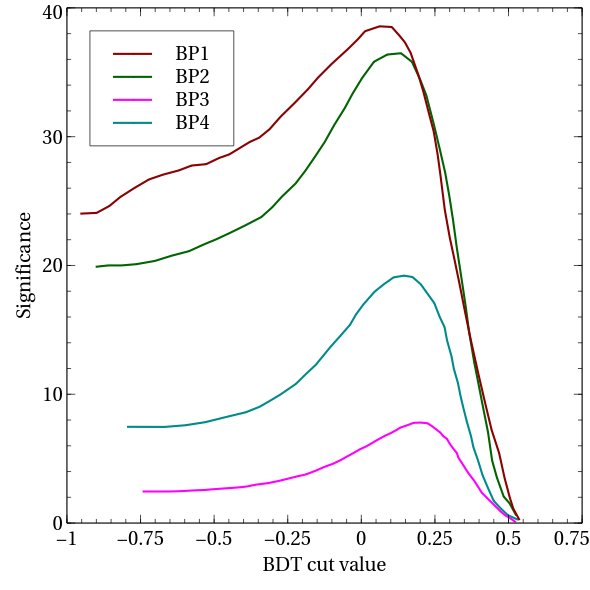}}}
\caption{ (a) ROC curves for chosen benchmark points for $1\tau_\ell + 1 \tau_h + \mET$ channel. (b) BDT-scores corresponding 
to BP1, BP2, BP3 for $ 1\tau_\ell + 1 \tau_h + \mET$ channel.}
\label{ROC-BDTScore-1l1taumet}
\end{figure}

We train the signal and background samples by adjusting the BDTD parameters (mentioned previously in subsection \ref{2lmet}) tabulated in Table \ref{BDT-param-1l1taumet}. The ROC curves and the variation of the significance with BDT score for this channel are depicted in Fig.\ref{ROC-BDTScore-1l1taumet}(a) and Fig.\ref{ROC-BDTScore-1l1taumet}(b) respectively. Fig.\ref{ROC-BDTScore-1l1taumet}(a) suggests that background rejection is the best for BP3 and BP4.

\begin{center}
\begin{table}[htb!]
\begin{tabular}{|c|c|c|c|}\hline
 Benchmark Point & Signal Yield & Background Yield & $\mathcal{L}_{5\sigma}$ (fb$^{-1}$)  \\ 
 & at 4000 fb$^{-1}$  &at 4000 fb$^{-1}$ &  \\ \hline 
BP1 & 3345 & 455 & 11 \\ \hline
BP2 &  1847 & 245 & 19\\ \hline
BP3 &  200 & 60 & 276\\ \hline
BP4 & 53 &  127 &  5114 \\ \hline
 \end{tabular}
\caption{The signal and background yields at 1 TeV ILC with 4000 fb$^{-1}$ integrated luminosity for BP1,BP2, 
BP3, BP4 for the $ e^+ e^- \rightarrow 1 \ell + 1 \tau_h +\mET$ channel after performing 
the BDTD analysis. }
\label{BDTD-1l1taumet}
\end{table}
\end{center}

It is read from Table \ref{BDTD-1l1taumet} that the $1\tau_\ell + 1\tau_h + \mET$
channel comes with a 5$\sigma$ discovery potential for $M_A$ up to 350 GeV (BP1-3) within 300 fb$^{-1}$ integrated luminosity.

\subsubsection{$2 \tau_h+ \mET$ final state}

We denote the hadronic decay products of the two $\tau$-leptons by $\tau_{h_1}$ and $\tau_{h_2}$. The only sizeable background in this case too is $e^+ e^- \to \tau^+ \tau^- + \mET$ since $e^+ e^- \to j j + \mET,~\ell^+ \ell^- + \mET$ have negligible effects, as explained in the previous subsection. The kinematic variables used for the BDTD analysis for this case are
\bea
&& M_{{\tau}_{{h_1}} {\tau}_{{h_2}}},~ \mET,~ {p_T}_{{\tau}_{{h_1}} {\tau}_{{h_2}}}^{\rm vect},~ {p_T}^{{\tau}_{{h_1}}},~ {M_T}^{\rm vis} ({{\tau}_{{h_1}}, {\tau}_{{h_2}}}),~ {\eta}_{{\tau}_{{h_1}}},~ \Delta {\eta}_{{\tau}_{{h_1}} {\tau}_{{h_2}}}, ~
  \Delta \phi_{{\tau}_{{h_1}} \mET}~,\nonumber \\
  && p_T^{{\tau}_{{h_1}} {\tau}_{{h_2}}},~\Delta \phi_{{\tau}_{{h_1}} {\tau}_{{h_2}}}, 
 \Delta R_{{\tau}_{{h_1}} {\tau}_{{h_2}}},~ \eta_{{\tau}_{{h_2}}},~ \Delta \phi_{{\tau}_{{h_2}} \mET}~,~ \phi_{{\tau}_{{h_1}}},~ \phi_{{\tau}_{{h_2}}}
\eea

\begin{figure}[htpb!]{\centering
\subfigure[]{
\includegraphics[height = 5.5 cm, width = 8 cm]{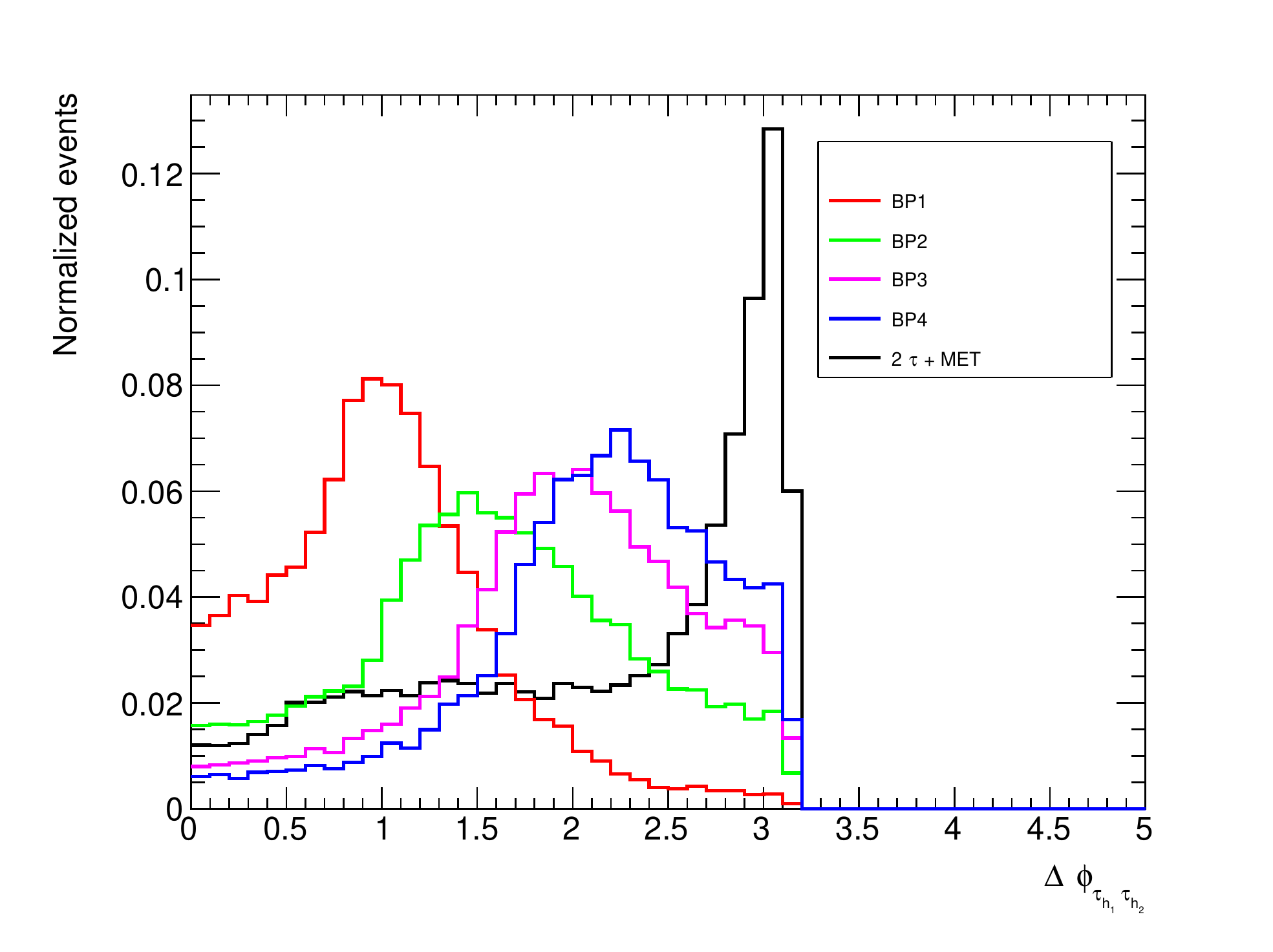}} 
\subfigure[]{
\includegraphics[height = 5.5 cm, width = 8 cm]{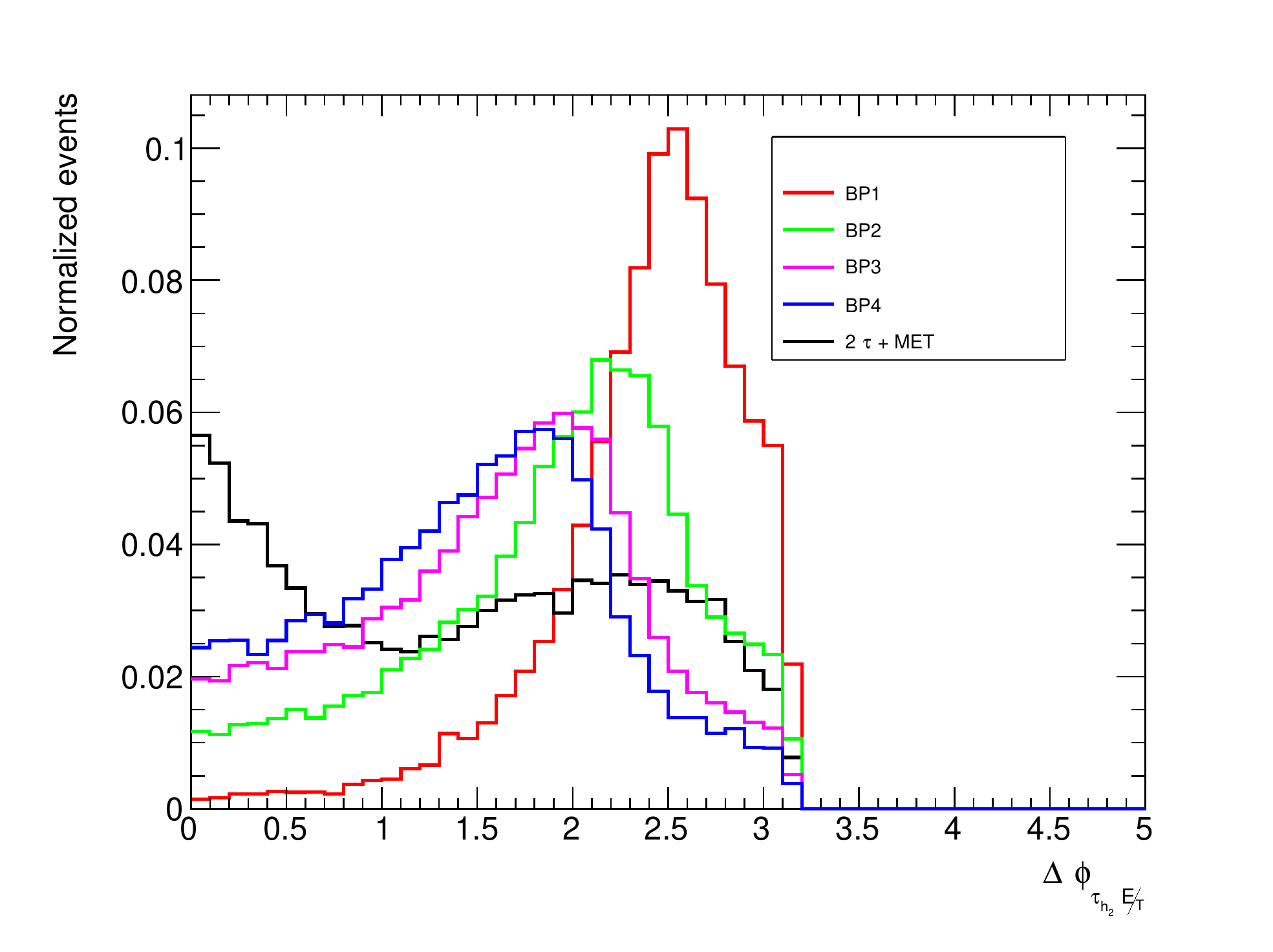}} \\
\subfigure[]{
\includegraphics[height = 5.5 cm, width = 8 cm]{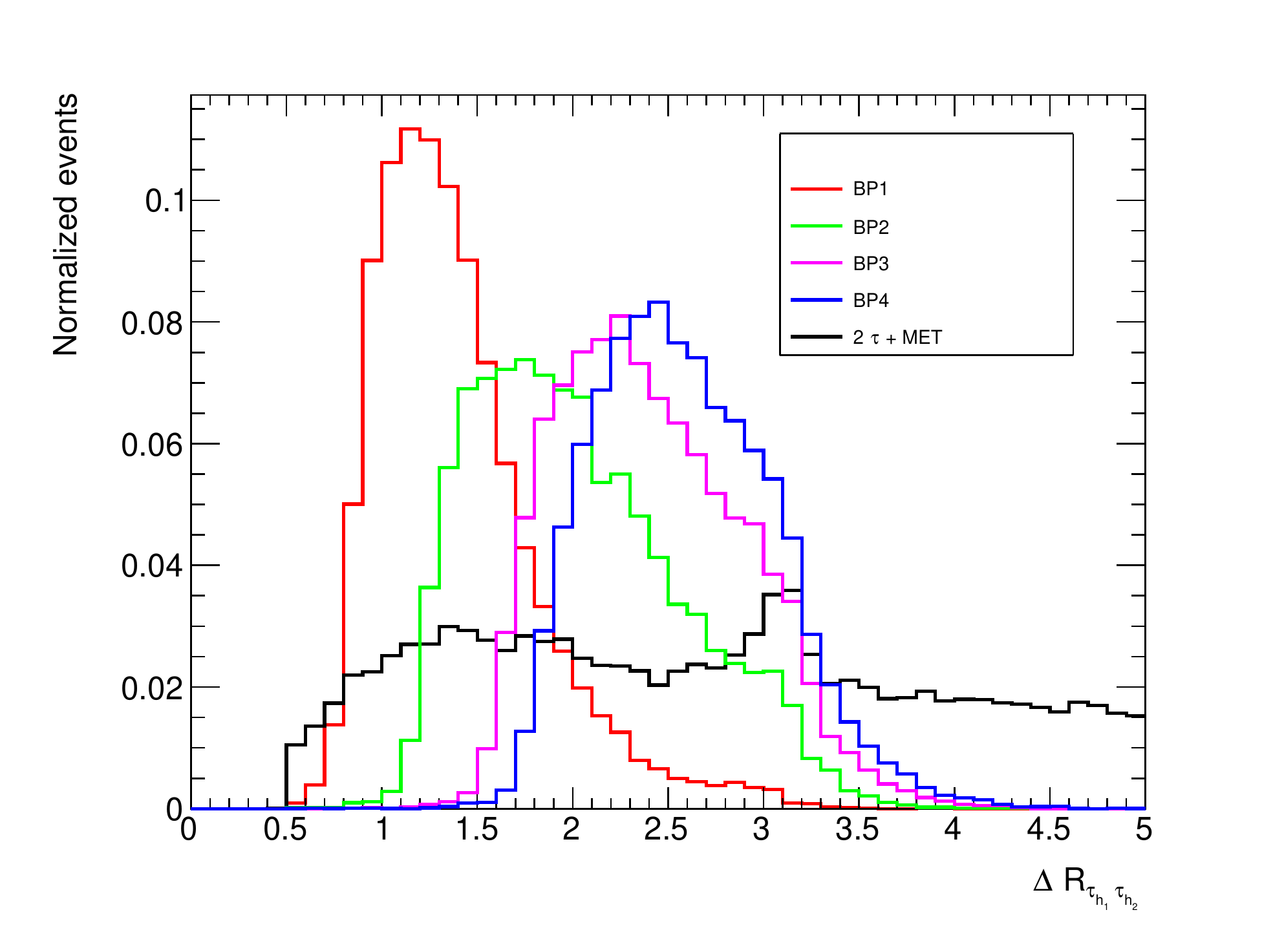}}
\subfigure[]{
\includegraphics[height = 5.5 cm, width = 8 cm]{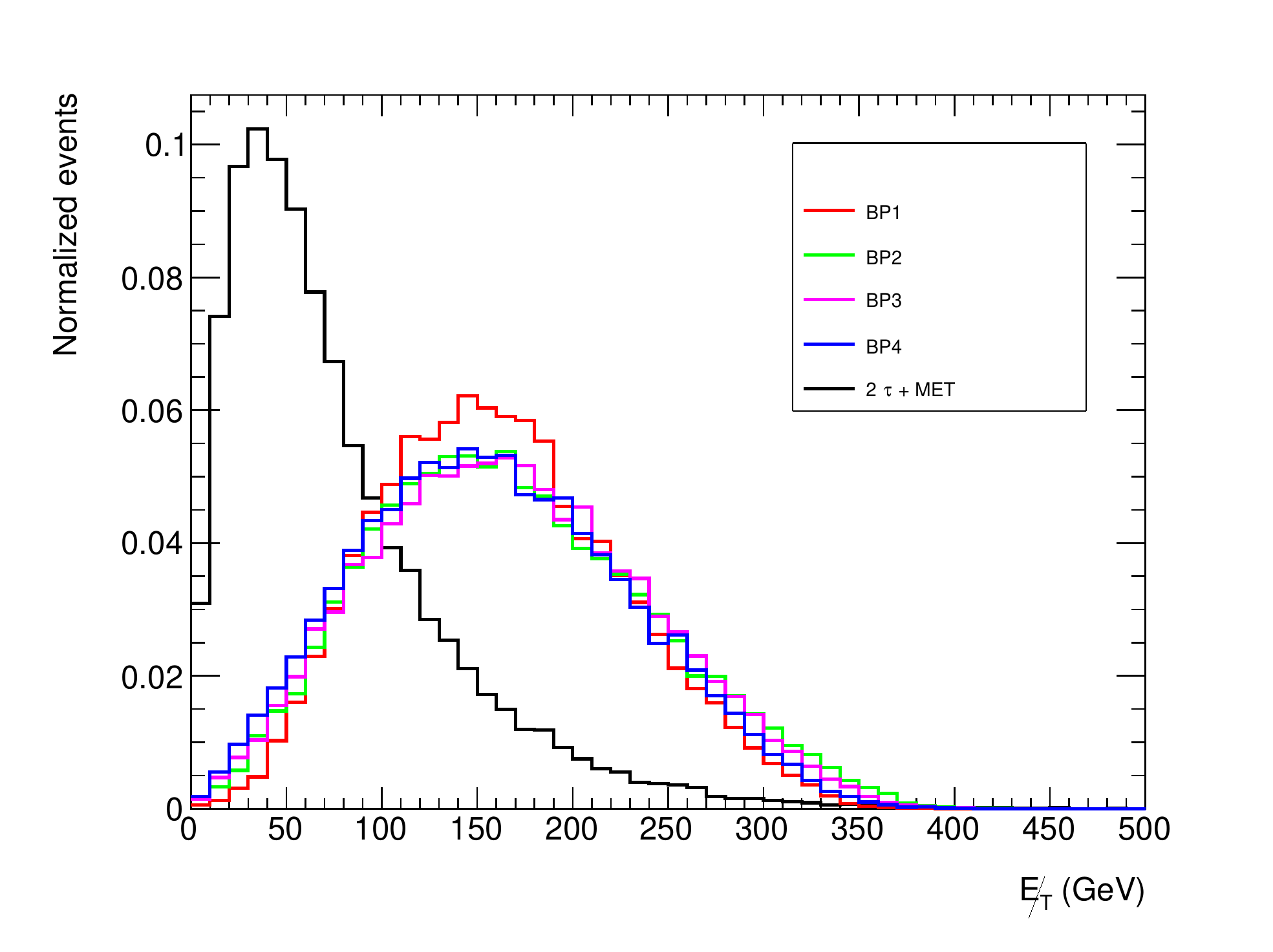}} \\
\subfigure[]{
\includegraphics[height = 5.5 cm, width = 8 cm]{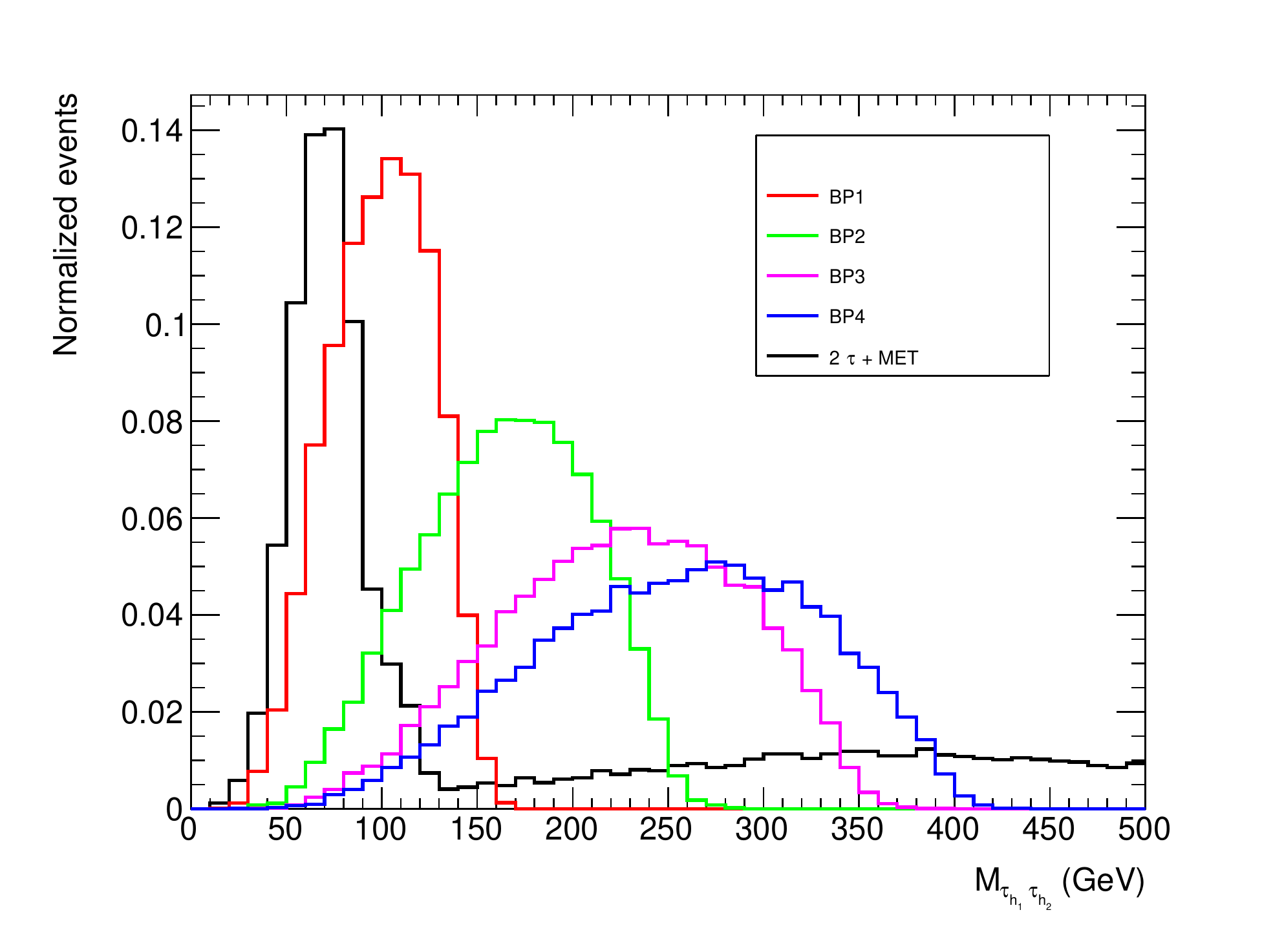}} 
\subfigure[]{
\includegraphics[height = 5.5 cm, width = 8 cm]{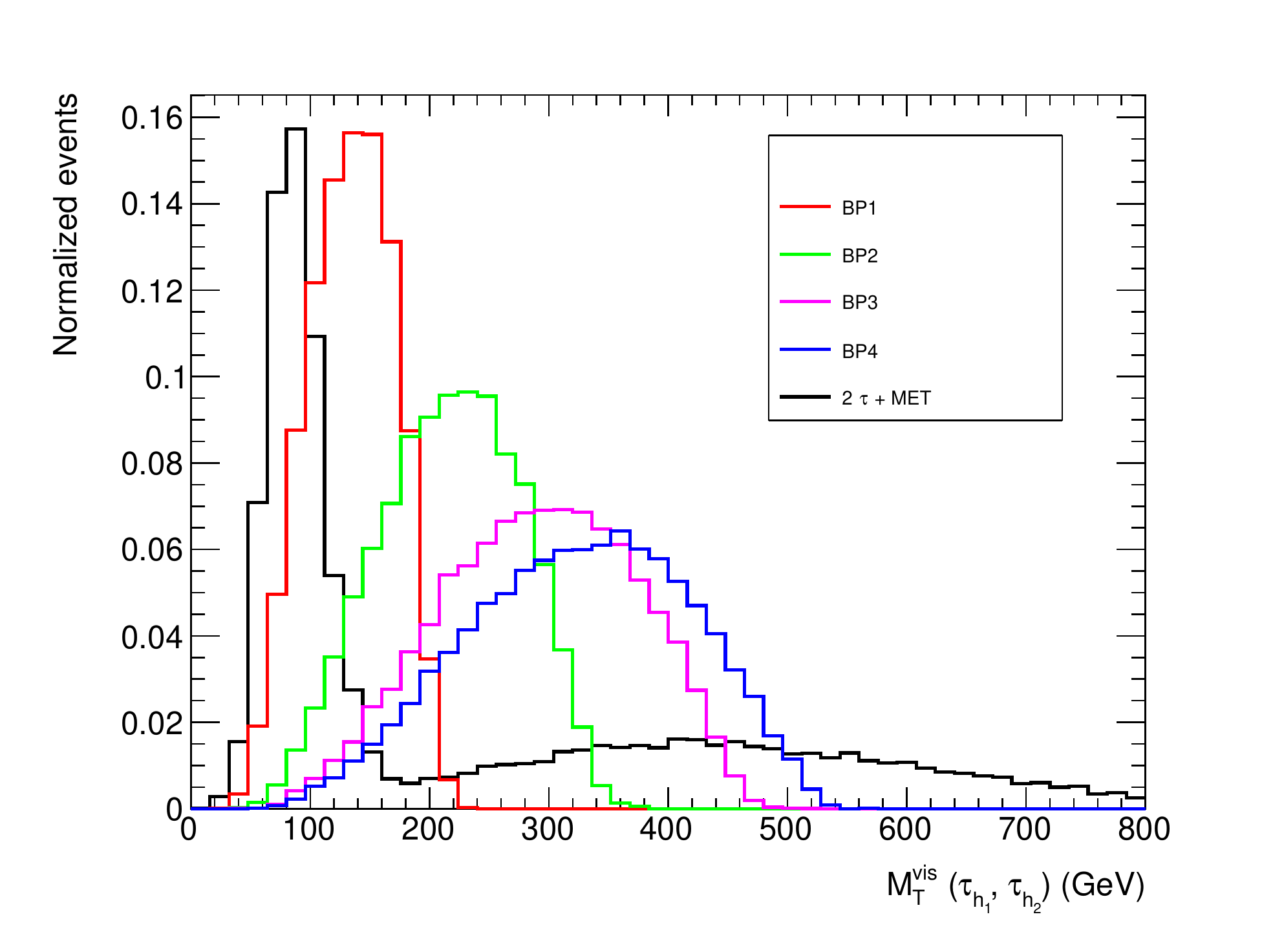}} 
} \\
\caption{ Normalized distributions for $\Delta \phi_{{\tau}_{{h_1}} {\tau}_{{h_2}}}, \Delta \phi_{{\tau}_{{h_2}} \mET}~,  \Delta R_{{\tau}_{{h_1}} {\tau}_{{h_2}}}, \mET, M_{{\tau}_{{h_1}} {\tau}_{{h_2}}},  {M_T}^{\rm vis} ({{\tau}_{{h_1}}, {\tau}_{{h_2}}})$ for $2 \tau_h + \mET$ channel at 1 TeV ILC.}
\label{distribution-2taumet-1}
\end{figure}

\begin{figure}[htpb!]{\centering
\subfigure[]{
\includegraphics[height = 5.5 cm, width = 8 cm]{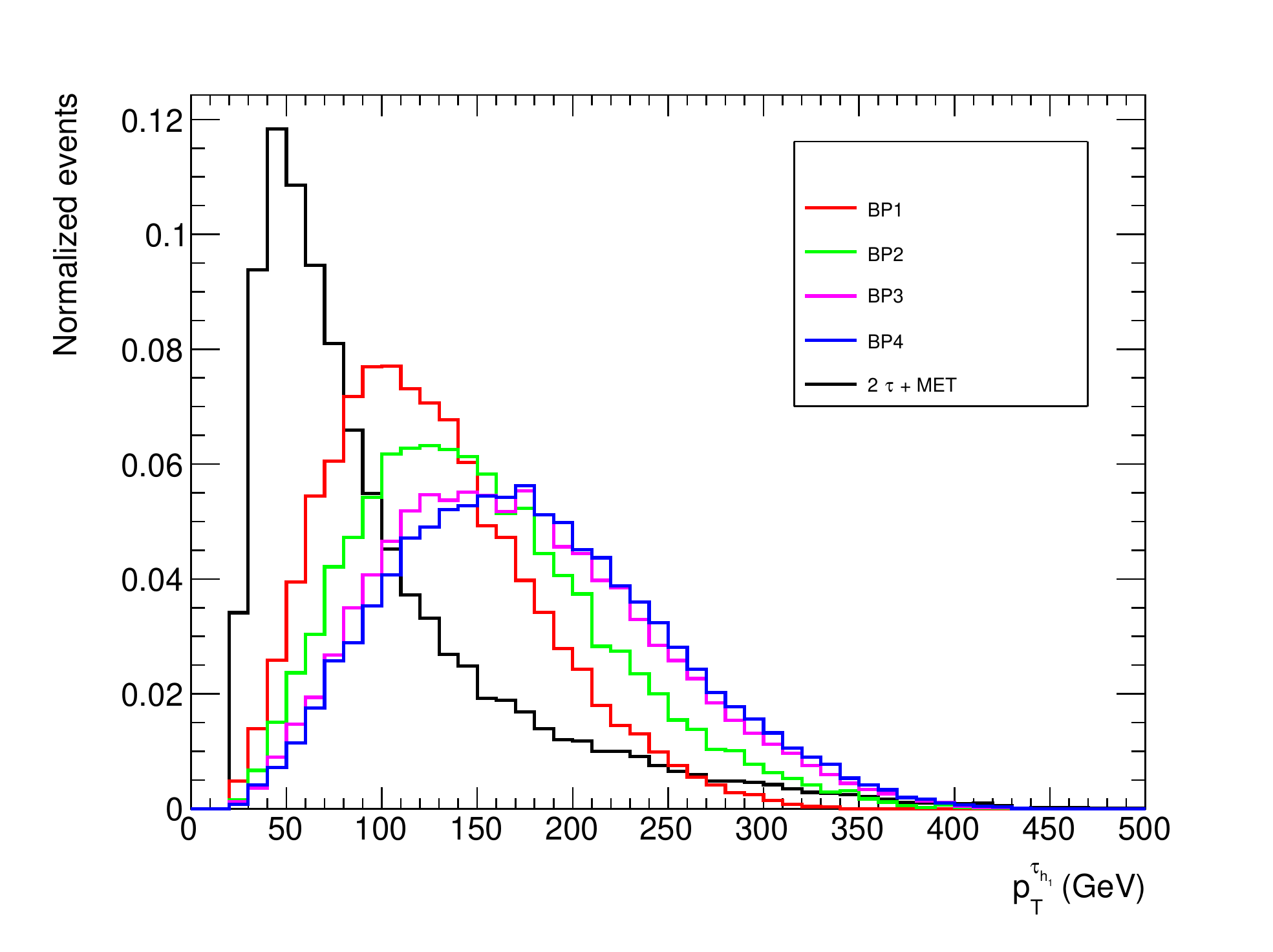}}
\subfigure[]{
\includegraphics[height = 5.5 cm, width = 8 cm]{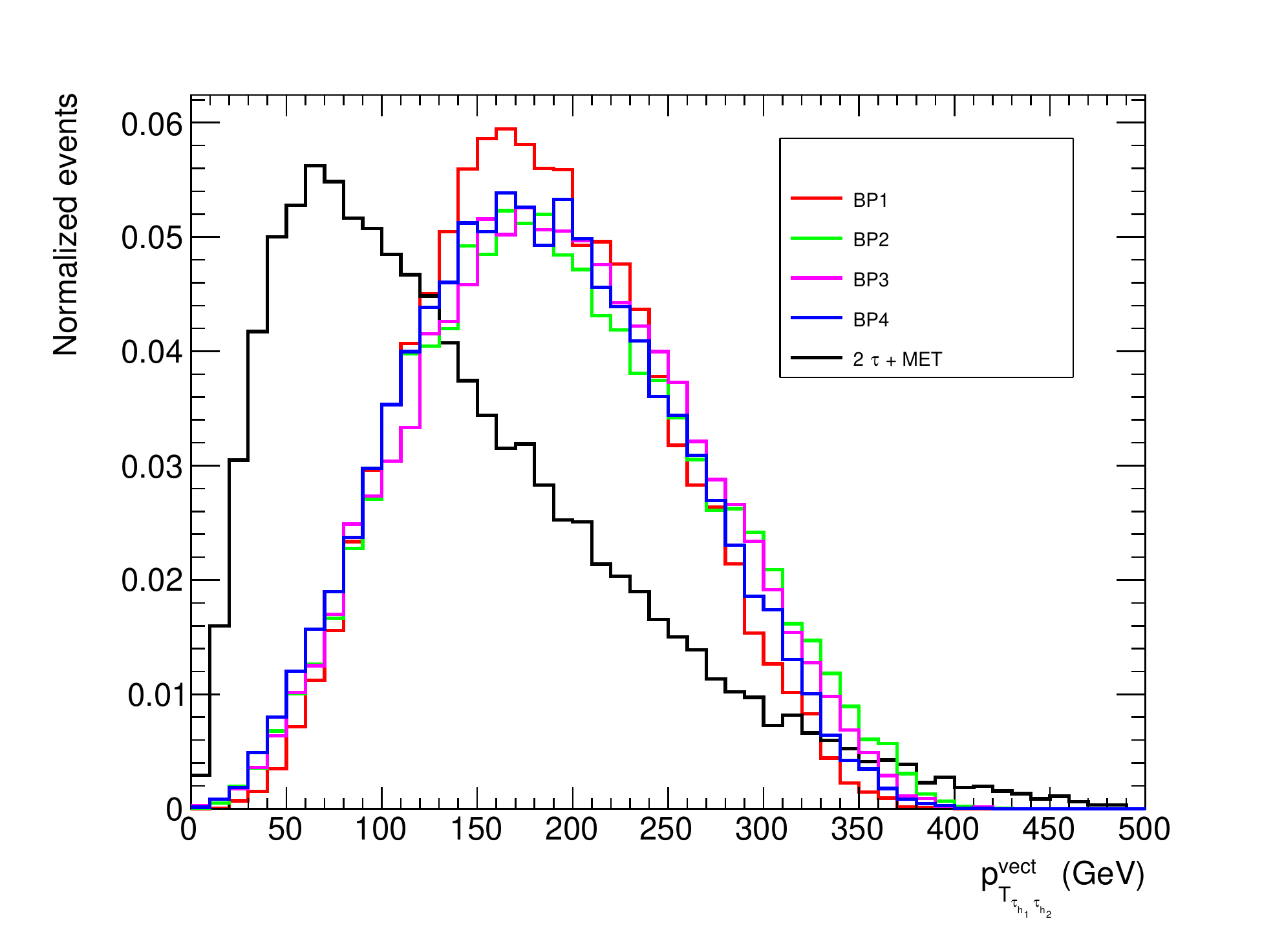}}
} \\
\caption{ Normalized distributions of ${p_T}_{{\tau}_{{h_1}}},{p_T}_{{\tau}_{{h_1}} {\tau}_{{h_2}}}^{\rm vect}$ for $2 \tau_h + \mET$ channel at 1 TeV ILC. }
\label{distribution-2taumet-2}
\end{figure}

Here $M_{{\tau}_{{h_1}} {\tau}_{{h_2}}}$ is the invariant mass of two $\tau$-jets in the final state. ${p_T}_{{\tau}_{{h_1}} {\tau}_{{h_2}}}^{\rm vect},~p_T^{{\tau}_{{h_1}} {\tau}_{{h_2}}},~ {p_T}^{{\tau}_{{h_1}}}$ are the vector and scalar sum of the transverse momenta of two $\tau$-jets, transverse momentum of leading $\tau$-jet respectively. Other variables are already defined earlier for different final states. Normalized distributions for $\Delta \phi_{{\tau}_{{h_1}} {\tau}_{{h_2}}}, \Delta \phi_{{\tau}_{{h_2}} \mET}~,  \Delta R_{{\tau}_{{h_1}} {\tau}_{{h_2}}}, \mET, M_{{\tau}_{{h_1}} {\tau}_{{h_2}}},  {M_T}^{\rm vis} ({{\tau}_{{h_1}}, {\tau}_{{h_2}}}), {p_T}_{{\tau}_{{h_1}}},{p_T}_{{\tau}_{{h_1}} {\tau}_{{h_2}}}^{\rm vect} $ are presented in Fig.\ref{distribution-2taumet-1}(a),\ref{distribution-2taumet-1}(b),\ref{distribution-2taumet-1}(c),\ref{distribution-2taumet-1}(d),\ref{distribution-2taumet-1}(e),\ref{distribution-2taumet-1}(f),\ref{distribution-2taumet-2}(a),\ref{distribution-2taumet-2}(b) respectively. The nature of the distributions of signals and backgrounds for each variable can be explained from the previous subsections which contain analysis of different final states arising out of the leptonic and/or hadronic decay modes of two $\tau$-leptons.

\begin{figure}[htpb!]{\centering
\subfigure[]{
\includegraphics[width=3in,height=2.45in]{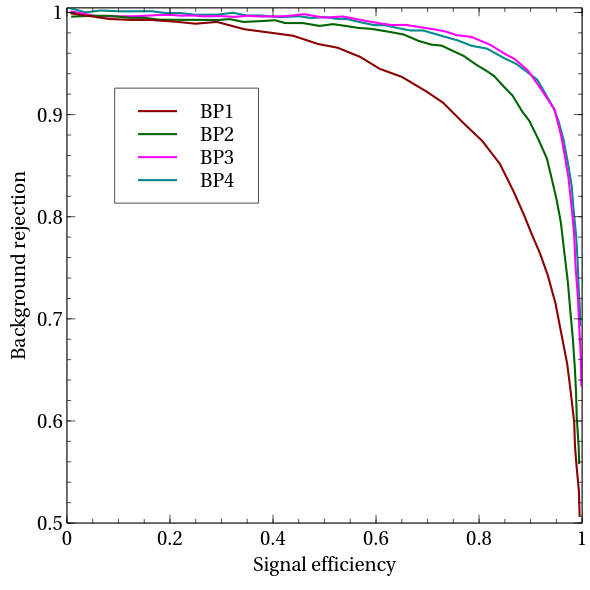}}
\subfigure[]{
\includegraphics[width=3.1in,height=2.46in]{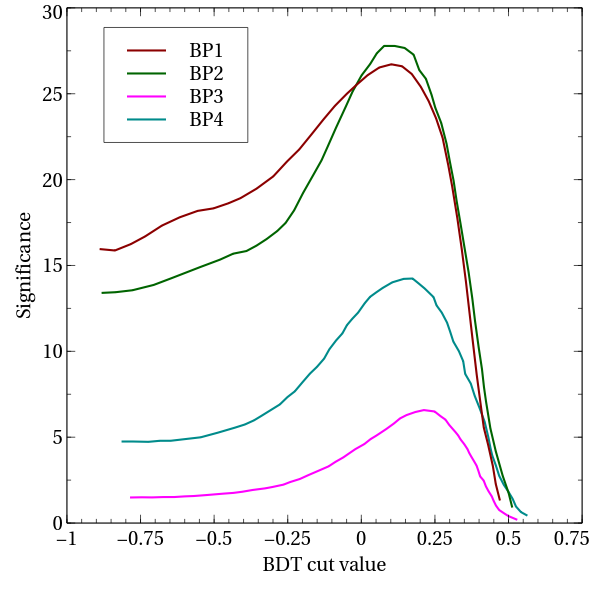}}}
\caption{ (a) ROC curves for chosen benchmark points for $2 \tau_h + \mET$ channel. (b) BDT-scores corresponding 
to BP1, BP2, BP3 for $2 \tau_h + \mET$ channel.}
\label{ROC-BDTScore-tautaumet}
\end{figure}

The signal and background KS-scores for each BP are shown in Table \ref{BDT-param-tautaumet} along with the corresponding tuned values of the BDTD variables. We also show the ROC curve and the significance vs. BDT cut-value plot in Fig.\ref{ROC-BDTScore-tautaumet}(a) and Fig.\ref{ROC-BDTScore-tautaumet}(b) respectively. It is read from the ROC curve that background-rejection in the least efficient for BP1. The efficiency enhances with increasing $M_A$ albeit BP3 and BP4 are close by in this regard.

\begin{table}[htpb!]
\begin{center}
\begin{tabular}{|c|c|c|c|c|c|}
\hline
 &  \hspace{5mm} {\texttt{NTrees}} \hspace{5mm} & \hspace{5mm} {\texttt{MinNodeSize}} \hspace{5mm} & \hspace{5mm} {\texttt{MaxDepth}}~~ \hspace{5mm} & \hspace{5mm} {\texttt{nCuts}} ~~\hspace{5mm} & \hspace{5mm} {\texttt{KS-score for}}~~\hspace{5mm}\\
 & & & & & {\texttt{Signal(Background)}} \\
\hline
\hline
\hspace{5mm} BP1 \hspace{5mm} & 120 & 3 \% & 2.0 & 50 & 0.013~(0.078) \\ \hline
\hspace{5mm} BP2 \hspace{5mm} & 120 & 3 \% & 2.0 & 50 & 0.508~(0.41) \\ \hline
\hspace{5mm} BP3 \hspace{5mm} & 120 & 3 \% & 2.0 & 55 & 0.346~(0.041) \\ \hline
\hspace{5mm} BP4 \hspace{5mm} & 120 & 4 \% & 2.0 & 40 & 0.065~(0.028) \\ \hline
\end{tabular}
\end{center}
\caption{Tuned BDT parameters for BP1, BP2, BP3, BP4 for the $2 \tau_h + \mET$ channel.}
\label{BDT-param-tautaumet}
\end{table}

\begin{center}
\begin{table}[htb!]
\begin{tabular}{|c|c|c|c|}\hline
 Benchmark Point & Signal Yield & Background Yield & $\mathcal{L}_{5\sigma}$ (fb$^{-1}$)  \\ 
 & at 4000 fb$^{-1}$  &at 4000 fb$^{-1}$ &  \\ \hline 
BP1 & 1356 & 120 & 22 \\ \hline
BP2 &  861 & 57 & 30\\ \hline
BP3 &  87 & 10 & 375\\ \hline
BP4 & 40 &  36 &  2979 \\ \hline
 \end{tabular}
\caption{The signal and background yields at 1 TeV ILC with 4000 fb$^{-1}$ integrated luminosity for BP1,BP2, 
BP3, BP4 for the $ e^+ e^- \rightarrow 2 \tau_h +\mET$ channel after performing 
the BDTD analysis.  }
\label{BDTD-tautaumet}
\end{table}
\end{center}

Following the BDTD analysis, we summarize the discovery prospects for the various BPs in Table \ref{BDTD-tautaumet}. For the $\tau^+\tau^-$ pair decaying fully hadronically, BP1-3 can be discovered at $5\sigma$ within 375 fb$^{-1}$. In fact, BP4 is also found within the reach of the proposed 4000 fb$^{-1}$ integrated luminosity.

We compare the efficacies of the $2\tau_\ell + \mET,1\tau_\ell + 1\tau_h + \mET,2\tau_h + \mET$ channels before closing the discussion of the lepton-specific (2+1)HDM.
It is readily seen that $2 \tau_\ell + \mET$ is the least promising among the three. And this is attributed to two reasons. First, the leptonic decay of a $\tau$ has a smaller branching fraction than the hadronic one. More importantly, identification efficiency of the leptons coming from $\tau$ decays is poor in a realistic collider environment. The other two channels are mutually competing. For both, the BDTD training is the most efficiently trained to reject background in case of BP4. The quality of training becomes inferior for the BPs with lower values of $M_A$. While the semileptonic channel is found to be more promising in case of BP1-3, the hadronic channel is more efficient in case of BP4.

Lastly, we also compare the performances of the LHC and the ILC in looking for an $A$ in the $2\tau_h + \mET$ final state. It is seen from \cite{Chakrabarty:2021ztf} that an $A$ of mass $\simeq$ 250 GeV can be observed at  
5$\sigma$ at the LHC when the integrated luminosity is around 3300 fb$^{-1}$. Therefore, the LHC discovery potential is considerably less compared to the ILC which predicts 
5$\sigma$ observability for $M_A \simeq 400$ GeV for an integrated luminosity around 3000 fb$^{-1}$. Therefore, this enhanced observability at the ILC is a clear upshot of the present analysis. And this is attributed to the fact that the hadronic background in a leptonic collider
is miniscule compared to in a hadronic collider. And the former therefore would be generically more efficient in detecting hadronic activity stemming from a BSM scenario, an example of which is the $e^+ e^- \to \eta_R \eta_I \to \eta_R \eta_R A \to \tau^+ \tau^- + \mET$ signal.

\subsection{Muon specific 2HDM}  

We present two new sample points (SPs) in Table \ref{tab:SP} for the muon-specific case from the corresponding allowed parameter region. It follows from the preceding discussions on the lepton-specific case that the most relevant background in the muon-specific case would be $e^+ e^- \to \ell^+ \ell^- + \mET$. 

\begin{table}
\centering
\begin{tabular}{ |c|c|c| } 
\hline
& SP1 & SP2  \\ \hline
$m_{12}$ & 22.8 GeV & 21.6 GeV    \\
tan$\b$ & 42.94 & 47.98    \\
$M_A$ & 153.28 GeV & 228.74 GeV  \\
$M_{\eta_I}$ & 292.0 GeV & 569.5 GeV     \\
$k_1$ & -1.74673 & -2.27451    \\
$\omega_1$ &  1.3069 &  -3.12903   \\
$\s_1$ & -4.20973 & -5.4915    \\
$\s_2$ & 4.32283 & 6.06956    \\
$\s_3$ & 6.14496 & -5.5669    \\
$\Delta a_\mu$ & 1.47372 & 1.38208    \\
$\sigma^{eff}_{SI}$ &  cm$^2$ &   cm$^2$    \\ \hline
BR$(\eta_I \to \eta_R A)$ & 0.971753 & 0.633975    \\ 
BR$(A \to \mu^+ \mu^-)$ & 0.998527 & 0.999054    \\ \hline
\end{tabular}
\caption{Benchmark points used for studying the discovery prospects of an $A$ in the muon-specific (2+1)HDM. The values for the rest of the masses are 
$M_H = M_{H^+} = 150$ GeV, $M_{\eta^+} = M_{\eta_R} + 1$ GeV = 100 GeV.}
\label{tab:SP}
\end{table}

\begin{figure}[htpb!]{\centering
\subfigure[]{
\includegraphics[height = 5.5 cm, width = 8 cm]{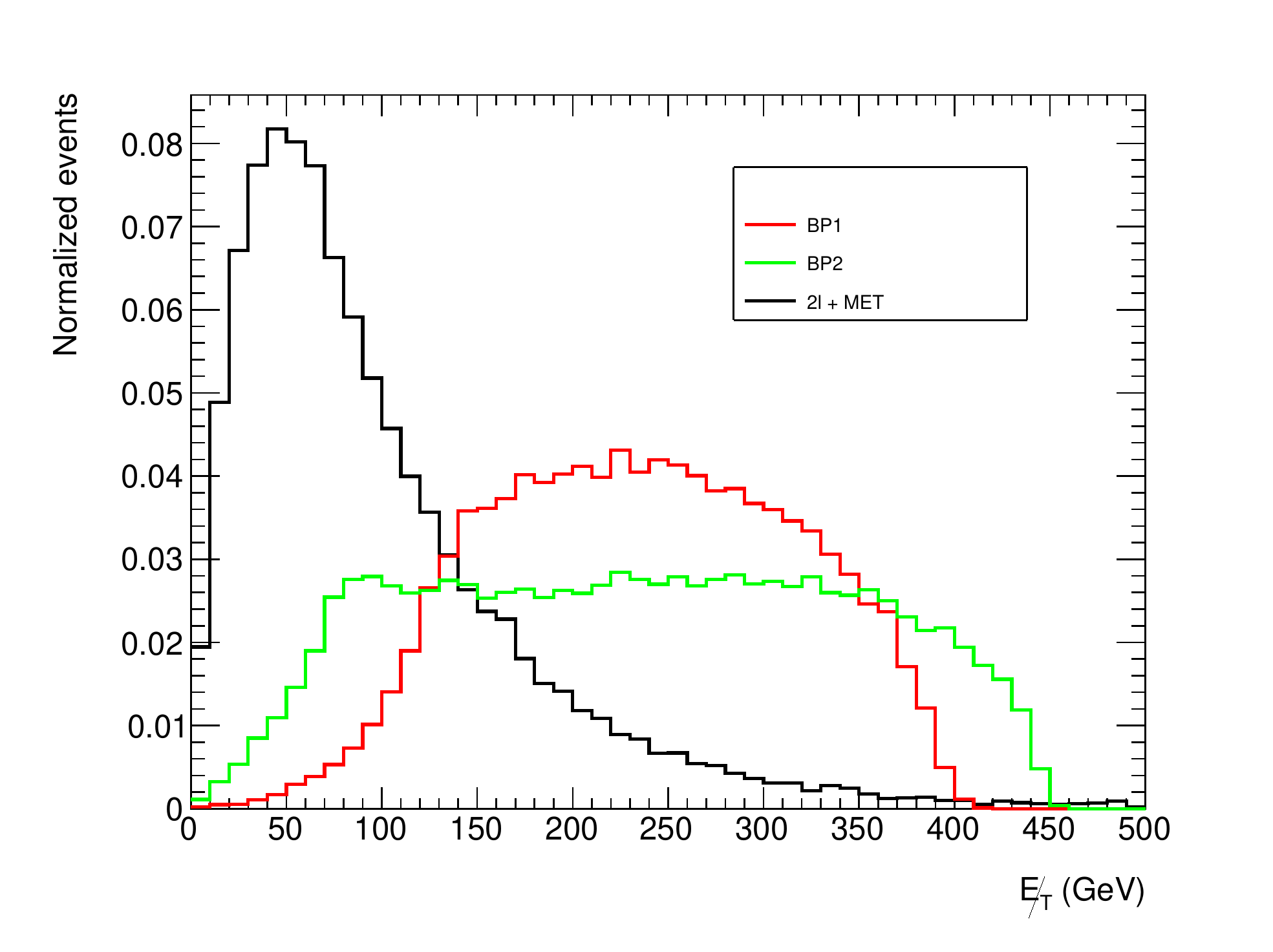}}
\subfigure[]{
\includegraphics[height = 5.5 cm, width = 8 cm]{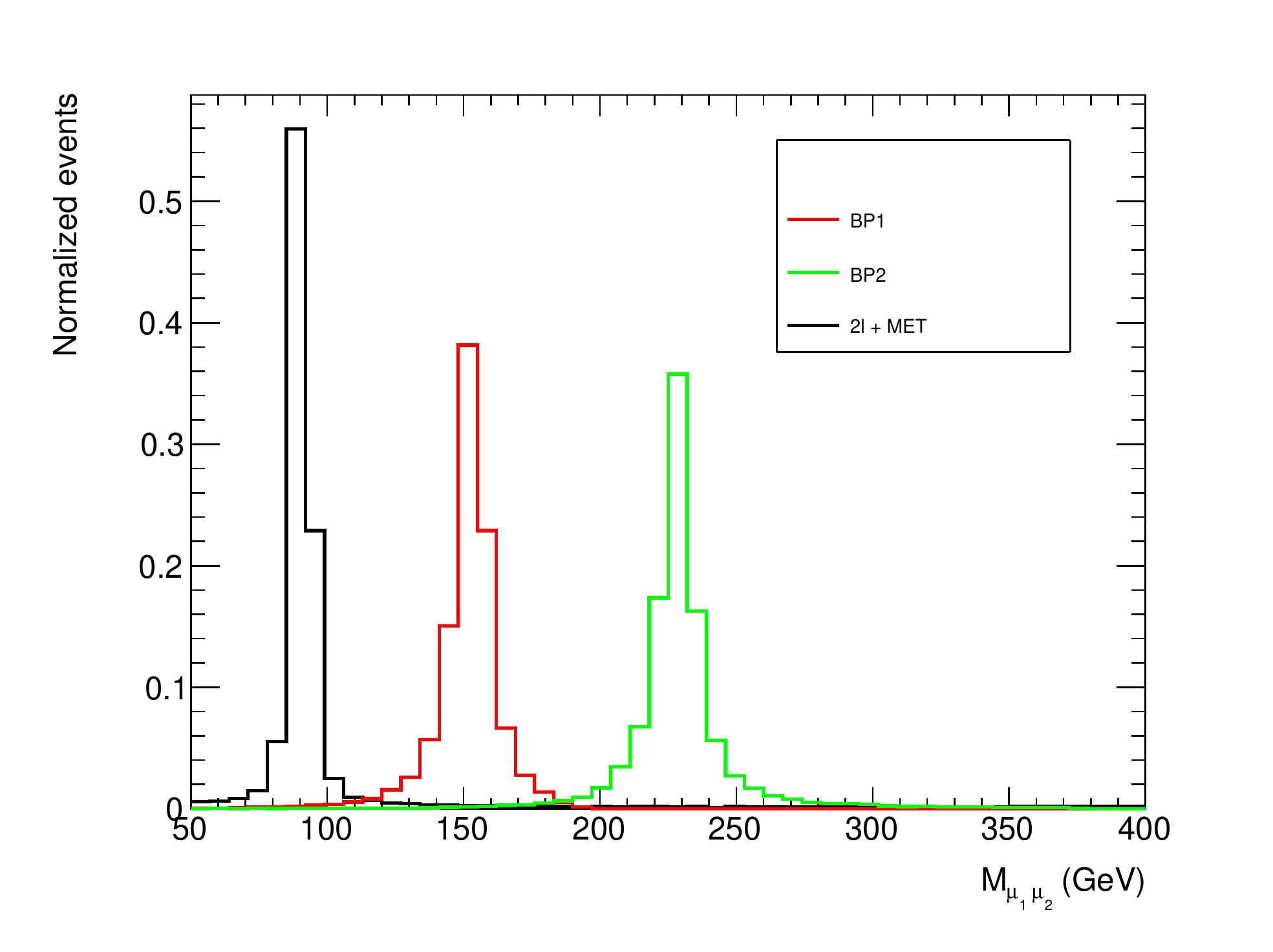}}
\subfigure[]{
\includegraphics[height = 5.5 cm, width = 8 cm]{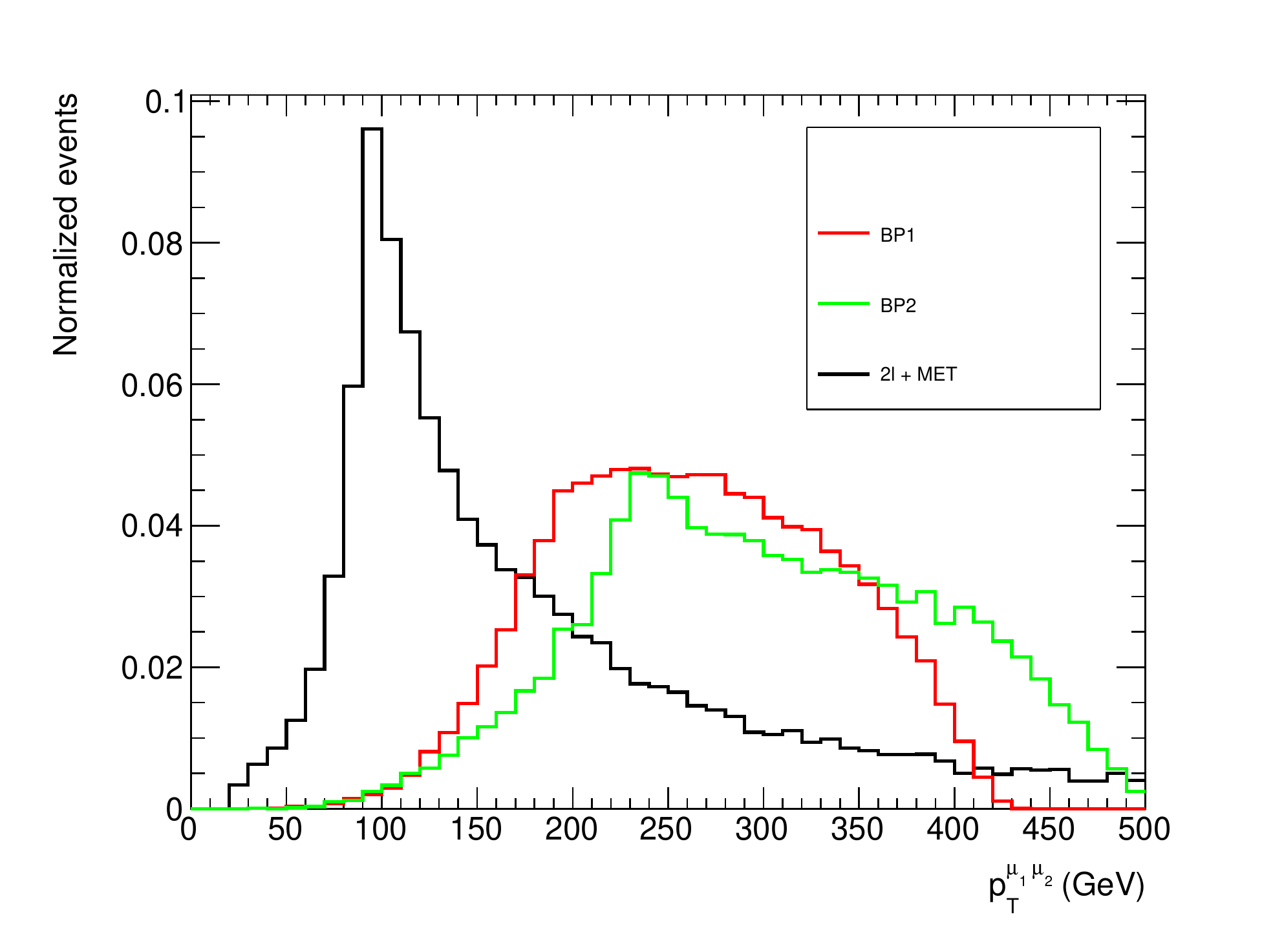}}
\subfigure[]{
\includegraphics[height = 5.5 cm, width = 8 cm]{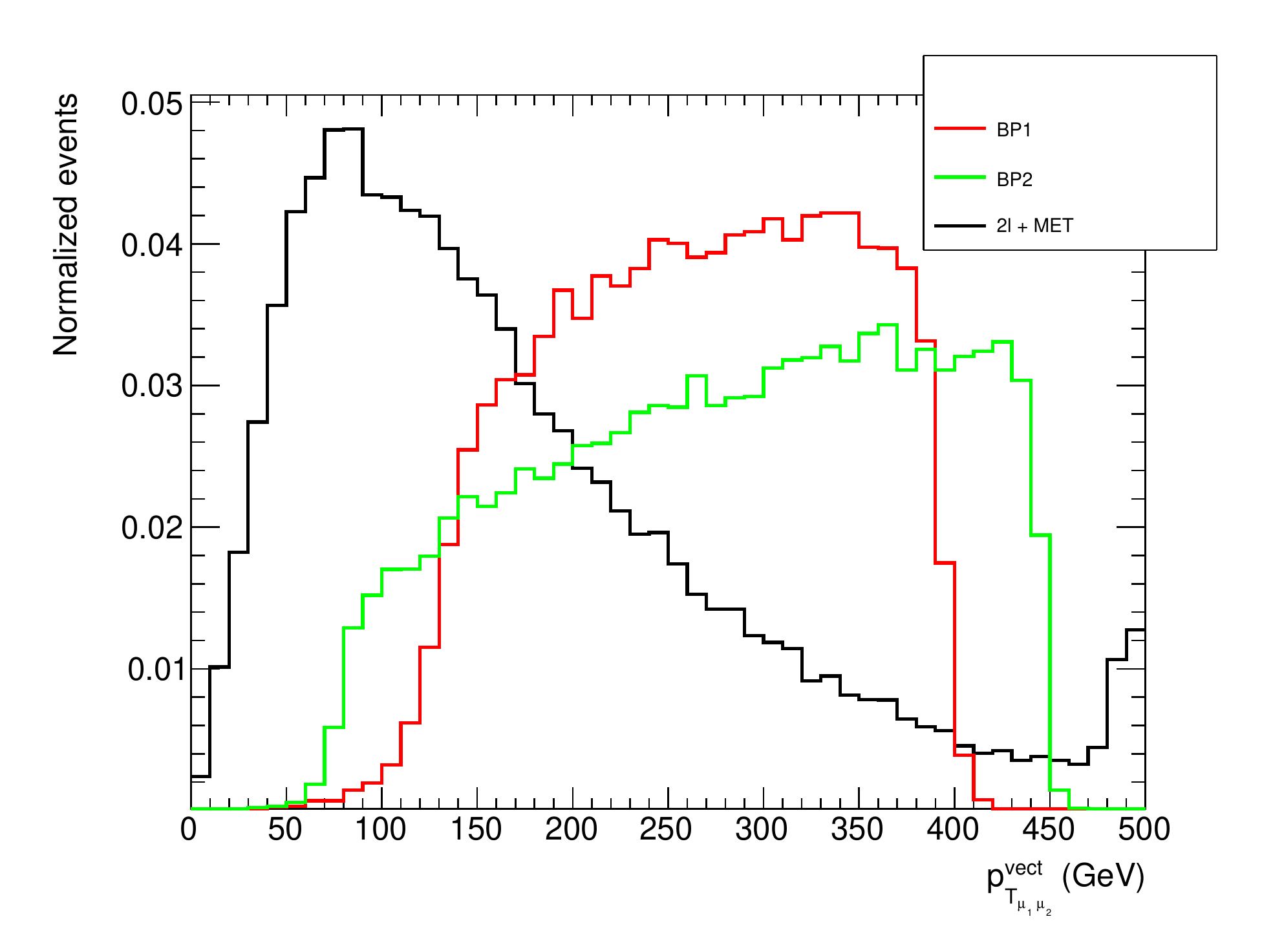}}
} \\
\caption{  The normalized distributions of $\mET, M_{\mu_1 \mu_2}, p_{T}^{\mu_1 \mu_2}, {p_T}_{\mu_1 \mu_2}^{\rm vect}$ for $ 2 \mu + \mET$ channel at 1 TeV.}
\label{distribution-2mumet}
\end{figure}

In Table \ref{tab:cs-SP}, we have tabulated the signal cross-sections (for SP1, SP2) along with the background cross-section for the polarization configurations P3. The normalized distributions of $\mET, M_{\mu_1 \mu_2}, p_{T}^{\mu_1 \mu_2}, {p_T}_{\mu_1 \mu_2}^{\rm vect}$ are shown in Figs. \ref{distribution-2mumet}(a), \ref{distribution-2mumet}(b), \ref{distribution-2mumet}(c), \ref{distribution-2mumet}(d), where $\mu_1$ and $\mu_2$ are $p_T$-ordered muons of the final state. The invariant mass of the di-muon pair for the signal BPs cleanly peaks around the corresponding $M_A$ values. The signals thus have practically no overlap with the background as far as $M_{\ell_1 \ell_2}$ is concerned. 
This is an important point of difference from the $\tau^+ \tau^- + \mET$ signal in the lepton-specific (2+1)HDM in which case
the invariant masses of the $\tau_{\ell_1}\tau_{\ell_2}$, $\tau_{\ell_1}\tau_{h_1}$ and $\tau_{h_1}\tau_{h_2}$ are not directly connected to $M_A$. The $\mET$-spectrum  of the signal BPs in the muon-specific case also differs from the lepton-specific case.
In the former, the only source of missing transverse energy is the DM particle 
$\eta_R$. And this implies a harder $\mET$-spectrum compared to the lepton-specific scenario wherein $\mET$ also draws contribution from the neutrinos coming from $\tau$ decay. Given the $\mu^+\mu^-$ pair comes directly from the pseudoscalar, it gets tagged far more efficiently as opposed to what it would be with the involvement of an intermediate $\tau_\ell \tau_\ell$ pair. In all, $M_{\mu_1 \mu_2}$ and $\mET$ are important observables to distinguish the signal from backgrounds for the muon-specific (2+1)HDM.

\begin{table}
\centering
\begin{tabular}{ |c | c | c | } 
\hline
Signal/Backgrounds & Process & Cross section (fb) for P3 \\ \hline
Signal & & \\
SP1 & $e^+ e^- \to \eta_R \eta_I \to \eta_R \eta_R A \to \mu^+\mu^- + \met$ & 9.74 \\
SP2 & 
 & 2.42 \\  \hline
Background  & $e^+ e^- \to 2\ell + \met$  & 89.34
 \\ \hline
\end{tabular}
\caption{Signal and background cross sections for muon-specific (2+1)HDM at the 1 TeV ILC.}
\label{tab:cs-SP}
\end{table}

\begin{table}[htpb!]
\begin{center}
\begin{tabular}{|c|c|c|c|c|c|}
\hline
 &  \hspace{5mm} {\texttt{NTrees}} \hspace{5mm} & \hspace{5mm} {\texttt{MinNodeSize}} \hspace{5mm} & \hspace{5mm} {\texttt{MaxDepth}}~~ \hspace{5mm} & \hspace{5mm} {\texttt{nCuts}} ~~\hspace{5mm} & \hspace{5mm} {\texttt{KS-score for}}~~\hspace{5mm}\\
 & & & & & {\texttt{Signal(Background)}} \\
\hline
\hline
\hspace{5mm} SP1 \hspace{5mm} & 110 & 4 \% & 2.0 & 55 & 0.401~(0.872) \\ \hline
\hspace{5mm} SP2 \hspace{5mm} & 110 & 4 \% & 2.0 & 55 & 0.9~(0.162) \\ \hline
\end{tabular}
\end{center}
\caption{Tuned BDT parameters for SP1, SP2 for the $2 \mu + \mET$ channel.}
\label{BDT-param-mumumet}
\end{table}

To evaluate the signal significance we use BDTD algorithm and the tuned BDTD parameters along with the KS-scores for signal and background are tabulated in Table \ref{BDT-param-mumumet}.

We use the following kinematic variables for the BDTD analysis.
\bea
M_{\mu_1 \mu_2},~p_{T}^{\mu_1},~p_{T}^{\mu_1 \mu_2},~ {p_T}_{\mu_1 \mu_2}^{\rm vect},~ \Delta R_{\mu_1 \mu_2},~ \mET, ~\eta_{\mu_1}, ~ \Delta \phi_{\mu_1,\mET},~\eta_{\mu_2},~\Delta \phi_{\mu_2,\mET},~ \phi_{\mu_2}, ~ \Delta \eta_{\mu_1 \mu_2}
\eea

\begin{figure}[htpb!]{\centering
\subfigure[]{
\includegraphics[width=3in,height=2.45in]{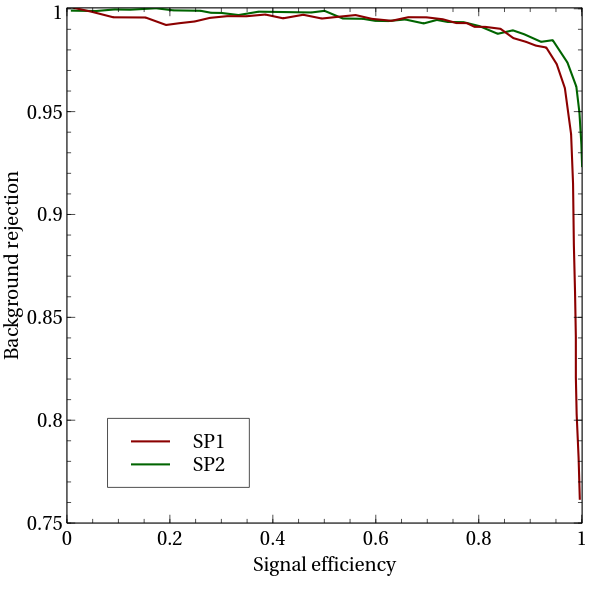}}
\subfigure[]{
\includegraphics[width=3.1in,height=2.46in]{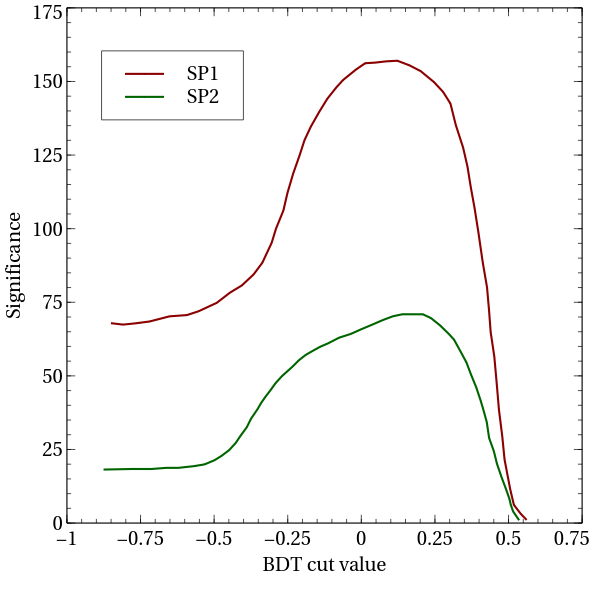}}}
\caption{ (a) ROC curves for chosen benchmark points for $2 \mu  + \mET$ channel. (b) BDT-scores corresponding 
to SP1, SP2 for $2 \mu + \mET$ channel.}
\label{ROC-BDTScore-mumumet}
\end{figure}

Once again, the definition of the variables should be clear from the notation. We have shown the ROC curve and the variation of significance with respect to BDT cut values in Figs.\ref{ROC-BDTScore-mumumet}(a), \ref{ROC-BDTScore-mumumet}(b) respectively for two benchmarks. After carrying out the BDTD analysis, the signal and the background yields at an integrated luminosity 4000 fb$^{-1}$ along with the required luminosity for obtaining $5\sigma$ significance are given in Table \ref{BDTD-mumumet}. The crucial points of differences between the  lepton- and muon-specific analyses implies that the latter should offer a much higher observability than the former. An inspection of Table \ref{BDTD-mumumet} reveals that the luminosity necessary to produce a $5\sigma$ significance is small. The $M_A \simeq$ 230 GeV in SP2 would require a mere 4 fb$^{-1}$ integrated luminosity to get discovered.

\begin{center}
\begin{table}[htb!]
\begin{tabular}{|c|c|c|c|}\hline
 Benchmark Point & Signal Yield & Background Yield & $\mathcal{L}_{5\sigma}$ (fb$^{-1}$)  \\ 
 & at 4000 fb$^{-1}$  &at 4000 fb$^{-1}$ &  \\ \hline 
SP1 & 29939 & 587 & $< 1$ \\ \hline
SP2 &  7336 & 458 & $\sim 4$ \\ \hline
 \end{tabular}
\caption{The signal and background yields at 1 TeV ILC with 4000 fb$^{-1}$ integrated luminosity for SP1,SP2 for the $ e^+ e^- \rightarrow 2 \mu +\mET$ channel after performing 
the BDTD analysis. }
\label{BDTD-mumumet}
\end{table}
\end{center}
\section{Summary and conclusions}
\label{conclusions}

We reprise the (2+1)HDM framework, that is, a 2HDM augmented with an additional scalar doublet. The scenario is endowed with a 
$\mathbb{Z}_2$ symmetry under which the additional doublet is negatively charged.
Thus, the neutral CP-even component of the 
same is rendered cosmologically stable and becomes a potential DM candidate. The contribution to the muon anomalous magnetic moment from the (2+1)HDM has been examined in detail in the previous studies. And it was shown that a muon $g-2$ in the observed ballpark is obtainable in the (2+1)HDM for a much heavier pseudoscalar $A$ that what it would be in the 2HDM. In this work, we look for signatures of such a setup at an $e^+ e^-$ collider operating at $\sqrt{s}$ = 1 TeV with polarized beams. In addition to the canonical lepton-specific Yukawa interactions, we also consider a muon-specific variant in this study. We find that the signal cascades $e^+ e^- \to \eta_R \eta_I \to \eta_R \eta_R A \to \tau^+ \tau^- + \mET$ and $e^+ e^- \to \eta_R \eta_I \to \eta_R \eta_R A \to \mu^+ \mu^- + \mET$ are promising to probe the pseudoscalar $A$ in the lepton- and muon-specific cases respectively.

We have put forth benchmark points that are carefully filtered after applying the relevant constraints. Such constraints include the theoretical restrictions of perturbative unitarity and stability conditions as well as the experimental limits from Higgs signal strengths, oblique parameters and dark matter direct detection. It is ensured that $M_{\eta_I} > M_{\eta_R} + M_A$  for all the benchmarks such that the decay mode $\eta_I \to \eta_R A$ is kinematically open. We have further chosen the polarization configuration ($P_{e^-},P_{e^+} = 80\%R, 30\%L$) in the study since it predicts the maximum signal-to-background ratio. Multivariate analyses are subsequently carried out using the BDTD algorithm to improve the signal significance.

We analyse all three possible decay possibilities of the $\tau^+ \tau^-$ pair, fully leptonic, semileptonic and fully hadronic. The semileptonic and fully hadronic modes predict overwhelmingly better observabilities over the fully leptonic mode. This is expected given the much higher efficiency of tagging a hadronic $\tau$ than a leptonic one. While the semileptonic and  the fully hadronic mode are show competing results, the latter fares better in case of $M_A \simeq$ 400 GeV (BP4), the heaviest pseudoscalar amongst in the benchmarks. For this case, a 5$\sigma$ discovery is expected for 3000 fb$^{-1}$ integrated luminosity. In contrast,
a similar statistical significance in case of the LHC is limited to $M_A <$ 250 GeV in the LHC. 

The $\mu^+ \mu^- + \mET$ channel is \emph{cleaner} final state offering the $\mu^+ \mu^-$ invariant mass as a handle to look for the $A$ directly. When such kinematics is combined with the sizeable production $\mu^+ \mu^- + \mET$ cross sections for the muon-specific signal benchmarks, this channel turns out to be generously promising. We have shown that $M_A$ up to $\simeq$ 230 GeV can be discovered at 5$\sigma$ for an integrated luminosity as low as 4 fb$^{-1}$. This further upholds the prospects of the $e^+ e^-$ machine to probe a leptophillic pseudoscalar.

%
%

%
%
%

\section{Appendix}
\subsection{Expressions of various contributions to $\Delta a_\mu$}
\label{App:A}
\begin{figure}
\centering
\subfigure[]{
\includegraphics[scale=0.37]{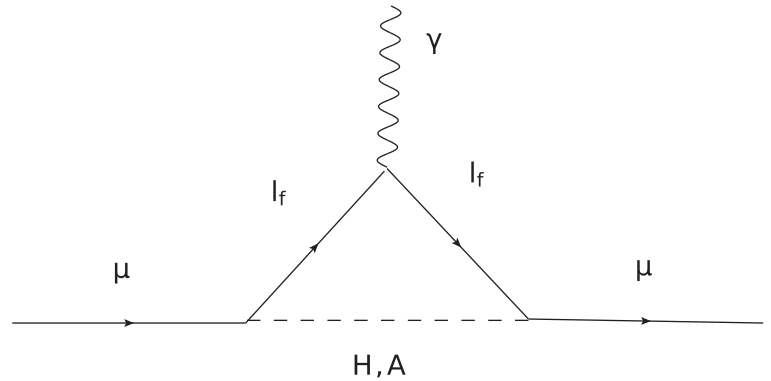}}
\subfigure[]{
\includegraphics[scale=0.34]{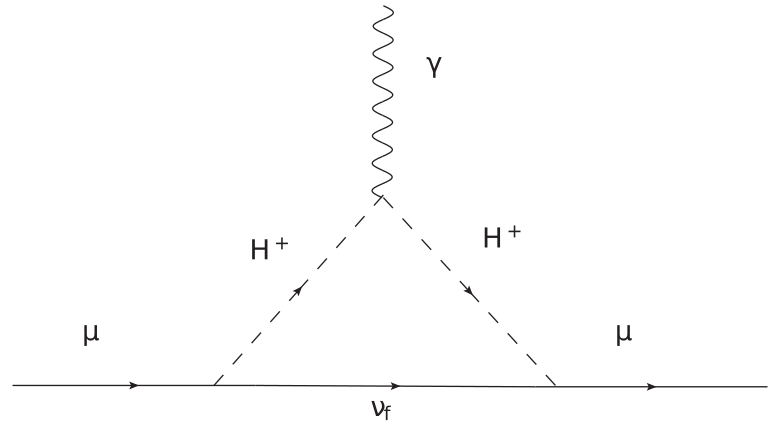}}
\caption{One loop contributions to $\Delta a_\mu$ from (a) $H,A$ and (b) $H^+$in the loop.}
\label{f:delmu_oneloop}
\end{figure}

The numerical expressions for the various 
$\Delta a_\mu$ contributions in the (2+1)HDM are given below. Here the loop order and the particle circulating in the loop are denoted by the superscript and the subscript respectively. The one-loop contributions (shown in Fig.\ref{f:delmu_oneloop}) at the alignment limit look like: 
\besub
\bea
{\Delta a_\mu}_{(H)}^{(1\text{loop})} &=& \frac{M_\mu^2}{8 \pi^2 v^2}  \left(\frac{M_\mu^2}{M_H^2}\right) \big(\xi_\mu^{H} \big)^2~ \int_{0}^{1} dx \frac{x^2(2-x)}{\left(\frac{M_\mu^2}{M_H^2}\right) x^2 -x + 1}, \\
{\Delta a_\mu}_{(A)}^{(1\text{loop})} &=& -\frac{M_\mu^2}{8 \pi^2 v^2}  \left(\frac{M_\mu^2}{M_A^2}\right) \big(\xi_\mu^{A} \big)^2~ \int_{0}^{1} dx \frac{x^3}{\left(\frac{M_\mu^2}{M_A^2}\right) x^2 -x + 1}, \\ 
{\Delta a_\mu}_{(H^+)}^{(1\text{loop})} &=& \frac{M_\mu^2}{8 \pi^2 v^2} \left(\frac{M_\mu^2}{M_{H^+}^2}\right)\big(\xi_\mu^A \big)^2~\int_{0}^{1} dx \frac{x^2(1-x)}{\left(\frac{M_\mu^2}{M_{H^+}^2}\right)x (1-x)-x}.
\eea
\eesub

Numerical evaluation shows that ${\Delta a_\mu}_{(H^+)}^{(1\text{loop})} < 0$. 

\begin{figure}
\centering
\subfigure[]{
\includegraphics[scale=0.43]{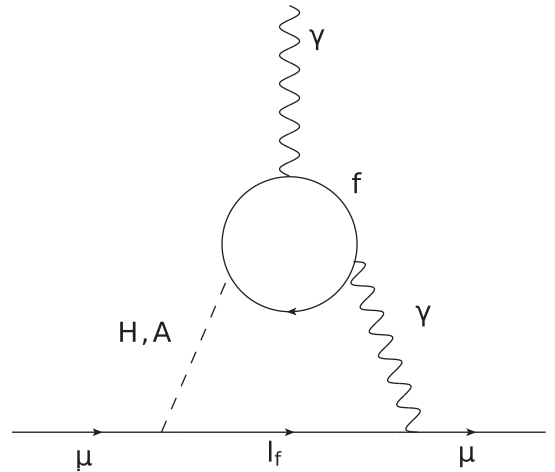}}
\subfigure[]{
\includegraphics[scale=0.11]{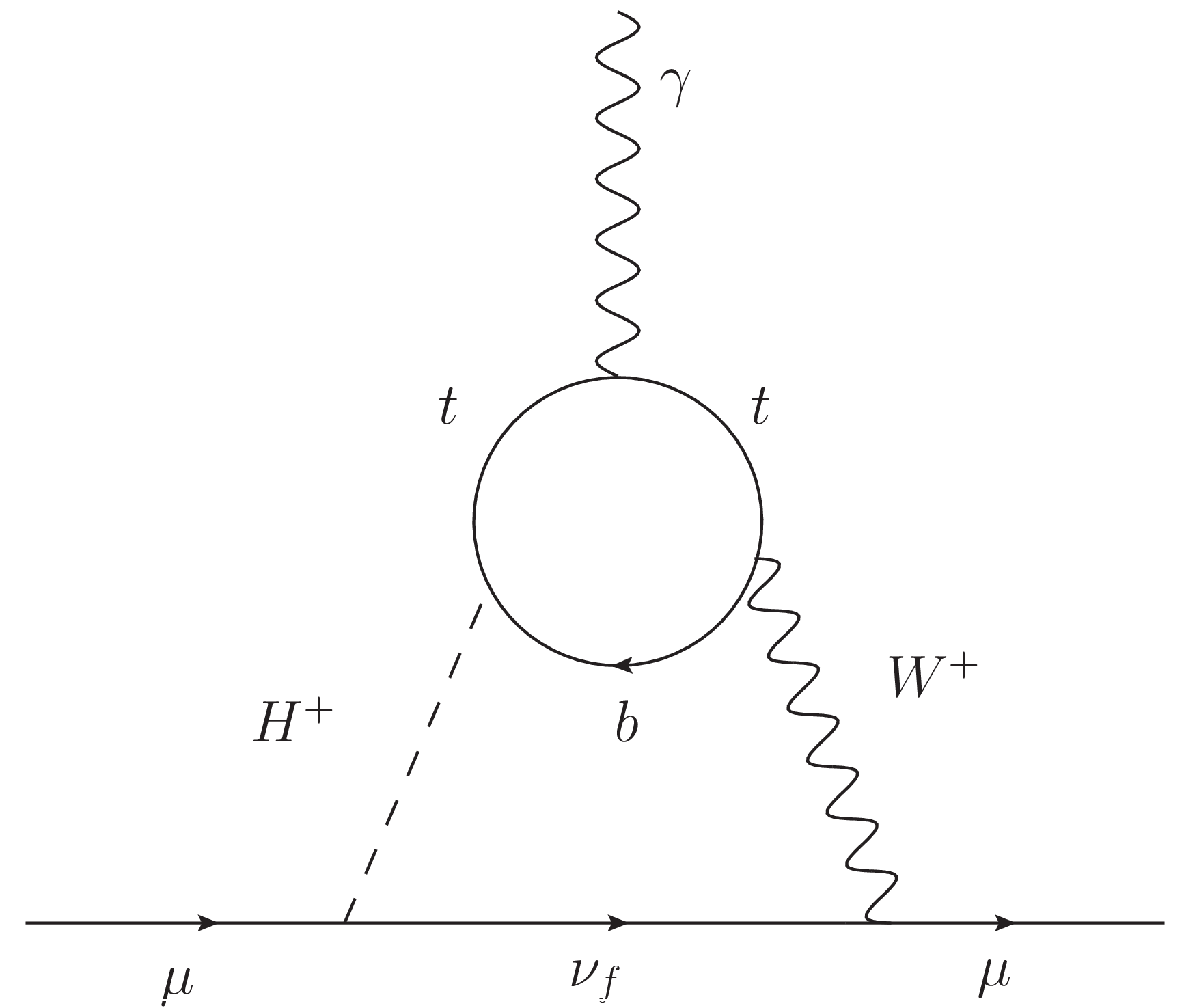}}
\caption{Two loop contributions to $\Delta a_\mu$ from the fermions through (a) an effective $\phi \gamma \gamma$ vertex with $\phi = H,A$ and (b) an effective $H^+ W^- \gamma$ vertex.}
\label{f:delmu_twoloop_f}
\end{figure}

Next let us list out all the relevant two-loop Barr-Zee topologies contributing to $\Delta a_\mu$. First we draw the Feynman diagrams featuring fermions in the one-loop in Fig.\ref{f:delmu_twoloop_f}(a),(b).

Expressions for the corresponding two-loop amplitudes are:
\besub
\bea
{\Delta a_\mu}_{\{f,H \gamma\gamma\}}^{(2\text{loop})} &=& \sum_{f} \frac{\alpha M_\mu^2}{4 \pi^3 v^2}~ N_C^f Q_f^2 \xi_f^{H} \xi_\mu^{H} \mathcal{F}^{(1)}\left(\frac{M_f^2}{M_H^2}\right) \\
{\Delta a_\mu}_{\{f,A \gamma\gamma\}}^{(2\text{loop})} &=& \sum_{f} \frac{\alpha M_\mu^2}{4 \pi^3 v^2}~ N_C^f Q_f^2 \xi_f^A \xi_\mu^A \mathcal{\tilde{F}}^{(1)}\left(\frac{M_f^2}{M_A^2}\right), \\
{\Delta a_\mu}_{\{f,~H^+ W^- \gamma\}}^{(2\text{loop})} &=& \frac{\alpha M_\mu^2 N_t |V_{tb}|^2}{32 \pi^3 s_w^2 v^2 (M_{H^+}^2 - M_W^2)} \int_{0}^{1} dx \left[Q_t x + Q_b(1-x)\right] \nonumber \\
&&
\times \left[\xi^A_d \xi^A_\mu M_b^2 x(1-x) + \xi^A_u \xi^A_\mu M_t^2 x(1+x)\right] \nonumber \\
&& \times \left[\mathcal{G}\left(\frac{M_t^2}{M_{H^+}^2},\frac{M_b^2}{M_{H^+}^2},x\right) - \mathcal{G}\left(\frac{M_t^2}{M_W^2},\frac{M_b^2}{M_W^2},x\right)\right].
\eea
\eesub
Here, $N_C^f$ = 1(3) for leptons (quarks). 

Next we focus on the two-loop amplitudes with 2HDM scalars in the loops as shown in Fig.\ref{f:delmu_twoloop_2HDM}(a),(b) and corresponding amplitudes become :

\begin{figure}
\centering
\subfigure[]{
\includegraphics[scale=0.43]{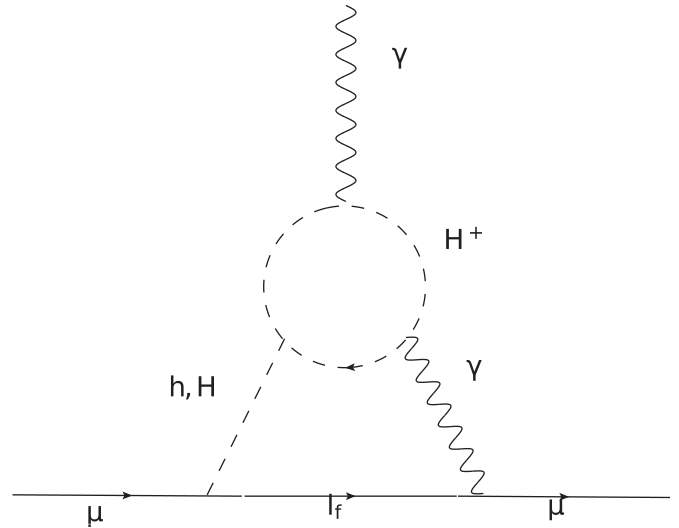}}
\subfigure[]{
\includegraphics[scale=0.43]{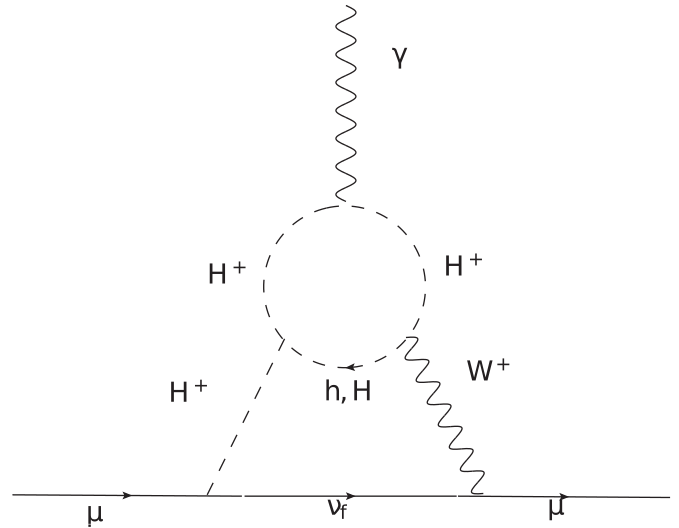}}
\caption{Two loop contributions to $\Delta a_\mu$ from the 2HDM scalars through (a) an effective $S \gamma \gamma$ vertex with $S = h, H$ and (b) an effective $H^+ W^- \gamma$ vertex.}
\label{f:delmu_twoloop_2HDM}
\end{figure}
\besub
\bea
{\Delta a_\mu}_{\{S,~S \gamma\gamma\}}^{(2\text{loop})} &=&  \sum_{S = h, H} \frac{\alpha M_\mu^2}{8 \pi^3 M_{S}^2}~ \xi_\mu^{S}~ \lambda_{S H^+ H^-}\mathcal{F}^{(2)}\left(\frac{M_{H^+}^2}{M_{S}^2}\right), \\
{\Delta a_\mu}_{\{S,~H^+ W^-\gamma\}}^{(2\text{loop})} &=& \frac{\alpha M_\mu^2 }{64 \pi^3 s_w^2 (M_{H^+}^2 - M_W^2)} \sum_{S = h, H} \xi^{S}_\mu ~\lambda_{S H^+ H^-} \int_{0}^{1} dx~x^2 (x-1) \nonumber \\
&&\times \left[\mathcal{G}\left(1,\frac{M_{S}^2}{M_{H^+}^2},x\right) - \mathcal{G}\left(\frac{M_{H^+}^2}{M_W^2},\frac{M_{S}^2}{M_W^2},x\right)\right].
\eea
\eesub
Finally, we depict the contributions from the inert scalars in loop in Fig.\ref{f:delmu_twoloop_inert}(a),(b), with corresponding contributions : 
\begin{figure}
\centering
\subfigure[]{
\includegraphics[scale=0.32]{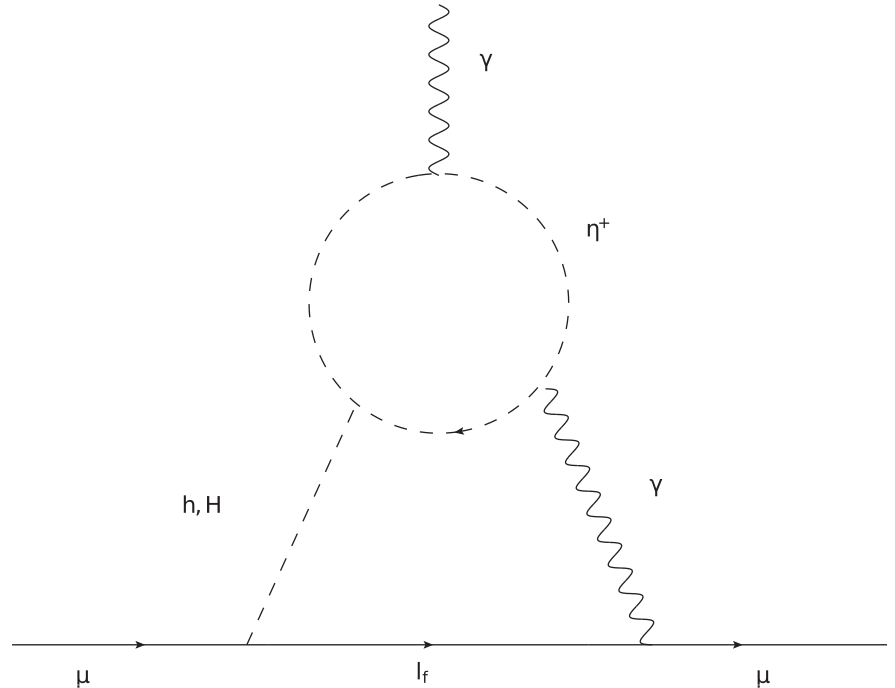}}
\subfigure[]{
\includegraphics[scale=0.32]{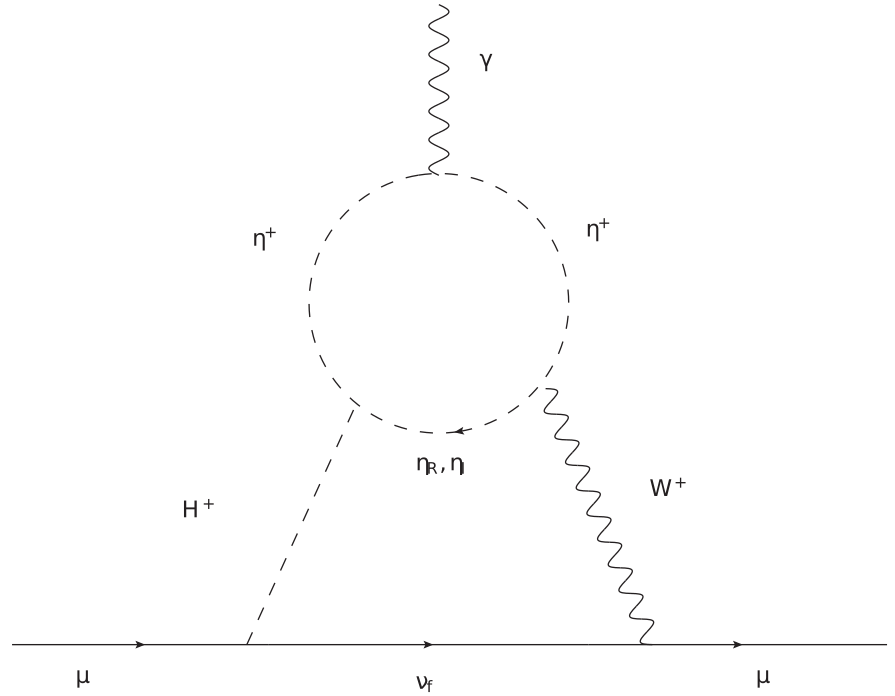}}
\caption{Two loop contributions to $\Delta a_\mu$ from the inert scalars through (a) an effective $S \gamma \gamma$ vertex with $S = h, H$and (b) an effective $H^+ W^- \gamma$ vertex.}
\label{f:delmu_twoloop_inert}
\end{figure}
\besub
\bea
{\Delta a_\mu}_{\{\eta,~S \gamma\gamma \}}^{(2\text{loop})} &=&  \sum_{S = h, H} \frac{\alpha M_\mu^2}{8 \pi^3 M_{\phi}^2}~ \xi_\mu^{S}~ \lambda_{S \eta^+ \eta^-}\mathcal{F}^{(2)}\left(\frac{M_{\eta^+}^2}{M_{S}^2}\right), \\
{\Delta a_\mu}_{\{\eta,~H^+ W^-\gamma \}}^{(2\text{loop})} &=& \frac{\alpha M_\mu^2}
{64 \pi^3 s_w^2 (M_{H^+}^2 - M_W^2)}  
\xi_\mu^A  ~ \lambda_{ H^+ \eta^- \eta_R} \int_{0}^{1} dx~x^2 (x-1) \nonumber \\
&&\times \left[\mathcal{G}\left
(\frac{M_{\eta^+}^2}{M_{H^+}^2},\frac{M_{\eta_R}^2}{M_{H^+}^2},x\right) - \mathcal{G}\left(\frac{M_{\eta^+}^2}{M_W^2},\frac{M_{\eta_R}^2}{M_W^2},x\right)\right] \\ \nonumber 
&&
+ \frac{\alpha M_\mu^2}{64 \pi^3 s_w^2 (M_{H^+}^2 - M_W^2)} \xi^A_\mu  ~ 
\lambda_{ H^+ \eta^- \eta_I} \int_{0}^{1} dx x^2 (x-1) \nonumber \\
&&\times \left[\mathcal{G}\left(\frac{M_{\eta^+}^2}{M_{H^+}^2},\frac{M_{\eta_I}^2}{M_{H^+}^2},x\right) - \mathcal{G}\left(\frac{M_{\eta^+}^2}{M_W^2},\frac{M_{\eta_I}^2}{M_W^2},x\right)\right].
\eea
\eesub

The functions $\mathcal{F}^{(1)} (z), \tilde{\mathcal{F}}^{(1)} (z), \mathcal{F}^{(2)} (z)$ and $\mathcal{G}(z^a,z^b,x)$ can be defined as,
\besub
\bea
\mathcal{F}^{(1)} (z) &=& \frac{z}{2} \int_{0}^{1} dx \frac{2x(1-x)-1}{z-x(1-x)} ~{\rm ln} \left(\frac{z}{x(1-x)}\right), \\
\tilde{\mathcal{F}}^{(1)} (z) &=& \frac{z}{2} \int_{0}^{1} dx \frac{1}{z-x(1-x)} ~{\rm ln} \left(\frac{z}{x(1-x)}\right), \\
\mathcal{F}^{(2)} (z) &=& \frac{1}{2} \int_{0}^{1} dx \frac{x(1-x)}{z-x(1-x)} ~{\rm ln} \left(\frac{z}{x(1-x)}\right), \\
\mathcal{G}(z^a,z^b,x) &=& \frac{{\rm ln} \left(\frac{z^a x + z^b (1-x)}{x(1-x)}\right)}{x(1-x) - z^a x - z^b (1-x)}.
\eea
\eesub

\section{ACKNOWLEDGEMENTS}

IC acknowledges support
from DST, India, under grant number IFA18-PH214 (INSPIRE Faculty Award). NC acknowledges financial support from DST, India, under grant number IFA19-PH237 (INSPIRE Faculty Award).

\bibliographystyle{JHEP}
\bibliography{ref_2plus1_ILC} 

\providecommand{\href}[2]{#2}\begingroup\raggedright\begin{thebibliography}{10}

\bibitem{Peskin:1995ev}
M.~E. Peskin and D.~V. Schroeder, \emph{{An Introduction to quantum field
  theory}}.
\newblock Addison-Wesley, Reading, USA, 1995.

\bibitem{Blum:2013xva}
T.~Blum, A.~Denig, I.~Logashenko, E.~de~Rafael, B.~L. Roberts, T.~Teubner
  et~al., \emph{{The Muon (g-2) Theory Value: Present and Future}},
  \href{http://arxiv.org/abs/1311.2198}{{\tt 1311.2198}}.

\bibitem{RBC:2018dos}
{\scshape RBC, UKQCD} collaboration, T.~Blum, P.~A. Boyle, V.~G\"ulpers,
  T.~Izubuchi, L.~Jin, C.~Jung et~al., \emph{{Calculation of the hadronic
  vacuum polarization contribution to the muon anomalous magnetic moment}},
  \href{http://dx.doi.org/10.1103/PhysRevLett.121.022003}{\emph{Phys. Rev.
  Lett.} {\bf 121} (2018) 022003}, [\href{http://arxiv.org/abs/1801.07224}{{\tt
  1801.07224}}].

\bibitem{Keshavarzi:2018mgv}
A.~Keshavarzi, D.~Nomura and T.~Teubner, \emph{{Muon $g-2$ and $\alpha(M_Z^2)$:
  a new data-based analysis}},
  \href{http://dx.doi.org/10.1103/PhysRevD.97.114025}{\emph{Phys. Rev. D} {\bf
  97} (2018) 114025}, [\href{http://arxiv.org/abs/1802.02995}{{\tt
  1802.02995}}].

\bibitem{Davier:2019can}
M.~Davier, A.~Hoecker, B.~Malaescu and Z.~Zhang, \emph{{A new evaluation of the
  hadronic vacuum polarisation contributions to the muon anomalous magnetic
  moment and to $\mathbf{\boldsymbol\alpha(m_Z^2)}$}},
  \href{http://dx.doi.org/10.1140/epjc/s10052-020-7792-2}{\emph{Eur. Phys. J.
  C} {\bf 80} (2020) 241}, [\href{http://arxiv.org/abs/1908.00921}{{\tt
  1908.00921}}].

\bibitem{Aoyama:2020ynm}
T.~Aoyama et~al., \emph{{The anomalous magnetic moment of the muon in the
  Standard Model}},
  \href{http://dx.doi.org/10.1016/j.physrep.2020.07.006}{\emph{Phys. Rept.}
  {\bf 887} (2020) 1--166}, [\href{http://arxiv.org/abs/2006.04822}{{\tt
  2006.04822}}].

\bibitem{Colangelo:2018mtw}
G.~Colangelo, M.~Hoferichter and P.~Stoffer, \emph{{Two-pion contribution to
  hadronic vacuum polarization}},
  \href{http://dx.doi.org/10.1007/JHEP02(2019)006}{\emph{JHEP} {\bf 02} (2019)
  006}, [\href{http://arxiv.org/abs/1810.00007}{{\tt 1810.00007}}].

\bibitem{Hoferichter:2019mqg}
M.~Hoferichter, B.-L. Hoid and B.~Kubis, \emph{{Three-pion contribution to
  hadronic vacuum polarization}},
  \href{http://dx.doi.org/10.1007/JHEP08(2019)137}{\emph{JHEP} {\bf 08} (2019)
  137}, [\href{http://arxiv.org/abs/1907.01556}{{\tt 1907.01556}}].

\bibitem{Melnikov:2003xd}
K.~Melnikov and A.~Vainshtein, \emph{{Hadronic light-by-light scattering
  contribution to the muon anomalous magnetic moment revisited}},
  \href{http://dx.doi.org/10.1103/PhysRevD.70.113006}{\emph{Phys. Rev. D} {\bf
  70} (2004) 113006}, [\href{http://arxiv.org/abs/hep-ph/0312226}{{\tt
  hep-ph/0312226}}].

\bibitem{Hoferichter:2018kwz}
M.~Hoferichter, B.-L. Hoid, B.~Kubis, S.~Leupold and S.~P. Schneider,
  \emph{{Dispersion relation for hadronic light-by-light scattering: pion
  pole}}, \href{http://dx.doi.org/10.1007/JHEP10(2018)141}{\emph{JHEP} {\bf 10}
  (2018) 141}, [\href{http://arxiv.org/abs/1808.04823}{{\tt 1808.04823}}].

\bibitem{Blum:2019ugy}
T.~Blum, N.~Christ, M.~Hayakawa, T.~Izubuchi, L.~Jin, C.~Jung et~al.,
  \emph{{Hadronic Light-by-Light Scattering Contribution to the Muon Anomalous
  Magnetic Moment from Lattice QCD}},
  \href{http://dx.doi.org/10.1103/PhysRevLett.124.132002}{\emph{Phys. Rev.
  Lett.} {\bf 124} (2020) 132002}, [\href{http://arxiv.org/abs/1911.08123}{{\tt
  1911.08123}}].

\bibitem{ParticleDataGroup:2020ssz}
{\scshape Particle Data Group} collaboration, P.~A. Zyla et~al., \emph{{Review
  of Particle Physics}},
  \href{http://dx.doi.org/10.1093/ptep/ptaa104}{\emph{PTEP} {\bf 2020} (2020)
  083C01}.

\bibitem{Muong-2:2006rrc}
{\scshape Muon g-2} collaboration, G.~W. Bennett et~al., \emph{{Final Report of
  the Muon E821 Anomalous Magnetic Moment Measurement at BNL}},
  \href{http://dx.doi.org/10.1103/PhysRevD.73.072003}{\emph{Phys. Rev. D} {\bf
  73} (2006) 072003}, [\href{http://arxiv.org/abs/hep-ex/0602035}{{\tt
  hep-ex/0602035}}].

\bibitem{Muong-2:2021ojo}
{\scshape Muon g-2} collaboration, B.~Abi et~al., \emph{{Measurement of the
  Positive Muon Anomalous Magnetic Moment to 0.46 ppm}},
  \href{http://dx.doi.org/10.1103/PhysRevLett.126.141801}{\emph{Phys. Rev.
  Lett.} {\bf 126} (2021) 141801}, [\href{http://arxiv.org/abs/2104.03281}{{\tt
  2104.03281}}].

\bibitem{Branco:2011iw}
G.~C. Branco, P.~M. Ferreira, L.~Lavoura, M.~N. Rebelo, M.~Sher and J.~P.
  Silva, \emph{{Theory and phenomenology of two-Higgs-doublet models}},
  \href{http://dx.doi.org/10.1016/j.physrep.2012.02.002}{\emph{Phys. Rept.}
  {\bf 516} (2012) 1--102}, [\href{http://arxiv.org/abs/1106.0034}{{\tt
  1106.0034}}].

\bibitem{Broggio:2014mna}
A.~Broggio, E.~J. Chun, M.~Passera, K.~M. Patel and S.~K. Vempati,
  \emph{{Limiting two-Higgs-doublet models}},
  \href{http://dx.doi.org/10.1007/JHEP11(2014)058}{\emph{JHEP} {\bf 11} (2014)
  058}, [\href{http://arxiv.org/abs/1409.3199}{{\tt 1409.3199}}].

\bibitem{Cao:2009as}
J.~Cao, P.~Wan, L.~Wu and J.~M. Yang, \emph{{Lepton-Specific Two-Higgs Doublet
  Model: Experimental Constraints and Implication on Higgs Phenomenology}},
  \href{http://dx.doi.org/10.1103/PhysRevD.80.071701}{\emph{Phys. Rev. D} {\bf
  80} (2009) 071701}, [\href{http://arxiv.org/abs/0909.5148}{{\tt 0909.5148}}].

\bibitem{Wang:2014sda}
L.~Wang and X.-F. Han, \emph{{A light pseudoscalar of 2HDM confronted with muon
  g-2 and experimental constraints}},
  \href{http://dx.doi.org/10.1007/JHEP05(2015)039}{\emph{JHEP} {\bf 05} (2015)
  039}, [\href{http://arxiv.org/abs/1412.4874}{{\tt 1412.4874}}].

\bibitem{Han:2015yys}
T.~Han, S.~K. Kang and J.~Sayre, \emph{{Muon $g-2$ in the aligned two Higgs
  doublet model}}, \href{http://dx.doi.org/10.1007/JHEP02(2016)097}{\emph{JHEP}
  {\bf 02} (2016) 097}, [\href{http://arxiv.org/abs/1511.05162}{{\tt
  1511.05162}}].

\bibitem{Ilisie:2015tra}
V.~Ilisie, \emph{{New Barr-Zee contributions to $\mathbf{(g-2)_\mu}$ in
  two-Higgs-doublet models}},
  \href{http://dx.doi.org/10.1007/JHEP04(2015)077}{\emph{JHEP} {\bf 04} (2015)
  077}, [\href{http://arxiv.org/abs/1502.04199}{{\tt 1502.04199}}].

\bibitem{Abe:2015oca}
T.~Abe, R.~Sato and K.~Yagyu, \emph{{Lepton-specific two Higgs doublet model as
  a solution of muon g \ensuremath{-} 2 anomaly}},
  \href{http://dx.doi.org/10.1007/JHEP07(2015)064}{\emph{JHEP} {\bf 07} (2015)
  064}, [\href{http://arxiv.org/abs/1504.07059}{{\tt 1504.07059}}].

\bibitem{Crivellin:2015hha}
A.~Crivellin, J.~Heeck and P.~Stoffer, \emph{{A perturbed lepton-specific
  two-Higgs-doublet model facing experimental hints for physics beyond the
  Standard Model}},
  \href{http://dx.doi.org/10.1103/PhysRevLett.116.081801}{\emph{Phys. Rev.
  Lett.} {\bf 116} (2016) 081801}, [\href{http://arxiv.org/abs/1507.07567}{{\tt
  1507.07567}}].

\bibitem{Chun:2016hzs}
E.~J. Chun and J.~Kim, \emph{{Leptonic Precision Test of Leptophilic
  Two-Higgs-Doublet Model}},
  \href{http://dx.doi.org/10.1007/JHEP07(2016)110}{\emph{JHEP} {\bf 07} (2016)
  110}, [\href{http://arxiv.org/abs/1605.06298}{{\tt 1605.06298}}].

\bibitem{Cherchiglia:2016eui}
A.~Cherchiglia, P.~Kneschke, D.~St\"ockinger and H.~St\"ockinger-Kim,
  \emph{{The muon magnetic moment in the 2HDM: complete two-loop result}},
  \href{http://dx.doi.org/10.1007/JHEP10(2021)242}{\emph{JHEP} {\bf 01} (2017)
  007}, [\href{http://arxiv.org/abs/1607.06292}{{\tt 1607.06292}}].

\bibitem{Han:2018znu}
X.-F. Han, T.~Li, L.~Wang and Y.~Zhang, \emph{{Simple interpretations of lepton
  anomalies in the lepton-specific inert two-Higgs-doublet model}},
  \href{http://dx.doi.org/10.1103/PhysRevD.99.095034}{\emph{Phys. Rev. D} {\bf
  99} (2019) 095034}, [\href{http://arxiv.org/abs/1812.02449}{{\tt
  1812.02449}}].

\bibitem{Chun:2019oix}
E.~J. Chun, J.~Kim and T.~Mondal, \emph{{Electron EDM and Muon anomalous
  magnetic moment in Two-Higgs-Doublet Models}},
  \href{http://dx.doi.org/10.1007/JHEP12(2019)068}{\emph{JHEP} {\bf 12} (2019)
  068}, [\href{http://arxiv.org/abs/1906.00612}{{\tt 1906.00612}}].

\bibitem{Dey:2021pyn}
A.~Dey, J.~Lahiri and B.~Mukhopadhyaya, \emph{{Muon g-2 and a type-X two Higgs
  doublet scenario: some studies in high-scale validity}},
  \href{http://arxiv.org/abs/2106.01449}{{\tt 2106.01449}}.

\bibitem{Chowdhury:2017aav}
D.~Chowdhury and O.~Eberhardt, \emph{{Update of Global Two-Higgs-Doublet Model
  Fits}}, \href{http://dx.doi.org/10.1007/JHEP05(2018)161}{\emph{JHEP} {\bf 05}
  (2018) 161}, [\href{http://arxiv.org/abs/1711.02095}{{\tt 1711.02095}}].

\bibitem{Wang:2018hnw}
L.~Wang, J.~M. Yang, M.~Zhang and Y.~Zhang, \emph{{Revisiting lepton-specific
  2HDM in light of muon $g−2$ anomaly}},
  \href{http://dx.doi.org/10.1016/j.physletb.2018.11.045}{\emph{Phys. Lett. B}
  {\bf 788} (2019) 519--529}, [\href{http://arxiv.org/abs/1809.05857}{{\tt
  1809.05857}}].

\bibitem{Chun:2017yob}
E.~J. Chun, S.~Dwivedi, T.~Mondal and B.~Mukhopadhyaya, \emph{{Reconstructing a
  light pseudoscalar in the Type-X Two Higgs Doublet Model}},
  \href{http://dx.doi.org/10.1016/j.physletb.2017.09.037}{\emph{Phys. Lett. B}
  {\bf 774} (2017) 20--25}, [\href{http://arxiv.org/abs/1707.07928}{{\tt
  1707.07928}}].

\bibitem{Chun:2018vsn}
E.~J. Chun, S.~Dwivedi, T.~Mondal, B.~Mukhopadhyaya and S.~K. Rai,
  \emph{{Reconstructing heavy Higgs boson masses in a type X two-Higgs-doublet
  model with a light pseudoscalar particle}},
  \href{http://dx.doi.org/10.1103/PhysRevD.98.075008}{\emph{Phys. Rev. D} {\bf
  98} (2018) 075008}, [\href{http://arxiv.org/abs/1807.05379}{{\tt
  1807.05379}}].

\bibitem{CMS:2018qvj}
{\scshape CMS} collaboration, A.~M. Sirunyan et~al., \emph{{Search for an
  exotic decay of the Higgs boson to a pair of light pseudoscalars in the final
  state of two muons and two $\tau$ leptons in proton-proton collisions at $
  \sqrt{s}=13 $ TeV}},
  \href{http://dx.doi.org/10.1007/JHEP11(2018)018}{\emph{JHEP} {\bf 11} (2018)
  018}, [\href{http://arxiv.org/abs/1805.04865}{{\tt 1805.04865}}].

\bibitem{Chakrabarty:2021ztf}
N.~Chakrabarty, \emph{{Muon g-2 in a Type-X 2HDM assisted by inert scalars:
  probing at the LHC}},  \href{http://arxiv.org/abs/2112.13126}{{\tt
  2112.13126}}.

\bibitem{Johansen:2015nxa}
A.~R. Johansen and M.~Sher, \emph{{Electron/muon specific two Higgs doublet
  model at $e^{+}e^-$ colliders}},
  \href{http://dx.doi.org/10.1103/PhysRevD.91.054021}{\emph{Phys. Rev. D} {\bf
  91} (2015) 054021}, [\href{http://arxiv.org/abs/1502.00516}{{\tt
  1502.00516}}].

\bibitem{Kajiyama:2013sza}
Y.~Kajiyama, H.~Okada and K.~Yagyu, \emph{{Electron/Muon Specific Two Higgs
  Doublet Model}},
  \href{http://dx.doi.org/10.1016/j.nuclphysb.2014.08.009}{\emph{Nucl. Phys. B}
  {\bf 887} (2014) 358--370}, [\href{http://arxiv.org/abs/1309.6234}{{\tt
  1309.6234}}].

\bibitem{Abe:2017jqo}
T.~Abe, R.~Sato and K.~Yagyu, \emph{{Muon specific two-Higgs-doublet model}},
  \href{http://dx.doi.org/10.1007/JHEP07(2017)012}{\emph{JHEP} {\bf 07} (2017)
  012}, [\href{http://arxiv.org/abs/1705.01469}{{\tt 1705.01469}}].

\bibitem{10.1093/ptep/ptaa104}
P.~D. Group, P.~A. Zyla, R.~M. Barnett, J.~Beringer, O.~Dahl, D.~A. Dwyer
  et~al., \emph{{Review of Particle Physics}},
  \href{http://dx.doi.org/10.1093/ptep/ptaa104}{\emph{Progress of Theoretical
  and Experimental Physics} {\bf 2020} (08, 2020) },
  [\href{http://arxiv.org/abs/https://academic.oup.com/ptep/article-pdf/2020/8/083C01/34673722/ptaa104.pdf}{{\tt
  https://academic.oup.com/ptep/article-pdf/2020/8/083C01/34673722/ptaa104.pdf}}].

\bibitem{Aghanim:2018eyx}
{\scshape Planck} collaboration, N.~Aghanim et~al., \emph{{Planck 2018 results.
  VI. Cosmological parameters}},
  \href{http://dx.doi.org/10.1051/0004-6361/201833910}{\emph{Astron.
  Astrophys.} {\bf 641} (2020) A6},
  [\href{http://arxiv.org/abs/1807.06209}{{\tt 1807.06209}}].

\bibitem{Semenov:2008jy}
A.~Semenov, \emph{{LanHEP: A Package for the automatic generation of Feynman
  rules in field theory. Version 3.0}},
  \href{http://dx.doi.org/10.1016/j.cpc.2008.10.012}{\emph{Comput. Phys.
  Commun.} {\bf 180} (2009) 431--454},
  [\href{http://arxiv.org/abs/0805.0555}{{\tt 0805.0555}}].

\bibitem{Belanger:2018ccd}
G.~B\'elanger, F.~Boudjema, A.~Goudelis, A.~Pukhov and B.~Zaldivar,
  \emph{{micrOMEGAs5.0 : Freeze-in}},
  \href{http://dx.doi.org/10.1016/j.cpc.2018.04.027}{\emph{Comput. Phys.
  Commun.} {\bf 231} (2018) 173--186},
  [\href{http://arxiv.org/abs/1801.03509}{{\tt 1801.03509}}].

\bibitem{Aprile:2018dbl}
{\scshape XENON} collaboration, E.~Aprile et~al., \emph{{Dark Matter Search
  Results from a One Ton-Year Exposure of XENON1T}},
  \href{http://dx.doi.org/10.1103/PhysRevLett.121.111302}{\emph{Phys. Rev.
  Lett.} {\bf 121} (2018) 111302}, [\href{http://arxiv.org/abs/1805.12562}{{\tt
  1805.12562}}].

\bibitem{Behnke:2013lya}
H.~Abramowicz et~al., \emph{{The International Linear Collider Technical Design
  Report - Volume 4: Detectors}},  \href{http://arxiv.org/abs/1306.6329}{{\tt
  1306.6329}}.

\bibitem{Alloul:2013bka}
A.~Alloul, N.~D. Christensen, C.~Degrande, C.~Duhr and B.~Fuks,
  \emph{{FeynRules 2.0 - A complete toolbox for tree-level phenomenology}},
  \href{http://dx.doi.org/10.1016/j.cpc.2014.04.012}{\emph{Comput. Phys.
  Commun.} {\bf 185} (2014) 2250--2300},
  [\href{http://arxiv.org/abs/1310.1921}{{\tt 1310.1921}}].

\bibitem{Alwall:2014hca}
J.~Alwall, R.~Frederix, S.~Frixione, V.~Hirschi, F.~Maltoni, O.~Mattelaer
  et~al., \emph{{The automated computation of tree-level and next-to-leading
  order differential cross sections, and their matching to parton shower
  simulations}}, \href{http://dx.doi.org/10.1007/JHEP07(2014)079}{\emph{JHEP}
  {\bf 07} (2014) 079}, [\href{http://arxiv.org/abs/1405.0301}{{\tt
  1405.0301}}].

\bibitem{Frederix:2018nkq}
R.~Frederix, S.~Frixione, V.~Hirschi, D.~Pagani, H.~S. Shao and M.~Zaro,
  \emph{{The automation of next-to-leading order electroweak calculations}},
  \href{http://dx.doi.org/10.1007/JHEP11(2021)085}{\emph{JHEP} {\bf 07} (2018)
  185}, [\href{http://arxiv.org/abs/1804.10017}{{\tt 1804.10017}}].

\bibitem{Sjostrand:2014zea}
T.~Sjöstrand, S.~Ask, J.~R. Christiansen, R.~Corke, N.~Desai, P.~Ilten et~al.,
  \emph{{An Introduction to PYTHIA 8.2}},
  \href{http://dx.doi.org/10.1016/j.cpc.2015.01.024}{\emph{Comput. Phys.
  Commun.} {\bf 191} (2015) 159--177},
  [\href{http://arxiv.org/abs/1410.3012}{{\tt 1410.3012}}].

\bibitem{deFavereau:2013fsa}
{\scshape DELPHES 3} collaboration, J.~de~Favereau, C.~Delaere, P.~Demin,
  A.~Giammanco, V.~Lemaître, A.~Mertens et~al., \emph{{DELPHES 3, A modular
  framework for fast simulation of a generic collider experiment}},
  \href{http://dx.doi.org/10.1007/JHEP02(2014)057}{\emph{JHEP} {\bf 02} (2014)
  057}, [\href{http://arxiv.org/abs/1307.6346}{{\tt 1307.6346}}].

\bibitem{Hocker:2007ht}
A.~Hocker et~al., \emph{{TMVA - Toolkit for Multivariate Data Analysis}},
  \href{http://arxiv.org/abs/physics/0703039}{{\tt physics/0703039}}.

\bibitem{Adhikary:2020cli}
A.~Adhikary, N.~Chakrabarty, I.~Chakraborty and J.~Lahiri, \emph{{Probing the
  $H^\pm W^{\mp } Z$ interaction at the high energy upgrade of the LHC}},
  \href{http://dx.doi.org/10.1140/epjc/s10052-021-09335-x}{\emph{Eur. Phys. J.
  C} {\bf 81} (2021) 554}, [\href{http://arxiv.org/abs/2010.14547}{{\tt
  2010.14547}}].

\bibitem{Hou:2022nyh}
W.-s. Hou, R.~Jain and C.~Kao, \emph{{Searching for extra Higgs bosons via
  $pp\to H,A\to \tau\mu, \tau\tau $ at the Large Hadron Collider}},
  \href{http://arxiv.org/abs/2202.04336}{{\tt 2202.04336}}.

\bibitem{Ellis:1987xu}
R.~K. Ellis, I.~Hinchliffe, M.~Soldate and J.~J. van~der Bij, \emph{{Higgs
  Decay to tau+ tau-: A Possible Signature of Intermediate Mass Higgs Bosons at
  the SSC}}, \href{http://dx.doi.org/10.1016/0550-3213(88)90019-3}{\emph{Nucl.
  Phys. B} {\bf 297} (1988) 221--243}.

\end{thebibliography}\endgroup
\end{document}